\documentclass[prb,twocolumn,superscriptaddress]{revtex4-2}
\usepackage{graphicx}% Include figure files
\usepackage{subfigure}
\usepackage{dcolumn}% Align table columns on decimal point
\usepackage{bm}% bold math
\usepackage{color}
\usepackage{amsmath, amsthm, amssymb,stmaryrd}
\usepackage{braket}
\usepackage[polish,french,english]{babel}
\usepackage{natbib}
\usepackage{eqnarray}

\newcommand{\hto}[1]{Ho$_2$Ti$_2$O$_7${#1}}
\newcommand{\dto}[1]{Dy$_2$Ti$_2$O$_7${#1}}

\usepackage{gensymb}
\usepackage{amssymb}
\usepackage{epstopdf}

\begin{document}

\title{Tunable critical correlations in kagome ice}
\author{A. A. Turrini}
\affiliation{Laboratory for Neutron Scattering and Imaging, Paul Scherrer Institut, 5232 Villigen PSI, Switzerland}
\affiliation{Department of Quantum Matter Physics (DQMP), 24 Quai Ernest-Ansermet, CH-1211 Gen{\`e}ve 4, Switzerland}

\author{A. Harman-Clarke}
\affiliation{London Centre for Nanotechnology and Department of Physics and Astronomy, University College London, 17-19 Gordon Street, London WC1H 0AH, United Kingdom}
\affiliation{Universit\'e de Lyon, ENS de Lyon, Universit\'e Claude Bernard, CNRS, Laboratoire de Physique, F-69342 Lyon, France}

\author{T. Fennell}
\email{tom.fennell@psi.ch}
\affiliation{Laboratory for Neutron Scattering and Imaging, Paul Scherrer Institut, 5232 Villigen PSI, Switzerland}

\author{I. G. Wood}
\affiliation{Department of Earth Sciences, University College London, WC1E 6BT London, United Kingdom}

\author{P. Henelius} 
\affiliation{Department of Physics, Royal Institute of Technology, SE-106 91 Stockholm, Sweden}
\affiliation{Faculty of Science and Engineering,  \r{A}bo Akademi University, \r{A}bo, Finland}

\author{S. T. Bramwell}
\affiliation{London Centre for Nanotechnology and Department of Physics and Astronomy, University College London, 17-19 Gordon Street, London WC1H 0AH, United Kingdom}

\author{P. C. W. Holdsworth}
\affiliation{Universit\'e de Lyon, ENS de Lyon, Universit\'e Claude Bernard, CNRS, Laboratoire de Physique, F-69342 Lyon, France}

\date{\today}
                                            
\begin{abstract}

 We present a comprehensive experimental and theoretical study of the kagome ice Coulomb phase, that explores the fine tuning of critical correlations by applied field, temperature and crystal orientation. The continuous modification of algebraic correlations is observed by polarised neutron scattering experiments and is found to be well described by numerical simulations of an idealised model. We further clarify the thermodynamics of field tuned Kasteleyn transitions and demonstrate some dramatic finite size scaling properties that depend on how topological string defects wind around the system boundaries.  We conclude that kagome ice is a remarkable example of a critical and topological state in a real system that may be subject to fine experimental control.

%The kagome ice state is a two dimensional critical state of algebraic spin correlations formed by the application of a moderate magnetic field along the cubic $[1 1 1]$ direction of a pyrochlore spin ice. Tilts of the field away from perfect alignment allow for tuning of these algebraic correlations by variations of tilt angles, field or temperature, leading to symmetry-sustaining Kasteleyn transitions. We present a comprehensive study of the tunable critical correlations in kagome ice  by means of numerical simulations on an idealised model, polarized neutron scattering experiments on ${\rm Ho_2Ti_2O_7}$, and thermodynamic arguments, all of which we compare with existing analytical theory. Although we find some discrepancies with the latter, the essential physics of kagome ice is clearly identified, with continuous modification of algebraic correlations leading to the drift and anisotropic scaling of diffuse scattering features with respect to the strength and tilt of the magnetic field, and the temperature. Our results particularly highlight the role of topological string defects in the critical state and reveal some dramatic finite size scaling properties that depend on how these strings wind round the system boundaries. We further discuss the breakdown of the algebraic phase by the excitation of magnetic monopoles and by other departures from ideality, such as dipole interactions and demagnetizing fields. In general we find that kagome ice is a remarkable example of a critical and topological state in a real system that may be subject to fine experimental control. 
\end{abstract}

\pacs{}
\maketitle

\section{\label{sec:Introduction}Introduction}

\begin{figure*}
\centering
\includegraphics[width=0.7\textwidth,trim=1 1 1 1]{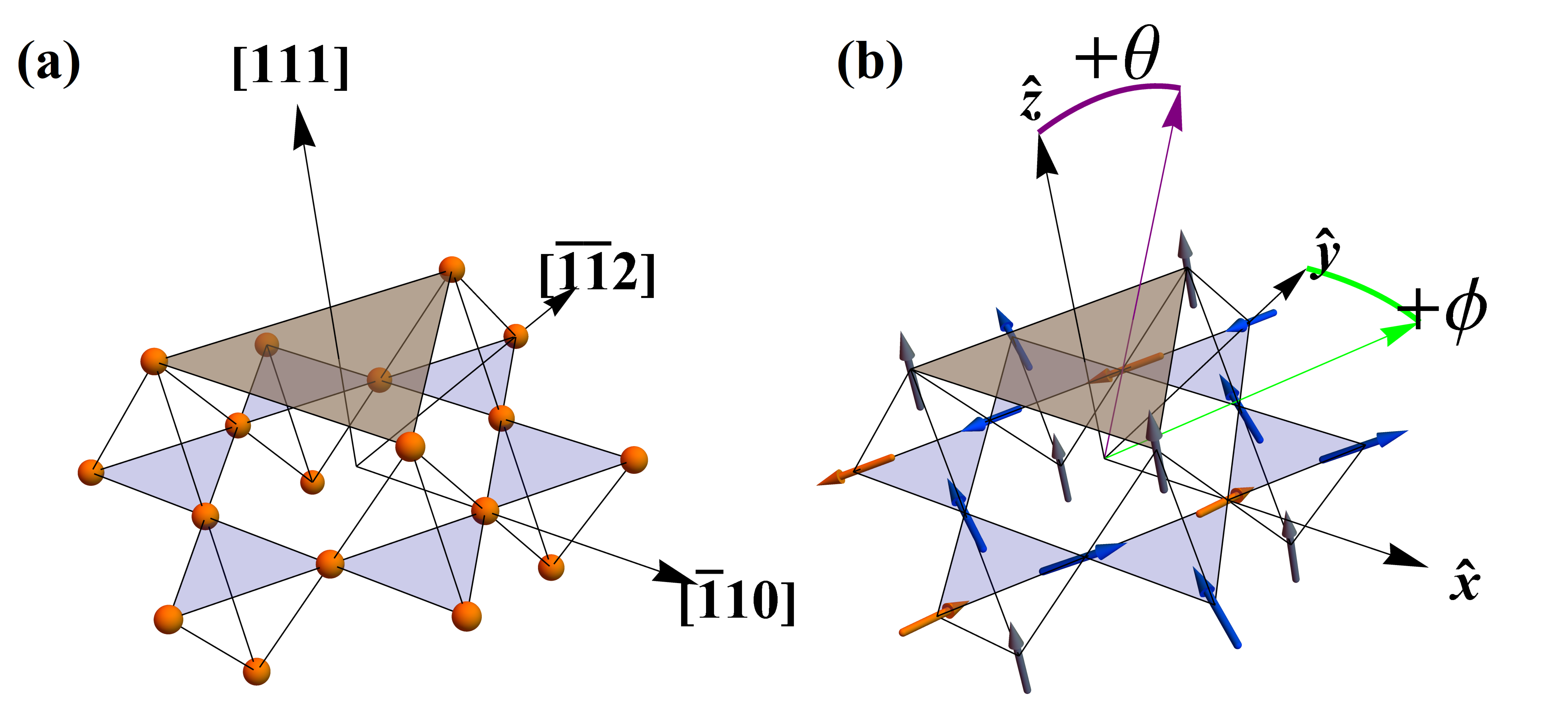}
\caption{\label{fig:KI} (a) A section of the pyrochlore lattice showing the alternating kagome (blue) and triangular (gray) lattices and associated crystallographic axes.  (b) An example of an ice-rule satisfying configuration of spins shows the field-pinned spins on the triangular layers (grey) and spins on the kagome layer that are oriented favorably (blue) and unfavorably (orange) with respect to the applied field.  The three-dimensional ice rule (`two-in--two-out' on each tetrahedron) and two-dimensional (`two-in--one-out' or vice versa on each triangle) local ordering rule are both obeyed. The angles $\theta > 0$ and $\phi > 0$ are shown in purple and green respectively.
}
\end{figure*}

%Tunable two dimensional magnetic and superconducting systems -- from high conductivity graphene layers~\cite{Novoselov2004} to WSe$_2$ quantum emitters~\cite{Hei2015,Koperski2015,Chakraborty2015} -- are highly sought after in condensed matter {\color{cyan}{1) are they? why?}}. A magnetic system with tunable two dimensional correlations can be realized in the geometrically frustrated spin ice pyrochlores \dto{} and \hto{}.  {\color{cyan}{2) What quantities are tunable in those other systems and how, if at all, is kagome ice related or relevant to them?}}.

\subsection{Context}

Kagome ice~\cite{Harris1998,Matsuhira2002} is a quasi-two dimensional magnetic state with finite configurational entropy and algebraic correlations, which is formed when a magnetic field is applied along the cubic $[111]$ direction of a spin ice like \hto{} and \dto{}. Theory~\cite{Moessner2003} predicts kagome ice to be a topologically constrained Coulomb phase that, through small tilts of the applied field, can be tuned toward lines of unconventional Kasteleyn transitions, with associated anisotropic algebraic scaling.  The physics of kagome ice is extremely rich and subtle, exemplifying departures from the usual Landau-Ginzburg-Wilson (LGW) paradigm of continuous phase transitions in magnetism, towards alternative paradigms of topological constraint~\cite{Macdonald2011}, hitherto only observed in soft matter~\cite{Nagle1973}. Phase transitions and scaling in such topologically constrained systems are of great interest as they present new challenges to theory~\cite{Alet2006,Powell:2011hda,baxter}.  Yet experimental model systems are quite scarce and, in this sense, kagome ice, being a very clean and well defined magnetic state that is easily controlled by applied field, is a most valuable example.

Over the years there have been extensive experimental investigations of the thermodynamic properties of kagome ice \cite{Matsuhira2002,Sakakibara2003, Fukazawa2002, Hiroi2003111, Aoki2004} as well as some neutron scattering studies of correlations~\cite{Tabata2006,Fennell2007,Kadowaki2009}. Despite this, and detailed analytical studies~\cite{Moessner2003}, the understanding of kagome ice has significant gaps. In this paper, we aim to complete the characterisation of static correlations in kagome ice by means of a direct confrontation between theory, experiment and numerical simulation. In particular, we present polarized neutron scattering experiments in the static approximation that we compare with our own thermodynamics and model simulations, as well as with the existing analytical predictions of Moessner and Sondhi~\cite{Moessner2003}. In this way, we are able to elucidate some new properties of kagome ice, and to subject the many theoretical predictions of Ref. [\onlinecite{Moessner2003}] to a detailed test against simulation and experiment. We broadly confirm the theoretical picture, adding further structure to the predictions of Ref. [\onlinecite{Moessner2003}].  

\subsection{Description of kagome ice}

In a spin ice, Ising-like magnetic moments point along the local easy axis directions {--} the body diagonals of the tetrahedra of the pyrochlore lattice or  $\langle1 1 1\rangle$ directions of the cubic unit cell {--} and their interaction energy is minimised by ensuring that the magnetic moments obey an ice rule, i.e. two spins point in and two point out of each tetrahedron (`2-in--2-out').  This condition is equivalent to the ice rule that governs proton disorder in water-ice~\cite{Bernal1933,Pauling1935}. It creates a three dimensional Coulomb phase~\cite{Bramwell1998,Henley2010,Fennell2009}, a classical spin liquid with emergent $U(1)$ symmetry and associated dipolar correlations.   

Referring to Fig.~\ref{fig:KI}, when a magnetic field of moderate strength is applied along the $[1 1 1]$ direction, it pins one quarter of the spins, which occupy the vertices of triangular $[111]$ planes.  The remaining three quarters of the spins occupy the vertices of kagome lattices, which are stacked alternately with the triangular planes.  Because these spins make a shallow angle with the field and have a lesser Zeeman energy than the pinned spin, the ice rule can compete with the field, such that one spin per triangle of the kagome plane has a component opposing the field, as also illustrated in Fig.~\ref{fig:KI}.  A subset of ice rule states with reduced entropy~\cite{Hiroi2003111} is selected, and a magnetization plateau develops at 2/3 of the eventual magnetization, which signals the kagome ice state  ~\cite{Harris1998,Matsuhira2002,Sakakibara2003}.    
When the field is strong enough to overwhelm the ice rule, the remaining field-opposing spin is reversed, forming a unique, ice rule breaking configuration with `3-in(out)--1-out(in)' on every tetrahedron.

The characteristic magnetization plateau~\cite{Matsuhira2002,Sakakibara2003,Fukazawa2002}, reduced residual entropy, and entropy release at plateau termination~\cite{Aoki2004} were originally identified in \dto{}, and the magnetization plateau was also identified in \hto{}~\cite{Fennell2007,Krey2012}.  An interesting aspect of kagome ice is the liquid-gas like critical point~\cite{Sakakibara2003} that separates plateau termination by a first-order phase transition at low temperature from a more gradual crossover at higher temperature.   The existence of this was first rationalized by the monopole theory of excitations in spin ices~\cite{Castelnovo2008}, in which it can be identified as the critical end point  of the monopole crystallisation transition \cite{Raban:2019bk}.  Differences in the exact ratio of competing exchange and dipolar interactions result in a critical field that is somewhat lower in \dto{} ($H_c\approx0.9$ T) than in \hto{} ($H_c\approx1.6$ T).

\subsection{Coulomb phase and Kasteleyn transition}

%As in the case of spin ice in zero field we can distinguish a near-neighbour model of kagome ice in which spins are coupled ferromagnetically and where ice-rule breaking defects carry topological charge but have no magnetic charge; from one which includes dipole interactions between the moments with the result that the topological charge becomes dressed with magnetic charge.
As in the case of spin ice in zero field, we can distinguish a near-neighbour model of kagome ice in which spins are coupled ferromagnetically and where ice-rule breaking defects carry no magnetic charge, from one which includes dipole interactions, which leads to a magnetic charge on each defect.
We will refer to both classes of topological defects as magnetic monopoles. In fact the physics discussed in the present  paper is almost entirely that of  the vacuum for such defects, which is the Coulomb phase. Hence we concentrate on the near neighbour model with ice rule breaking defects suppressed. One key signature of the Coulomb phase is the appearance of pinch points in the diffuse neutron scattering pattern of kagome ice, as observed in \hto{}~\cite{Fennell2007} and \dto{}~\cite{Tabata2006,Kadowaki2009}.  These pinch points occur at the zone center for the kagome lattice rather than that of the pyrochlore lattice, where the pinch points for spin ice~\cite{Fennell2009} in zero field occur. This difference indicates the change from a three to a two-dimensional Coulomb phase.

The topological nature of Coulomb phases~\cite{Huse2003} leads to unconventional phase transitions~\cite{Alet2006,Powell2008PRB,Powell2008PRL}.
A particular example is the Kasteleyn transition, originally predicted for dimers on the honeycomb lattice~\cite{Kasteleyn1963,Huse2003,Laeuchli:2008cd} (which form a Coulomb phase~\cite{Huse2003,Laeuchli:2008cd}), and observed experimentally, to a good approximation, in a lipid bilayer phase transition~\cite{Nagle1973}. Later, this transition was predicted to occur in both two-~\cite{Moessner2003,Kao2016} and three-~\cite{Jaubert2008,Jaubert2009,BrooksBartlett:2014kf,Baez2017} dimensional settings in spin ice.
% and simulated with a conserved monopole algorithm~\cite{Baez2017}.  

A finite concentration of monopoles destroys the topological phase transition and the associated thermodynamic singularities~\cite{Jaubert2009,Powell:2013ct,Baez2017}, so that the transition is formally unobservable if the energy scale for monopole creation is finite. However, if the monopole concentration is small enough, the asymptotic approach to the transition is observable, but unless the dynamics are non-local, such as in the worm Monte Carlo algorithm discussed below, a finite monopole concentration is necessary to maintain equilibrium~\cite{Baez2017}. In real systems a best compromise is required between a low monopole concentration and ergodic evolution. 
%In the real system the presence of a small concentration of ice-rule breaking defects will aid equilibration but will not significantly affect the correlations discussed. 

The connection between kagome ice and the honeycomb lattice dimer model was originally made by Moessner and Sondhi~\cite{Moessner2003}, who presented an analytical calculation of the spin-spin correlation functions in kagome ice, as well as a theory of the Kasteleyn transition.  A key prediction of the theory is that the Kasteleyn transition would be accompanied by unconventional scaling of the generalized susceptibility, manifesting as the movement of certain features in the diffuse neutron scattering structure factor.  Although some features of this theory have been observed~\cite{Fennell2007}, the detailed predictions of anisotropic scaling of algebraic correlations were not tested in previous work. Here we test them for the first time.   

The kagome ice phase also corresponds to the KII phase of `dipolar kagome spin ice' \cite{MoellerPRB2009,Chern2011}. For a discussion of this relationship, we refer the reader to Appendix \ref{Wills}.

\subsection{Plan of the paper}

The plan of the paper is as follows.  In section~\ref{sec:Kagome Ice Thermodynamics} we review the relevant parts of the theory of Ref.~[\onlinecite{Moessner2003}] and add to this our analysis of the thermodynamic and critical properties.  In section  \ref{sec:sim} we describe our numerical simulations of kagome ice and its Kasteleyn transitions and in Section \ref{sec:Experimental_methods} we describe our neutron scattering experiments and their  comparison with the  numerical simulations and the theory.  Our main findings are discussed in Section  \ref{sec:Discussion} and we draw conclusions in Section \ref{sec:Conclusion}.

%\begin{figure}
%\centering
%\subfigure[\ ]{
%\includegraphics[scale=0.38]{Graphics/Kagome_3D_Lattice.png}}
%\subfigure[\ ]{
%\includegraphics[scale=0.3]{Graphics/Kagome_3D_Ice_Rules.png}}
%\caption{\label{fig:KI} The pyrochlore lattice with alternating kagome and triangular lattices in yellow and gray (a), and one possible ice-rule satisfying configuration of spins (up in red and down in blue) (b){\color{cyan}{this figure will have to be clearer, I think less layers, and a third color for the pinned spins}}}
%\end{figure}

\section{\label{sec:Kagome Ice Thermodynamics}Theory}

\subsection{Kasteleyn Transition}

\subsubsection{The model}

Convenient unit vectors for describing kagome ice are defined (Fig.~\ref{fig:KI}) by the direction of the applied field, where the vertical direction is defined as $\hat{z} = [1 1 1]/\sqrt{3}$, and the two perpendicular horizontal directions spanning the kagome plane are $\hat{x} = [\bar{1} 1  0]/\sqrt{2}$ and $\hat{y} = [\bar{1} \bar{1} 2]/\sqrt{6}$. The field may be tilted away from $\hat{z}$ towards $\hat{y}$ by the angle $\theta$, and any rotation of the resulting in-plane field component from $\hat{y}$ toward $\hat{x}$ is quantified by the angle $\phi$.  In this work, all fields are applied along [$111$], selecting the $Z_2^+$ topological sector (as opposed to the [$\bar{1}\bar{1}\bar{1}$] and $Z_2^-$ respectively).  Our system is further defined such that `up' tetrahedra have the spin in the triangular lattice above the kagome plane relative to the applied magnetic field along $\hat{z}$, and `down' tetrahedra the opposite; a triangle in the kagome plane derives its up/down identity from its tetrahedron. 

The Zeeman energy of a spin in the presence of magnetic field $\vec B$ is $E^B = -\vec{\mu} \cdot \vec{B}$, where $\vec{\mu}$ is the single-ion magnetic moment. 
When the field is exactly aligned along the $[111]$ direction with $\vec{B}=B\hat{z}$, for the kagome plane spins $E^B=\mp \frac{1}{3}|\vec \mu| B$,  in the ratio $2:1$, 
so that the kagome ice microstates have equal probability.  Tilting the field away from $\hat{z}$ towards $\hat{y}$ by an angle $\theta>0$ gives a contribution to the Zeeman energy from the in-plane spin components, which singles out one sublattice in the kagome plane (labelled with index $\kappa =1$) as the prefered location for the field-opposing spin, while keeping the other sublattices ($\kappa =2, 3$) equivalent~\cite{Moessner2003,Fennell2007,Kao2016}. The tilt can be further generalised by rotating in the $x-y$ plane by an angle $\phi$, which further lowers the symmetry, distinguishing all three sublattices. 

\subsubsection{Spins and pseudo-spins}

Given that classical spin ice is built from discrete spin degrees of freedom, 
it often proves convenient to re-formulate the problem in the language of an Ising model, introducing pseudo-spin degrees of freedom, $\sigma_i=\pm 1$. 
Taking an up tetrahedron as the crystallographic basis, 
the four spins align with respect to the local axes $\vec d_0=\hat{z}=\frac{1}{\sqrt{3}}[1,1,1],\; \vec d_1 =\frac{1}{\sqrt{3}}[-1,-1,1],\; \vec d_2= \frac{1}{\sqrt{3}}[-1,1,-1],\; \vec d_3= \frac{1}{\sqrt{3}}[1,-1,-1]$, so that $\vec d_{\kappa}\cdot\vec d_0=-\frac{1}{3}$. In the convention that $\sigma_i=1$ corresponds to a spin pointing out, the pseudo-spin is defined%
\begin{equation}
    \sigma_i=\frac{\vec \mu_i\cdot \vec d_i}{|\vec \mu|}.
\end{equation}
In terms of these variables, spin ice maps to an Ising antiferromagnet with 2-in--2-out becoming `2-down--2-up' for pseudo-spins and the nearest neighbour model in zero field is the antiferromagnet studied by Anderson \cite{Anderson56}.

In the kagome ice problem, 2-in--1-out becomes 2-down--1-up for pseudo-spins. The Zeeman energy of a kagome plane spin $\kappa$, with field along the $\hat{z}$ axis, can be written $E^B_{\kappa}=-\sigma_{\kappa}\tilde{B}$, with $\tilde{B}=-\frac{|\vec \mu| B}{3}$ a pseudo-magnetic field in the reverse, $-\hat{z}$ direction. Hence, the three kagome plane spins map onto a kagome antiferromagnet in an external field. The pseudo-spin correlations can be accessed through the out-of-plane spin components which, as we discuss below, can be measured in the non-spin flip channel in polarised neutron scattering experiments.

\subsubsection{Dimer mapping}

\begin{figure}
\centering
\includegraphics[scale=0.325]{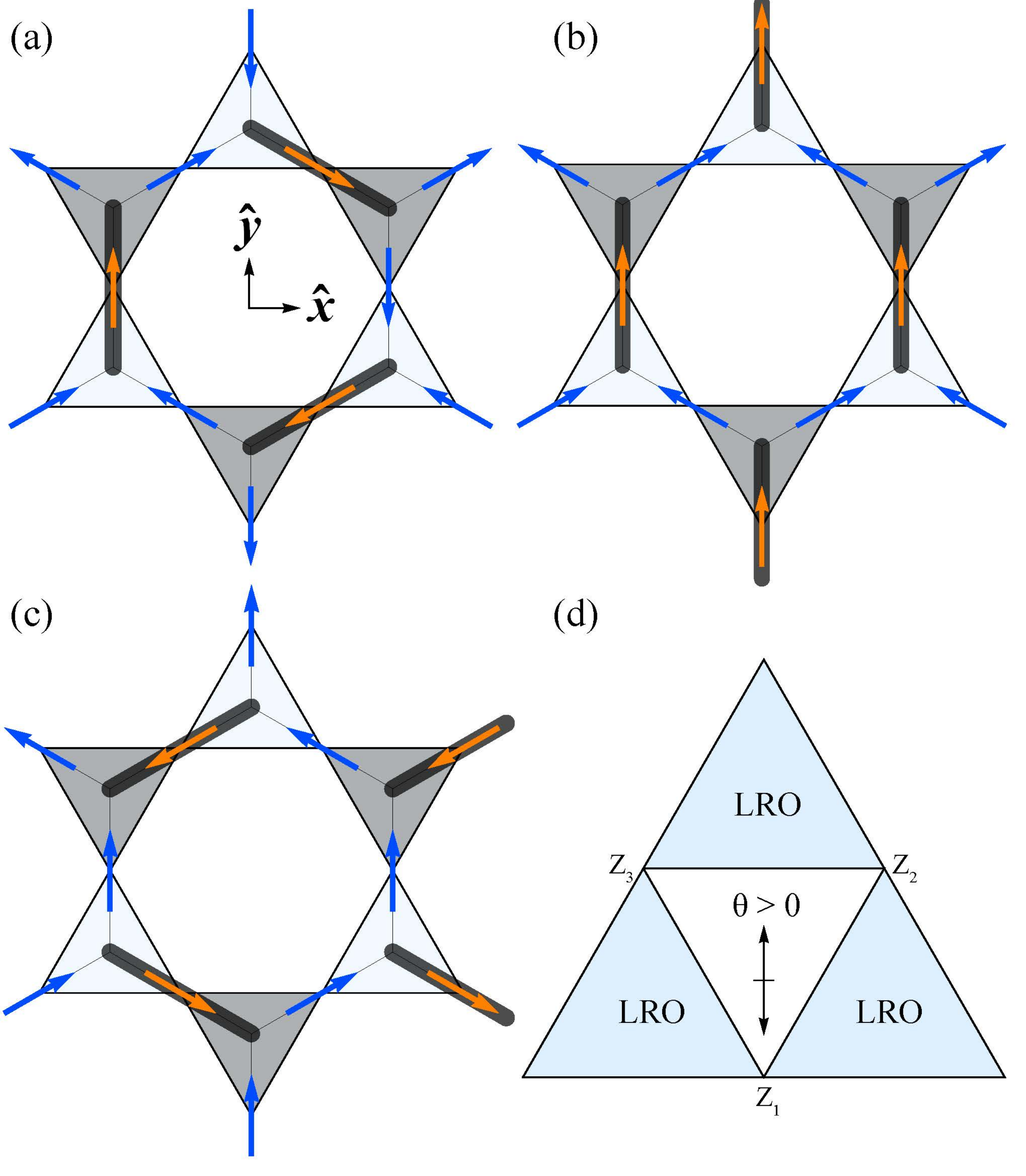}
\caption{\label{fig:DPD} (a-c) Kagome ice, showing the relationship of down spins (orange arrows) and dimers (dark grey rods), where down-triangles are shaded grey. (a) Disordered structure. (b) Long range order induced by a field tilted towards the $\hat{y}$ direction, (c) Partial order resulting from a tilted field perpendicular to the $\hat{y}$ direction, (d) the Kasteleyn phase diagram, where the central kagome ice state is surrounded by long range order depending on sublattice $\kappa=1,2,3$ selected by the tilted field.}
\end{figure}

Moessner and Sondhi's mapping to hardcore dimer configurations on a honeycomb lattice~\cite{Moessner2003} works as follows. The honeycomb lattice and the kagome lattice are a parent/medial pair, with sites of the kagome lattice at the mid-point of the links of the honeycomb lattice~\cite{Henley2010}.  A dimer is placed on a link of the honeycomb lattice located by a kagome site carrying a field-opposing spin, i.e. the outward pointing spin of an up triangle.  The kagome sublattice on which the dimer resides is specified by the index $\kappa$, as defined above. Fig.~\ref{fig:DPD} shows how, if the ice rules are obeyed, there is a single dimer per unit cell and no dimers can touch~\cite{Kasteleyn1963}.  The entropic phase of kagome ice is therefore a hardcore dimer liquid which has critical correlations and corresponds to the Coulomb phase for the spins.
%; a classical spin liquid with algebraic spin correlations.

As the dimers have no internal energy, the ``particle enthalpy" for this system is $ \langle\mathcal{H}\rangle=- \sum_{\kappa} \langle N_{\kappa}\rangle  \mu_{\kappa}$ with $\langle N_{\kappa}\rangle$ the mean number of dimers on sublattice $\kappa$ and $\mu_{\kappa}$ the relevant chemical potential. The $\mu_{\kappa}$ correspond to the change in the in-plane component of Zeeman energy of spin $\kappa$ when it is flipped to become the out-pointing spin of the triangle:
%Changing the Zeeman energy of the spins on a sublattice $\kappa$ is equivalent to changing the chemical potential $\mu_\kappa$ for dimers on this family of links and we can write:
%. Following according to field strength, Ising spin direction $\sigma_\kappa = \pm 1$, and tilt angle. Fixing the zero of energy for zero tilt: 
%
\begin{eqnarray} \label{eq:MuPhi}
%\begin{align}
      \mu_1  &=& \left(\frac{4\sqrt{2}|\vec{\mu}|B}{3}\right)\sin{\theta}\cos{\phi} \nonumber\\ 
      \mu_{2,3} &=&-\left(\frac{\sqrt{2}|\vec{\mu}|B}{3}\right) \sin{\theta}(\cos{\phi} \pm \sqrt{3}\sin{\phi}).
%\end{align}
\end{eqnarray}
These values are defined with respect to a large, positive and constant term which imposes the constraint that $N_1+N_2+N_3=\frac{N}{3}$, with $N$ the number of spins. The chosen sign giving the favourable placement on the first sublattice with $\mu_1>0$ is consistent with the standard conventions of thermodynamics and allows us to define a standard fugacity for dimer placement, $z_{\kappa} = e^{\mu_{\kappa}/k_BT}$.

%\begin{subequations} \label{eq:MuTheta}
%\begin{align}
%      \mu^B_1 & \propto -E_{z}\sigma_1(\cos{\theta} - 2\sqrt{2}\sin{\theta}) \\
%      \mu^B_{2,3} & \propto -E_{z}\sigma_{2,3}(\cos{\theta} + \sqrt{2}\sin{\theta})
%\end{align}
%\end{subequations}
%\begin{subequations} \label{eq:MuTheta}
%\begin{align}
%    \mu_1  \propto   E^B_1 & = (|\vec{\mu}| B/3) \sigma_1(\cos{\theta} - 2\sqrt{2}\sin{\theta}) \\
%     \mu_{2,3}  \propto  E^B_{2,3} & = (|\vec{\mu}| B/3) \sigma_{2,3}(\cos{\theta} + \sqrt{2}\sin{\theta})
%\end{align}
%\end{subequations}
A Kasteleyn transition to dimer alignment on sublattice $\kappa=1$ occurs when $z_1$ is equal to the sum of those for the other two sublattices~\cite{Kasteleyn1963,Moessner2003}: 
%
%\begin{equation}
$z_1 = z_2 + z_3$.
%\label{fugacity}
%\end{equation}
%
 The transition is from the dimer liquid phase to a dimer solid phase and the phase diagram, 
 illustrated in Fig.~\ref{fig:DPD}d, has 
 three-fold symmetry. In terms of the spins, the condition $z_1 = z_2 + z_3$ corresponds to a long-range ordered, ice-rule obeying state in which the field-opposing spins of the kagome plane are all located on sublattice $\kappa=1$ (Fig.~\ref{fig:DPD}b). 

\subsubsection{Phase Diagram}

 To discuss the phase diagram~\cite{Moessner2003}, it is convenient for us to define a scale free parameter 
\begin{equation} \label{eq:HDef}
     \Upsilon(B,T,\theta)  \equiv \frac{2\sqrt{2}}{\ln{2}} \frac{|\vec{\mu}|B \sin{\theta}}{k_BT}
\end{equation}
and to determine the value $\Upsilon_{\rm K}$ that this parameter takes at the Kasteleyn transition. The critical parameter $\Upsilon_{\rm K}$ will also be a function of the angle $\phi$ and will play an important role in our future discussions.  Moessner and Sondhi previously named this parameter $h$, but we have chosen $\Upsilon$ to avoid confusion with reciprocal space labelling, i.e. $(h,k,l)$.

The fugacity relation at the transition is derived in detail in Appendix~\ref{eqn3_deriv}, and can be written~\cite{HarmanClarke2012}
\begin{equation} \label{eq:KTCphi}
\cosh{\bigg(\frac{\Upsilon_{\rm K}\sin{(\phi)}\ln{2}}{\sqrt{3}}\bigg)} = 2^{\Upsilon_{\rm K}\cos{(\phi)}-1}.
\end{equation}
For $\phi=0$ this is solved for $\Upsilon_{\rm K}=1$ giving a surface of Kasteleyn transitions (denoted by subscript K) in the space of $B,T,\theta$:
\begin{equation} \label{eq:KTCtheta}
      k_{B} T_{\rm K} =  \frac{2\sqrt{2}}{\ln{2}}|\vec{\mu}|B_{\rm K} \sin{\theta_{\rm K}}.
\end{equation}
Usually two parameters will be fixed, typically $\theta, B$ (in addition to $\phi$), which then uniquely defines the transition point $T_{\rm K}$. As the field is rotated in the plane, an iterative solution can be found, with $\Upsilon_{\rm K}\rightarrow\infty$ for $\phi=\pm 60^{\circ}$, corresponding to the transition temperature falling to zero.   

The characteristic triangular form of the phase diagram with three lines of Kasteleyn transitions separated by discontinuous points thus reflects the symmetry of the kagome plane and rotation of the field between the equivalent $\langle11\bar{2}\rangle$  and $\langle1\bar{1}0\rangle$ axes, as shown in Fig.~\ref{fig:DPD}.  The ordered dimer phase and, therefore, the saturation magnetization, lies along a local $\hat{y}$-axis, even for arbitrary $\phi$. Hence, for such an arbitrary field tilt, the induced in-plane magnetisation follows the in-plane field direction at high temperature, but as the transition is approached it swings away from the field to order finally along one of the $\langle 1 1 \bar{2}\rangle$ axes. 

Given this three-fold symmetry, the case $\theta>0$, $\phi=\pm 60^{\circ}$ is equivalent to a tilt with $\theta<0$ and $\phi=0$. In this situation the dimer/field-opposing spin may occur with equal probability at sites with $\kappa=2,3$.  No Kasteleyn transition occurs because the degeneracy of the ice rule states is not fully lifted.  Although the entropy is reduced by the tilted field, the resulting state has chains of spins running across the kagome planes that need not be correlated with each other (see Fig.~\ref{fig:DPD}c).  These are the so-called $\beta$ chains found when the field is applied along a $\langle 110 \rangle$ type direction of a spin ice\cite{Hiroi2003110,Fennell2005}, or columnar order for dimers.  ($\alpha$ chains are also present, formed by the apical spins and the uniquely selected kagome sublattice where the spin cannot oppose the field.)  The dependence on $\theta$ can be seen to be asymmetric because in the pyrochlore lattice tilts with $\theta>0$ represent a tilt of the field toward the bisector of a tetrahedron face, where degeneracy is fully removed~\cite{Ruff2005}, while tilts with $\theta<0$ represent a tilt of the field toward a tetrahedron edge, where degeneracy is only partially removed~\cite{Hiroi2003110,Fennell2005}.

\subsection{Topological Excitations}

For a model system, with periodic boundaries, lying within the constrained manifold of states that satisfy the ice rules, the only allowed excitations are correlated spin flips around closed loops. These can be either short loops within the system, which do not change the magnetization, or long loops that span the entire system. The latter, which we call strings, are topological excitations. As a consequence, the magnetization defines the topological sector~\cite{Jaubert:2013jz}, and both are changed by flipping such strings. 

Starting from the ordered state, the only available excitations are the system-spanning strings, which cost an energy that scales linearly with the system size, $L$. As a result, the system remains completely frozen at low temperature.  However, the entropy introduced by a string also scales linearly with $L$, so that their introduction becomes favorable above a singular threshold, which is the Kasteleyn transition. The transition condition, Eqn.~\ref{eq:KTCphi}, follows directly by tracking the free energy cost, $\delta G$ of introducing a string into the ordered state. As the string passes through each unit cell, the change in Zeeman energy (or dimer number enthalpy) is $\delta \epsilon$ and the entropy creation, $\delta s$, so that $\delta G=L(\delta \epsilon -T\delta s)$, which is zero at the transition. The system can access different string configurations by flipping short loops of spins or dimers. When adding a second string, the two strings cannot pass through the same triangle. Consequently there is an entropic repulsion between strings and the free energy cost of adding the second string is slightly greater than the first ensuring that the transition is continuous rather than discontinuous. 

Such strings can be found in other problems of statistical mechanics, mapping onto world lines for hard core bosons undergoing Bose condensation at the transition \cite{Jaubert2008} or onto directed polymers \cite{Bhattacharjee1983}. In both cases the strings can be thought of as walkers making ballistic progress against the direction of ordering while diffusing in the $d-1$ dimensional plane perpendicular to this direction. Note that, in spin ice, the strings can, in principle, meander in three dimensions. In practice, the apical spin is considered to be firmly fixed so that we only consider string and loop excitations in isolated planes and we only simulate a single kagome plane, later returning to the relation with three dimensional loops in the discussion.

\subsection{Thermodynamics}

\subsubsection{Free energy}

As the magnetic ordering occurs along one of the local $\langle11\bar{2}\rangle$ axes we can  restrict the thermodynamic discussion to the case $\theta>0$, $\phi=0$ without loss of generality. Including a finite in-plane angle $\phi$ changes the finite size scaling properties at the transition, as shown in detail below, but the thermodynamics of the transition is captured by this constrained case. We define $M=\langle \frac{1}{|\vec \mu|}\sum_i \vec \mu_i\cdot \hat{y}\rangle$, the dimensionless in-plane component of the total magnetic moment, and its conjugate magnetic field variable ${H}=B|\vec \mu|\sin \theta$, the in-plane component of the applied field, in energy units.

All microstates forming the kagome ice manifold have the same internal energy so that the Helmholtz free energy, $F(M,T)=-TS(M)$ is purely determined by the system entropy. 
There are, however interactions, in the form of the hardcore dimer constraints and these are ultimately responsible for the phase transition but they do not appear directly in the phenomenology. Phase transitions driven only by entropy are actually not so rare for hard particle systems\cite{Alder60,Frenkel94}, but two things single out the Kasteleyn transition.  The first is that no symmetry, either microscopic or emergent, is broken at the transition, placing the transition outside the usual paradigm in which phase transitions and symmetry breaking in phase space go hand in hand.  The second is that the transition occurs for $S=0$ making it anisotropic, with zero fluctuations on the low temperature-high field side. 

Despite these particularities a complete thermodynamic description is possible. As in a paramagnet, the two  intensive variables $H, T$ collapse into a single thermodynamic variable~\cite{HarmanClarke2012}:
\begin{equation}\label{eq:Thermo}
\frac{H}{k_BT} = \frac{1}{k_BT}\frac{\partial F}{\partial M} = \frac{-1}{k_B}\frac{\partial S}{\partial M}.
\end{equation} 
In a paramagnet, the entropy approaches zero as the magnetisation saturates, but the slope $\frac{\partial S}{\partial M}$ is infinite, precluding a phase transition at finite temperature. Here, the entropy is also zero at saturation but the Kasteleyn phase transition for the finite ratio $(H/k_B T)_{\rm K}$ ensures that the entropy must go to zero with a finite slope.

\subsubsection{Landau-style expansion}

The asymmetry of the Kasteleyn transition has led previously to a classification lying between first and second order~\cite{Nagle1973}, although on the high entropy side the transition satisfies all the thermodynamic and phenomenological criteria of a second order transition. The honeycomb lattice dimer problem can  be solved exactly with the calculation of the partition function, all thermodynamic quantities and correlation functions, but it is useful to develop the phenomenology of the transition through the construction of a Landau-like free energy \cite{Jaubert:2009to,Bhattacharjee1983}. 

The Gibbs potential, $G^{\ast}=F-HM$, can be expanded in powers around the saturated moment $M_{\rm max}$. Introducing the dimensionless variable $m=\left(\frac{M_{\rm max}-M}{N}\right)$ and the dimensionless parameter $\eta=\left[\left(\frac{H}{k_{B}T}\right)_{\rm K} - \frac{H}{k_{B}T}\right]$ it follows that
\begin{equation}
\frac{G^{\ast}}{Nk_{B}T}=-\eta m +\frac{\alpha_2}{2}m^2+\frac{\alpha_3}{3}m^3 + \dots - \frac{\mu_0H M_{\rm max}}{Nk_{B}T},
\end{equation} 
where $\alpha_i$ are the parameters of the expansion of $G^{\ast}$ in $m$.
%Odd powers appear because the point about which the expansion is made is not a minimum in $G^{\ast}$.
Minimising with respect to $m$, the leading term of the Gibbs free energy is  $G\sim  -\eta^{1+\beta}$, with $\beta=1$ at this mean field level, while the exact solution~\cite{Kasteleyn1963} yields $\beta=1/2$ [\onlinecite{Moessner2003}]. The entropy, $S=-\frac{\partial G}{\partial T} \sim \eta^{\beta}\sim m$, indeed scales linearly with $M$ near the transition, consistent with the finite value for $(H/k_BT)_{\rm K}$. 

The magnetisation is singular around its maximum value, $m\sim \eta^{\beta}$ and, as there is a single intensive thermodynamic variable, the critical exponents for the susceptibility ($-\frac{\partial^2 G}{\partial B^2}$) and specific heat ($-T\frac{\partial^2 G}{\partial T^2}$), $\gamma$ and $\alpha$ are equal. As a consequence the Rushbrooke scaling relation $\alpha+2\beta +\gamma=2$ reduces to $\beta+\gamma=1$. This means that the one scaling dimension $\Delta=\gamma+\beta$ is unity, which excludes anomalous scaling, ensuring Gaussian exponents for dimensions below the upper critical dimension~\cite{Goldenfeld94}.

The singular free energy can be equated to the inverse correlation volume, $G \sim \xi_{x}^{-(d-1)}\xi_{y}^{-1}$ where $\xi_{y}\sim \eta^{-\nu_{y}}$ and $\xi_{x}\sim \eta^{-\nu_{x}}$ are the diverging correlation lengths parallel and perpendicular to the ordering direction. From this the modified hyperscaling relation, $1+\beta=(d-1)\nu_{x}+\nu_{y}$ follows~\cite{Goldenfeld94}. From the insertion of a single string with its ballistic and diffusive nature parallel and perpendicular to the ordering direction respectively, we can anticipate that $\nu_{y}=1$ and $\nu_{x}=1/2$, consistent with the absence of an anomalous scaling dimension. Putting the mean field value, $\beta=1$ into the hyperscaling relation gives an upper critical dimension of $d=3$ (Ref. [\onlinecite{Bhattacharjee1983,Jaubert2008}]) so that the two-dimensional problem is outside the mean field regime. Indeed the exact result yields $\beta=1/2, \gamma=1/2$. 

Hence, from this analysis one can conclude that singular part of the free energy satisfies the scaling and hyperscaling relations for a critical point for a one parameter system. 

\subsection{Correlation functions}\label{sec:corr_func}

% \textcolor{red}{I stopped here - to come, some bull about LGW paradigme and some more thermodynamics} prescribes a second order transition towards a long range ordered state where the slope $\frac{\partial F}{\partial M}$ is constant as $F \rightarrow 0$.  However, it is well known that Nagle classified the Kasteleyn transition as asymmetric, being continuous when approached from above and discontinuous when approached from below.

From the analytic solutions for the correlation function \cite{Moessner2003}, one finds the following expressions for the correlation lengths:
%
%Inverse correlation lengths along $\hat{x}$ and $\hat{y}$ were derived from the dimer correlation functions and the ratio of the activities ($Z_1/Z_{2,3} = 2^{-h}$):
%
\begin{subequations}\label{eq:InvCorrLength}
\begin{align}
\xi^{-1}_x & =  \arcsin{\sqrt{1 - 4^{\Upsilon - 1}}} \\
\xi^{-1}_y & =  \xi^{-1}_x \frac{2^{2 - \Upsilon}}{3}\sqrt{1 - 4^{\Upsilon - 1}},
\end{align}
\end{subequations}
which near the transition take the power law forms discussed above. Unusually, the development of diverging correlations lengths does not signal the onset of power law correlations. In the Coulomb phase, with zero tilted field, both spin and dimer correlation functions are already of dipolar form with, in two dimensions, correlations falling as $\sim 1/r^2$ at large distance, giving characteristic logarithmic divergences with system size, for the structure factors. The growing correlation lengths introduce an anisotropy to the correlation functions with in-plane distance $r$ replaced by an effective scale 
\begin{equation}
r^{'}=\sqrt{x^2+\left(\frac{\xi_y}{\xi_x}\right)^2 y^2}.
\end{equation}

Structure factors for both pseudo and real spins can be accessed by polarized neutron scattering (see below). As the transition is approached, the developing anisotropy causes  peaks in both structure factors to drift towards the Brillouin zone centre and to sharpen, arriving there as the transition is reached. Kagome ice therefore has field-tunable critical correlations, with the drift determining the ratio of the correlation lengths~\cite{Moessner2003}.

\section{Numerical simulations}\label{sec:sim}
 
 \subsection{Details of the simulations}
 
%Our numerical simulations of kagome ice were motivated by experimental observations of the kagome ice phase of \hto{}~\cite{Fennell2007,Fennell_unpub} and \dto{}~\cite{Tabata2006,Kadowaki2009}. Some of these~\cite{Fennell_unpub} suggested the observed structure factors were related to the analytical predictions of the Kasteleyn transition, but that the effect of generalized tilts (i.e. $\theta>0$ and $\phi\neq 0$) would be important to understand the experimental data in detail. To explore the effects of generalized tilts, we carried out Monte Carlo simulations of kagome ice.  

%\textcolor{red}{REPETITION: In the simulations, classical spins are considered to be constrained to the manifold of kagome spin ice states, neglecting all internal energy, as would be the case for the nearest neighbour spin ice or the dumbbell models of spin ice \cite{Castelnovo2008} in their vacuum states. Since the apical spin of each tetrahedron is firmly pinned along the field direction in the kagome ice plateau regime, the kagome planes are decoupled, opening the door to simulations of a single kagome plane in which the dynamics is restricted to the basal spins of the tetrahedra confined to the $Z_2^+$ sector.}

The Hamiltonian of two dimensional kagome ice is limited to the Zeeman energy term
\begin{equation}\label{eq:nnHam}
%\mathcal{H}=-J\sum_{ij}\vec{S}_i\cdot\vec{S}_j-\vec{H}\cdot\sum_i \vec{S}_i,
\mathcal{H}=-\vec{H}\cdot\sum_i \vec{S}_i.
\end{equation}  \label{H-constraint}
Here $\vec{S}_i$ is a dimensionless vector of length $S_{\perp}=\frac{2\sqrt{2}}{3}$, 
describing the component of the magnetic moment lying in the $x-y$ plane: $\vec \mu_i= |\vec \mu| \vec S_i +\mu_i^z \hat{z}$, and $\vec H$ is the field component in the plane for arbitrary $\phi$, again expressed in energy units. 

This system was updated using a worm algorithm, full details of which are given in Appendix~\ref{loop_appendix}.  In the following, we report simulations of diffuse neutron scattering from the magnetic moments with both in-plane and $\hat{z}$ components. In a polarized neutron scattering experiment, the in-plane spin components will be observed in the spin flip (SF) channel, while the $\hat{z}$ components, corresponding to the pseudo-spin variables, will be observed in the non-spin flip (NSF) channel. 

The majority of our simulations were carried out on a system of 8112 spins ($L\times L$ kagome unit cells with $L=52$) with periodic boundary conditions. For finite size scaling analysis the simulation sizes were extended to 99856 spins ($L=316$).  At the start of the simulation, the system was placed in an ordered state at zero temperature (as illustrated in Fig.~\ref{fig:DPD}b), and then evolved upward in temperature under constant field.  
The diffuse scattering maps were each generated from 100 independent spin configurations. 

%{\color{cyan}{Alexandra/Peter: 100 separate simulations each or separated by some number of moves?}}

\subsection{Simulated Kasteleyn transition}

\subsubsection{Neutron scattering map at $\theta = \phi = 0$}

 \begin{figure}
\includegraphics[width=0.4\textwidth]{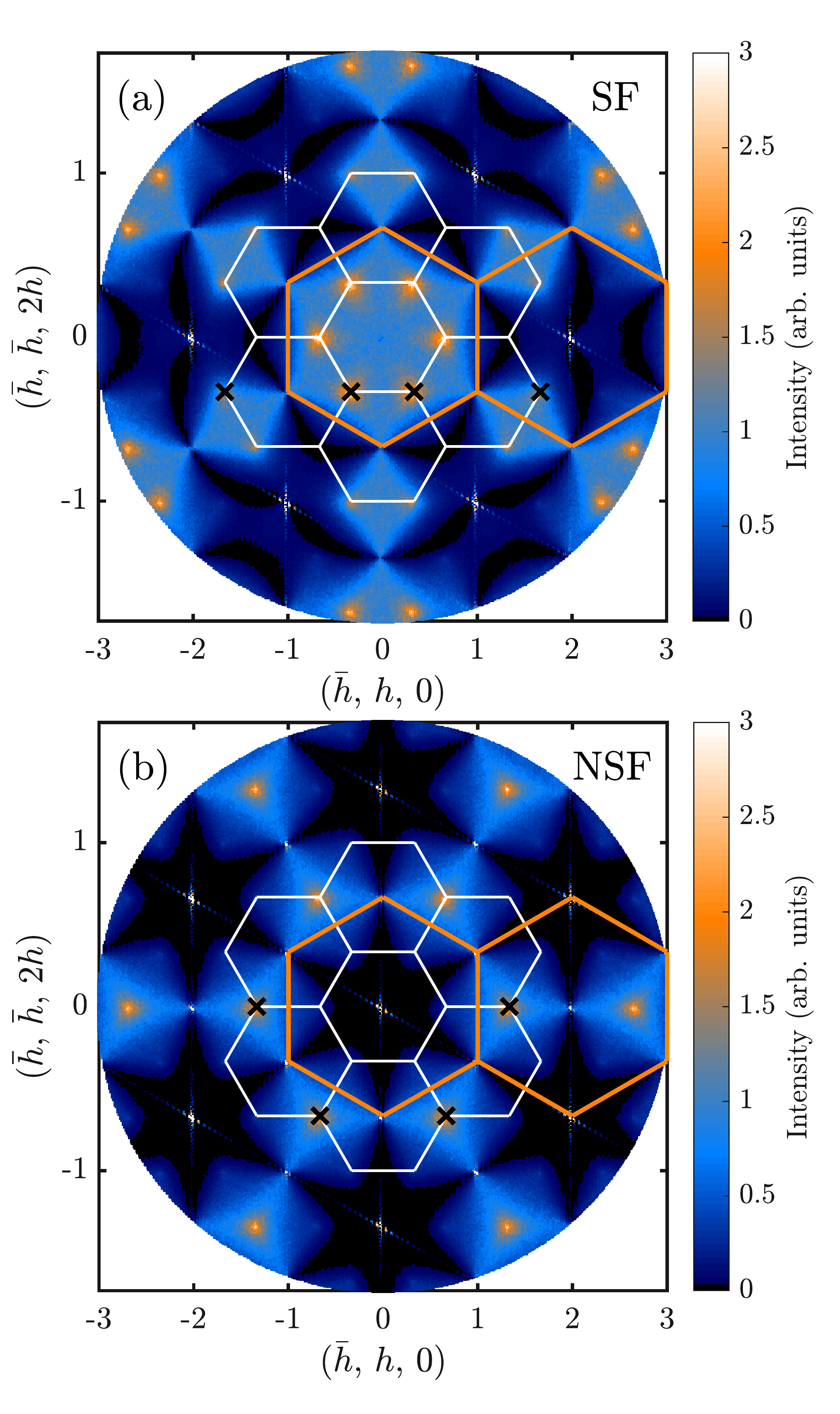}
\caption{Differential scattering cross section of kagome ice with $\theta=0$ and $\phi=0$ in the SF (a) and NSF (b) channels.  The six-fold symmetry of the kagome plane is notable and can be seen in the precise positioning of diffuse peaks on the kagome zone corners (white hexagons, orange hexagons are pyrochlore zones). 
In the paper we define $x=(\bar{h},h,0)$ and $y=(\bar{h},\bar{h},2h)$ and   
%Points in the plane can be obtained as linear combinations of these two principal orthogonal axes, for example the upward slanting direction equivalent to $(\bar{h},\bar{h},2h)$ is $(h,\bar{2h},h)$ and the downward slanting equivalent direction is $(2h,\bar{h},\bar{h})$.}
refer to points in the plane both by their components $(h,k,l)$, and by the values of $q_x$ and $q_y$ (in reciprocal lattice units). 
Crosses mark peaks whose positions were tracked and are discussed in the text.}
\label{fig:untilted}
\end{figure}

We first examine the manifestations of the Kasteleyn transition in the diffuse neutron scattering structure factor as predicted in Ref. [\onlinecite{Moessner2003}].

Fig.~\ref{fig:untilted} illustrates the scattering patterns for $\theta=0$. These are
six-fold symmetric pinch point patterns characteristic of the kagome ice Coulomb phase. They includes sharp Bragg peaks at the $(2,2,0)$ points signalling the partial all-in-all-out order~\cite{BrooksBartlett:2014kf}. As the tilt is applied, the in-plane field lowers this symmetry to two-fold with the development of arms of more intense scattering either side of the ordering direction \cite{Powell:2015ut}.  In Fig.~\ref{ahc_fig1} we show the general features of the pseudo-spin (NSF) and non-collinear spin (SF) structure factors, and an example of their evolution in tilted field with $\theta>0$ and $\phi=0$.

Our numerical results are consistent with experiment (see below) but show differences with the analytical predictions of Moessner and Sondhi~\cite{Moessner2003}. In the pseudo-spin or NSF structure factor a single peak appears at $(\frac{\bar{4}}{3},\frac{4}{3},0)$, ($q_x=4/3$, $q_y=0$) as compared to the  group of three peaks located symmetrically around $(\bar{1},1,0)$ ($q_x=1$, $q_y=0$) and including $(\frac{\bar{2}}{3},\frac{2}{3},0)$ ($q_x=2/3$, $q_y=0$) predicted in Ref.~[\onlinecite{Moessner2003}] (i.e. peaks at $(8\pi/3,0)$ as against $(4\pi/3,0)$ respectively, in the coordinates of Ref. [\onlinecite{Moessner2003}]).

%We suggest that these detailed differences might indicate, a sign error in the calculation of the structure factor in Ref. [\onlinecite{Moessner2003}], because other aspects of the calculations of Ref. [\onlinecite{Moessner2003}] are well confirmed by our simulations. 

\subsubsection{Drift of peaks  at $\theta \ne 0, \phi = 0$}\label{sec:drift_sim}

\begin{figure*}
\includegraphics[width=0.9\textwidth]{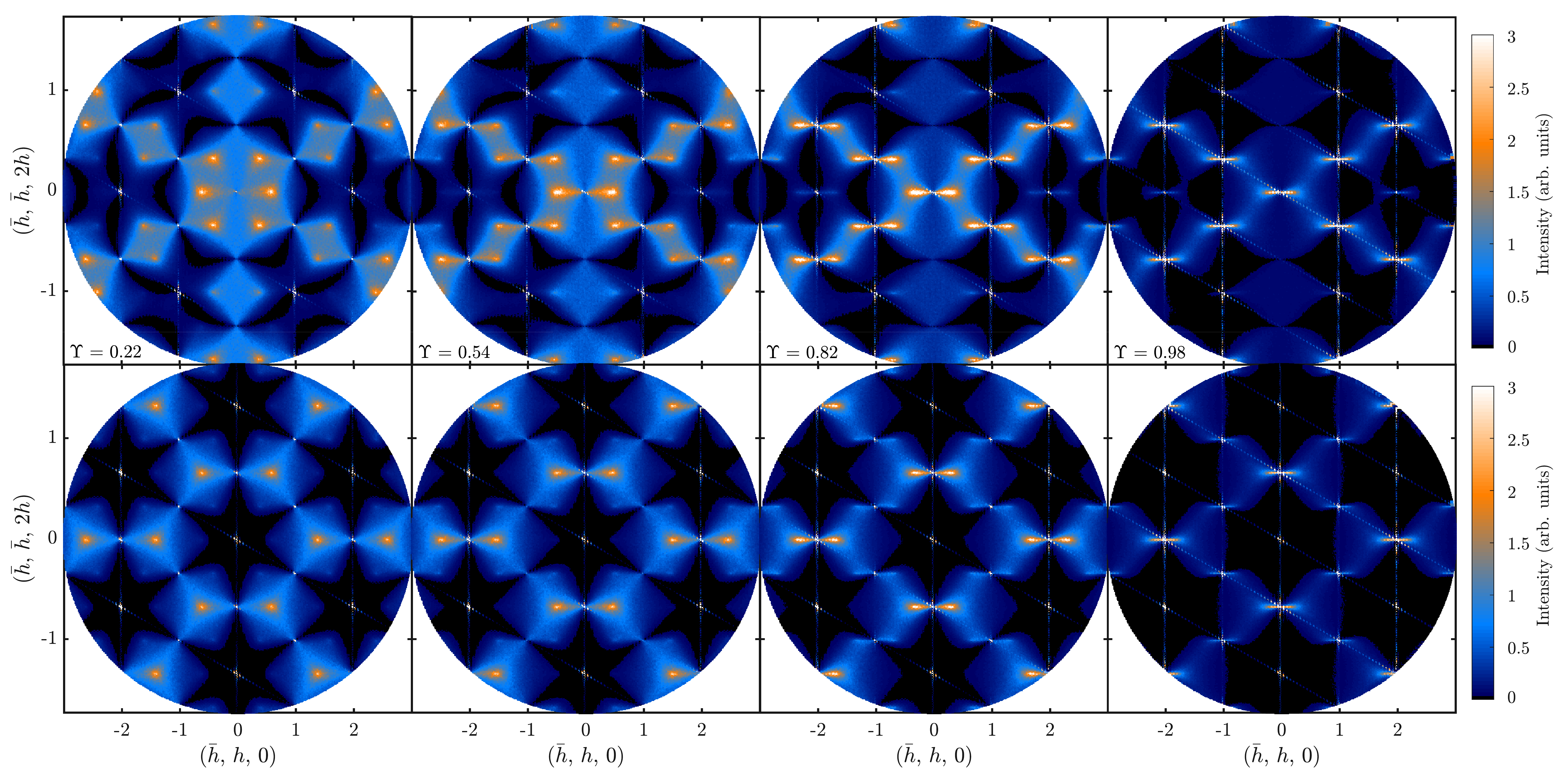}
\caption{\label{ahc_fig1} Simulated evolution of the diffuse neutron scattering as the Kasteleyn transition is approached.  The top row of panels shows the scattering in the spin flip channel (SF) and the bottom row shows the non-spin flip channel (NSF) as a function of the parameter $\Upsilon(B, T, \theta)$ in Eqn.~\ref{eq:HDef}.  When $\Upsilon$ is small, the breaking of the six-fold symmetry of the kagome ice scattering patterns to two-fold is only just apparent, but as $\Upsilon \rightarrow \Upsilon_{\rm K} = 1$ it becomes increasingly evident.  As $\Upsilon$ increases, diffuse peaks (orange), in both channels start close to Brillouin zone corners of the kagome lattice and drift in the $x$-direction toward the pinch points at kagome lattice Brillouin zone centers.}
\end{figure*}

\begin{figure}
\includegraphics[width=0.4\textwidth]{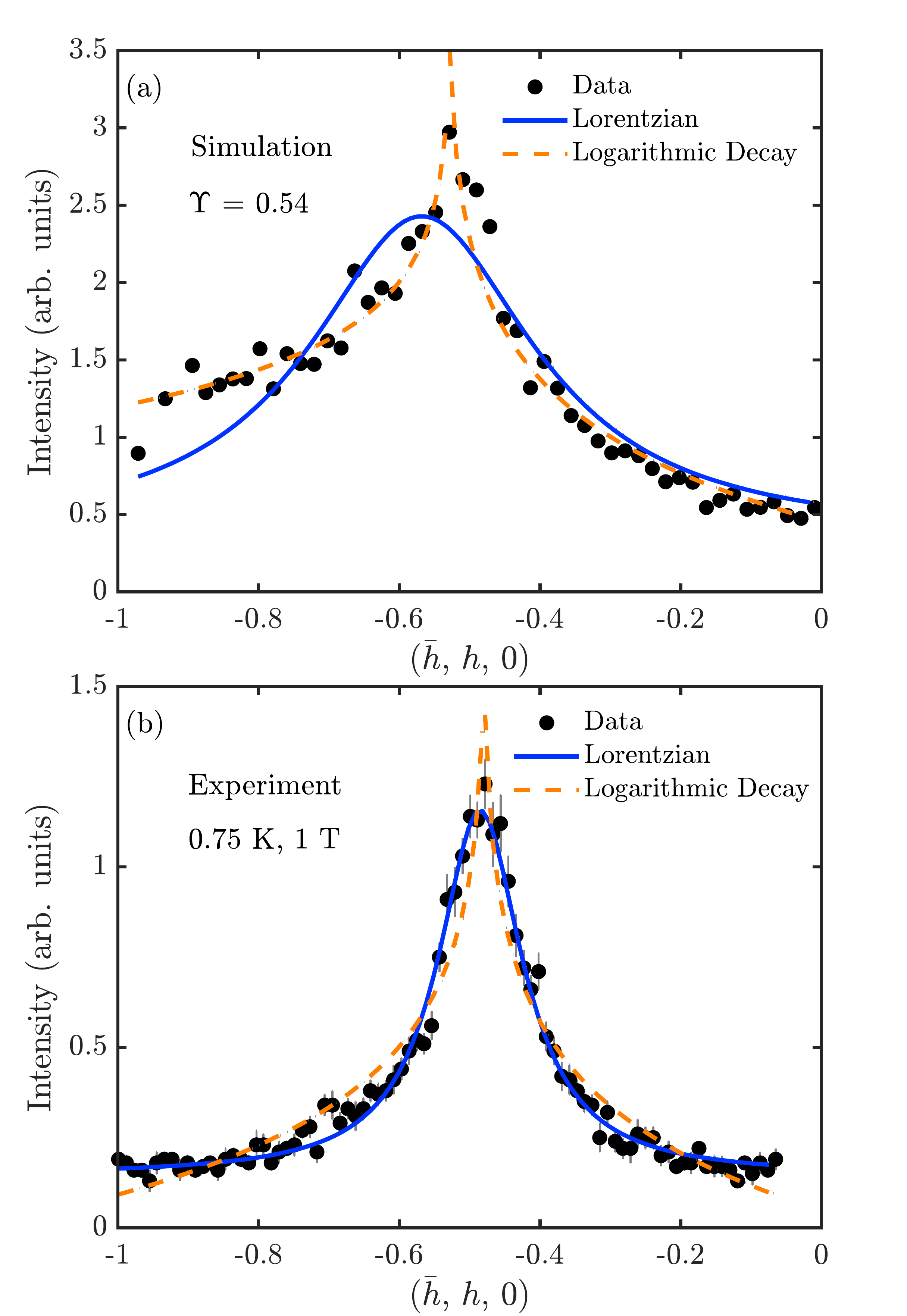}
 \caption{(a) Simulated logarithmic peak in the SF channel ($\Upsilon=0.54$) at approximately $(1,\frac{\bar{1}}{3},\frac{2}{3})$ ($q_x=-2/3$, $q_y=-1/3$) (see Fig.~\ref{ahc_fig1}). A comparison of fits of two logarithmic decay functions ($a \log(b-x)+c$) that intersect at infinity at the peak maximum with a fit to a single Lorentzian.  
 (b) For comparison we include the experimental data of Fig.~\ref{fig:KD2}, where this peak is marked by crosses. The estimated $\Upsilon$ for the experimental data is 0.45 for $\theta = 0.7^{\circ}$. The discrimination between logarithmic and Lorentzian is not so clear in experiment.}
    \label{log_peaks}
\end{figure}

Both the simulated spin and pseudo-spin structure factors confirm that the diffuse peaks drift~\cite{Moessner2003}, with increasing field tilt, from their initial locations close to the Brillouin zone boundaries, towards the pinch points at the zone center.  The intensity of the peaks (not illustrated here, see Ref.~[\onlinecite{HarmanClarke2012}]) is further confirmed to depend logarithmically on the size of the system as predicted~\cite{Moessner2003}, while the Bragg peak intensities scale linearly with the system size.  The peak shape (Fig.~\ref{log_peaks}) is also predicted to be logarithmic~\cite{Moessner2003}, and is well described by back-to-back logarithmic decay functions, which affords a superior description to a single Lorentzian (although the difference turns out to be less significant in experiment -- see Fig.~\ref{log_peaks} b). The pseudo-spin structure factor corresponds to that of an Ising antiferromagnet on the kagome lattice, constrained to the ``two-up-one-down'' sector of states by an external field.

\begin{figure}
\includegraphics[width=0.4\textwidth]{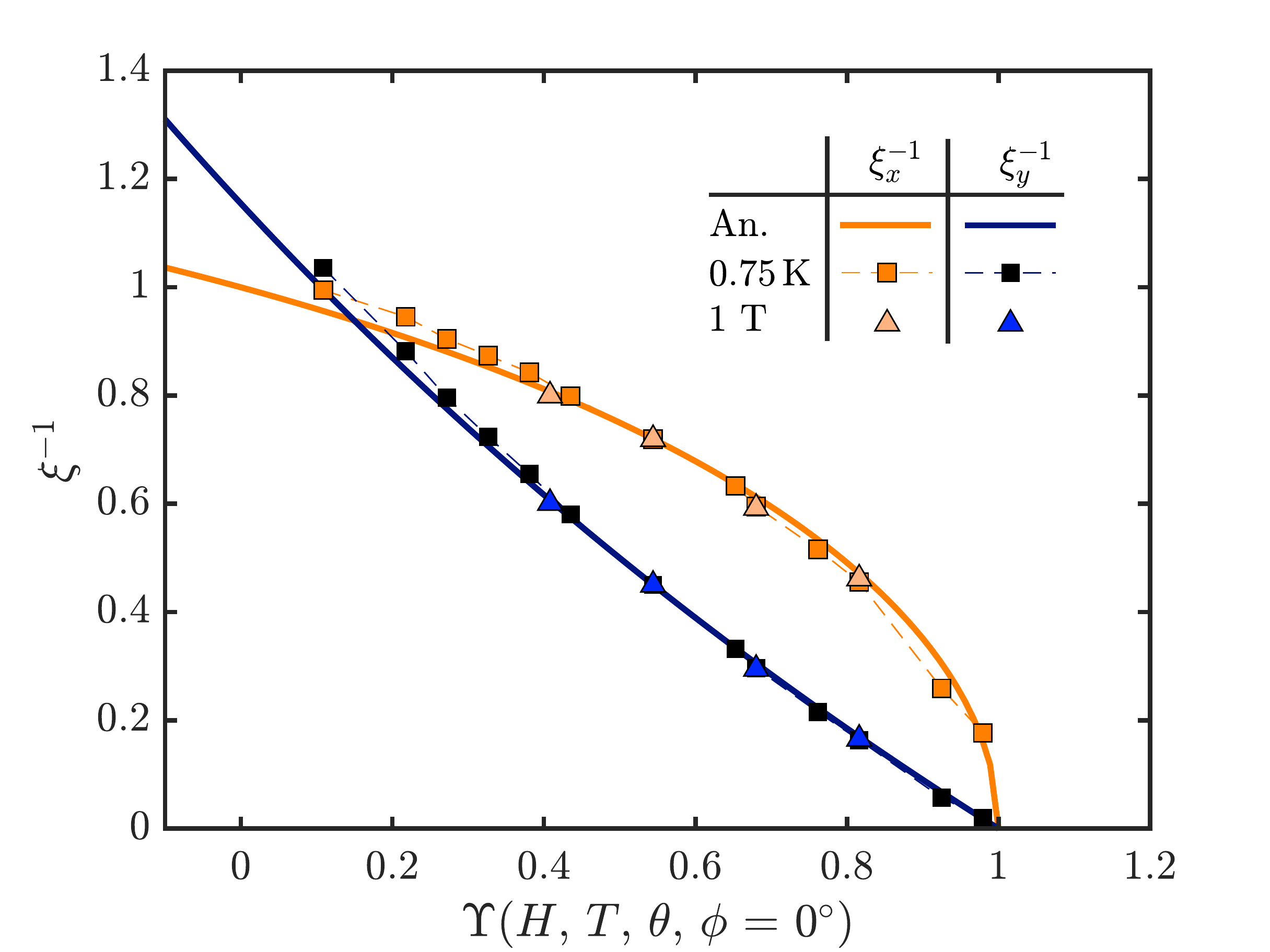}
    \caption{Simulated inverse correlation length $\xi^{-1}_x$ obtained by tracking the position of the logarithmic peak at $(\frac{\bar{4}}{3},\frac{4}{3},0)$ ($q_x=4/3$, $q_y=0$) (when $\Upsilon=0$) in the Monte Carlo simulation of the NSF channel, as well as $\xi^{-1}_y$ obtained from it, compared with the analytical predictions of Ref. [\onlinecite{Moessner2003}] (labelled `An.')
    }
    \label{fig:AHC_drift}
\end{figure}

\begin{figure}
\includegraphics[width=0.52\textwidth]{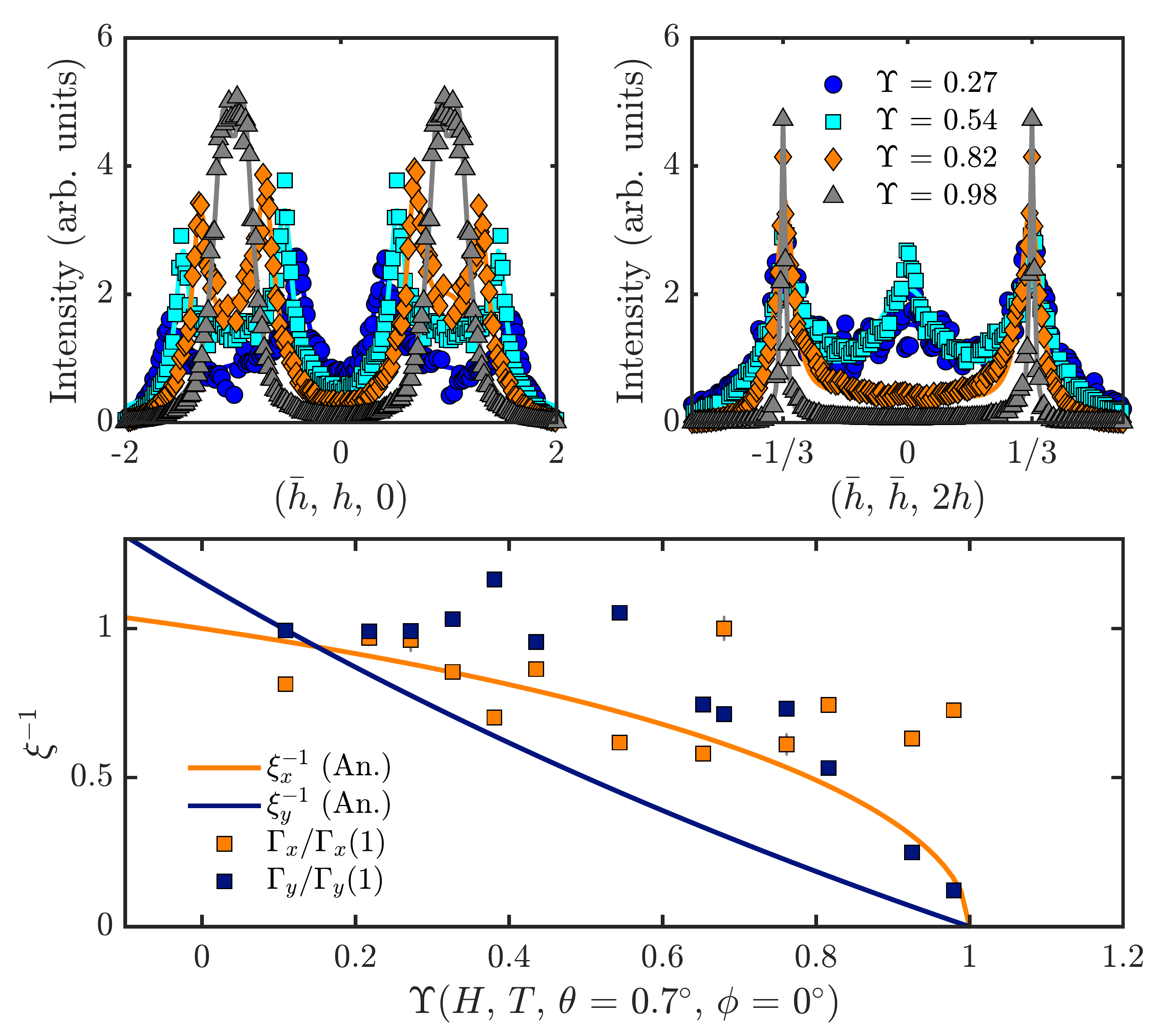}
    \caption{(a) Cuts through simulated data ($\phi = 0$) at several different values of $\Upsilon_{\rm K}$ approaching the Kasteleyn transition.  The cuts are perpendicular to $\hat{q}_y$, at  the average peak center for each of the four logarithmic peaks (which are fitted by 4 Lorentzians). (b) Cuts of simulated data at the same values of $\Upsilon_{\rm K}$ at the location on $\hat{q}_x$ of the average peak centers for two peaks that overlap with the experimental data. The central peak at small $\Upsilon_{\rm K}$ is another logarithmic peak that drifts towards a different kagome ice BZ center not found on this cut along $\hat{q}_x$.  (c) The average width $\Gamma$ of the four fitted Lorentzians in (a) as a function of $\Upsilon_{\rm K}$ for $x$ (cuts at constant $\hat{q}_y$) and $y$ (cuts at constant  $\hat{q}_x$), showing that the widths of the logarithmic peaks cut in either direction resemble  $\xi_x$ rather than $\xi_y$.}
    \label{fig:feature_widths}
\end{figure}

In Fig.~\ref{fig:AHC_drift} we show the inverse correlation length extracted by tracking the position of the logarithmic peak at $(\frac{\bar{4}}{3},\frac{4}{3},0)$ ($q_x=4/3$, $q_y=0$) in the NSF channel (i.e. the pseudo-spin correlation function). The peak positions are extracted from the simulated scattering pattern by applying the same peak tracking algorithm that we apply to the experimental data and describe in full in Appendix~\ref{peak_tracking_appendix}.  We see that, despite the peak appearing at a different reciprocal space position, the correlation length extracted from its drifting position as the Kasteleyn transition is approached is as predicted~\cite{Moessner2003}.   No feature of the scattering pattern was proposed in Ref. [\onlinecite{Moessner2003}] for the independent measurement of $\xi^{-1}_y$, and it can be seen in Fig.~\ref{ahc_fig1} that no peak does actually move along the $y$-direction when the tilted field is symmetric $(\theta>0,\phi=0)$.  However, $\xi^{-1}_y$ does depend on $\xi^{-1}_x$ and if the extracted values of $\xi^{-1}_x$ are used to obtain $\xi^{-1}_y$ as prescribed, we see that it does scale as predicted.  Usually inverse correlation lengths are measured using the width of features in reciprocal space.  We find that the width of the logarithmic peak along both $x$ and $y$, at the peak position, scales like $\xi^{-1}_x$, as shown in Fig.~\ref{fig:feature_widths}.

\subsection{Exploration of general tilt, $\phi\ne0$}

In this section, we identify some topological consequences of general tilt, $\phi \ne 0$. 

\subsubsection{Scattering function}

\begin{figure*}
\centering
\includegraphics[width=0.9\textwidth]{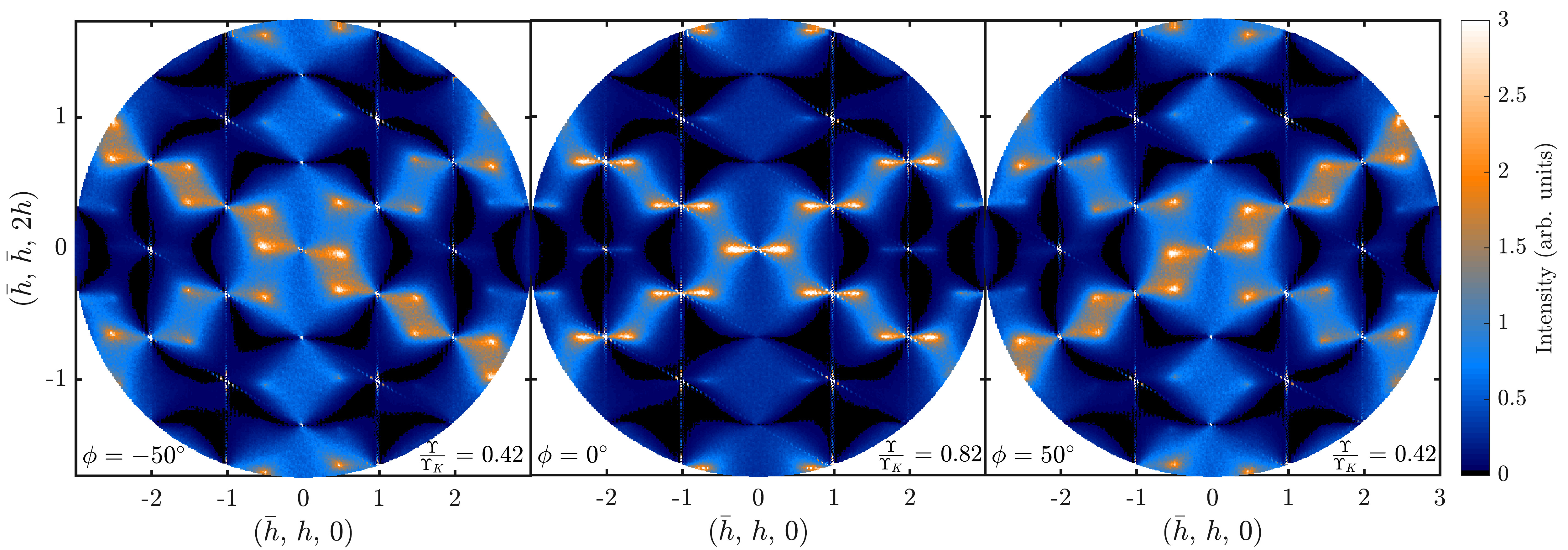}
    \caption{Simulated spin flip neutron scattering structure factor with angles $\phi =-50,0,50^{\circ}$ (left to right) corresponding to $\Upsilon/\Upsilon_{\rm K} = 0.42, 0.82, 0.42$ respectively (in the experimental system this would correspond to 0.75 K and 1.5 T for all values of $\phi$).
   % given to provide the reader an intuitive feel for how close each panel is to the Kasteleyn transition at $\frac{\Upsilon}{\Upsilon_{\rm K}} = 1$.   
   }
    \label{fig:StructFact}
\end{figure*}

\begin{figure*}
\centering
\includegraphics[width=0.9\textwidth]{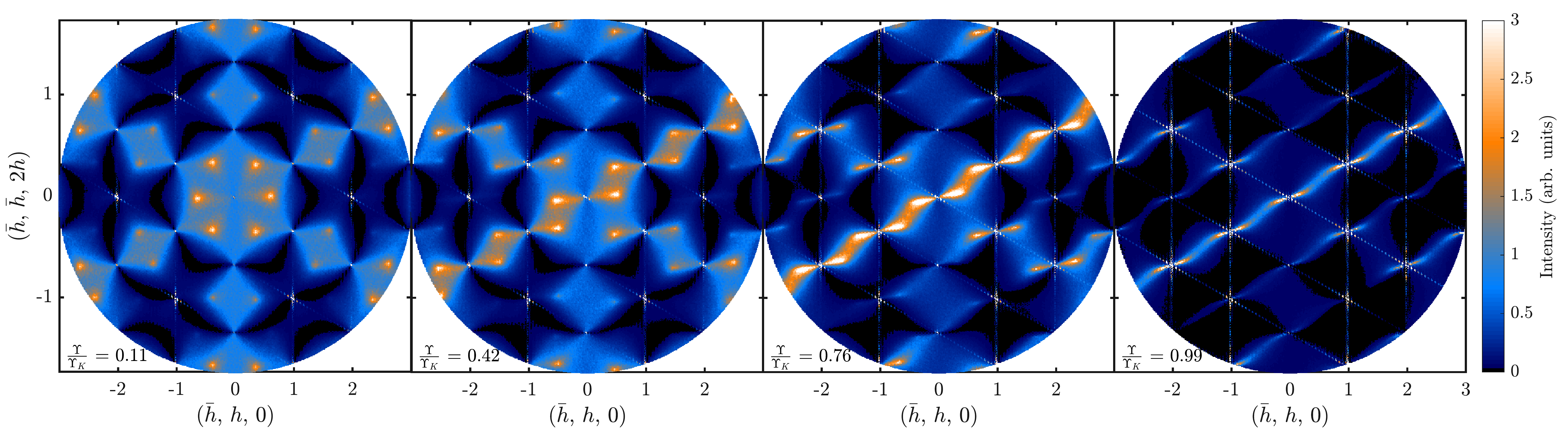}
    \caption{Simulated spin flip neutron scattering structure factor with an angle $\phi = 50^\circ$, as a function of $\frac{\Upsilon}{\Upsilon_{\rm K}}$ (Eqn. \ref{eq:KTCphi}) at  $\Upsilon/\Upsilon_{\rm K} = 0.11, 0.42, 0.76, 0.99$ (in the experimental system this would correspond to 0.75 K and 0.4, 1.5, 2.7, 3.5 T respectively). 
    \label{fig:TiltedStructFactDrift}}
\end{figure*}

Finite $\phi$ acts to further lower the symmetry for dimer placement in the kagome plane and hence further lowers the symmetry of the in-plane scattering function. The effects of this reduced symmetry are shown in Fig.~\ref{fig:StructFact}, where the two-fold symmetry observed for $\phi=0$, which is visible in Fig.~\ref{ahc_fig1}, is reduced, with the emergence of an `arm' of intense scattering on each side of the scattering plane. However, features still drift as the system evolves towards the Kasteleyn transition, as shown in Fig.~\ref{fig:TiltedStructFactDrift}.

\subsubsection{Finite Size Scaling of the Susceptiblity}

\begin{figure}
\centering
\includegraphics[width=0.4\textwidth]{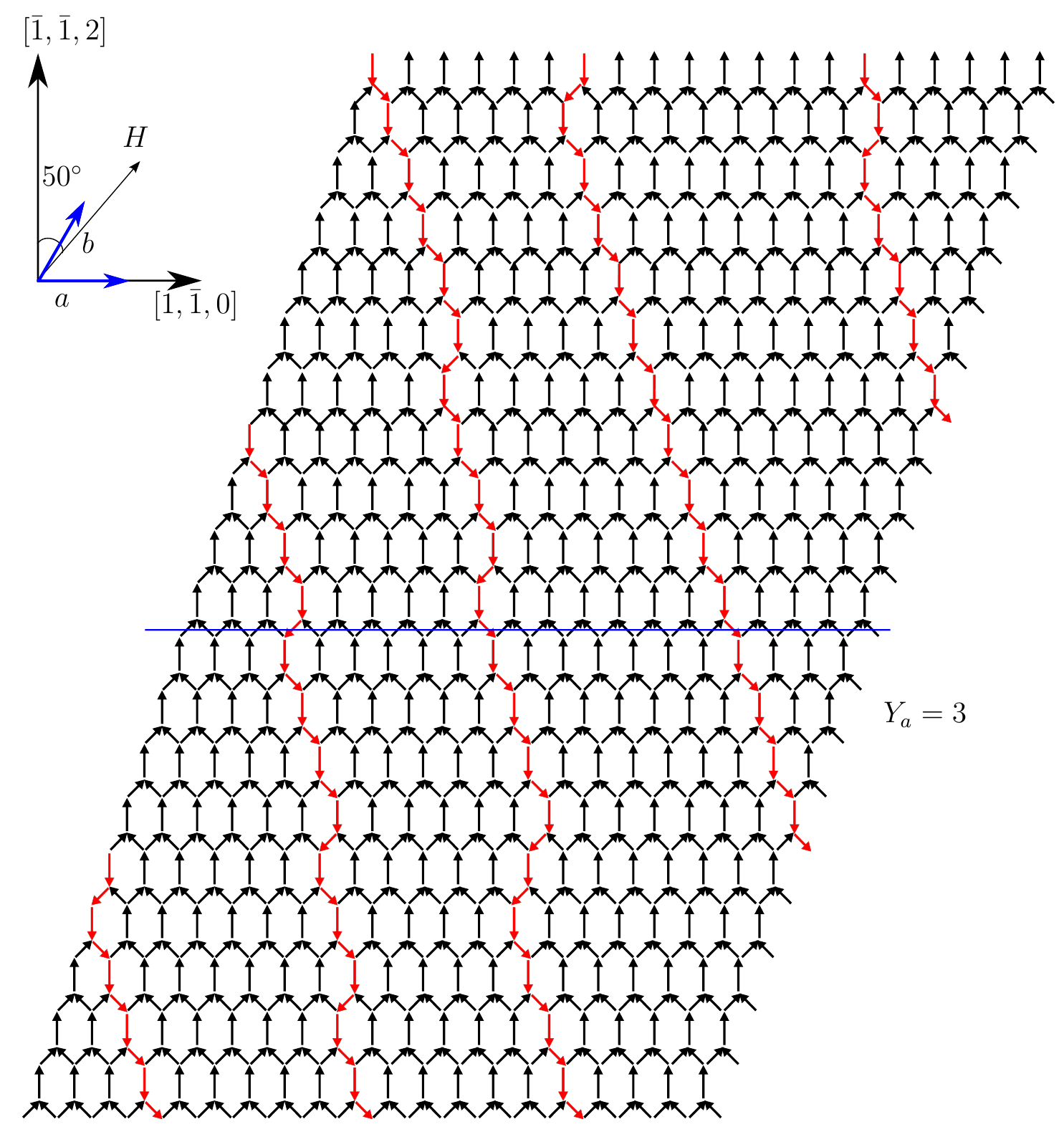}
    \caption{A snapshot of the kagome ice lattice taken from a simulation at $T = T_{\rm K}$ and $\phi= 50^{\circ}$. The lattice contains a single string (highlighted in red) which winds around the lattice through the periodic boundary conditions until it closes on itself. This snapshot shows the loop forming in a direction that is approximately perpendicular to the field direction as this minimises its unfavourable Zeeman interaction. The blue line parallel to the lattice vector is a trajectory for calculating a cut number which records the number of times a string is encountered in that direction.}
    \label{fig:adamfig55}
\end{figure}

Rotating the tilted field in the plane has spectacular consequences for the string insertion close to the transition. In Fig.~\ref{fig:adamfig55} we show a snapshot of the in-plane spins for $\phi=50^{\circ}$ in which we highlight the first string placed in the system. The string progresses through the system propagating in a mean direction, $\phi'<0$ which minimises the unfavourable Zeeman energy (see Appendix~\ref{loop_appendix}). This makes it  incompatible with the periodic boundaries, forcing it to make multiple loops of the system before closing on itself. The multiple passages are characterised by a cut variable $Y$, as shown in the figure. 

This approach to incommensurability has a dramatic effect on the finite size scaling properties of the susceptibility $\chi$, which following  the finite size scaling hypothesis can be written, close to the transition in the form
\begin{equation}
\chi=\eta^{-\gamma}\mathcal{G}\left(\frac{\xi_x}{RL},\frac{\xi_y}{L}\right),
\end{equation}
where $\mathcal{G}$ is a scaling function and $R=\left(\frac{1}{Y}\right)^{\frac{\nu_y}{\nu_x}}$ is an incommensurability function similar to the shape function studied in the context of directed polymers \cite{Bhattacharjee1983}. When $\phi=0$, as $\xi_y \gg \xi_x$, the scaling function should be dominated by the ballistic propagation of the strings in the $\hat{y}$ direction and consequently depend on the single variable $\xi_y/L$. This is achieved by setting $Y=0$ ($R=\infty$). 

In an incommensurate situation, as the string length becomes indeterminate, the diffusive evolution of the string in the perpendicular plane, with associated correlation length $\xi_x$ should dominate the finite size scaling. As the incommensurability factor evolves from $Y=0$ ($R = \infty$) to $Y=\infty$ ($R=0$), the scaling function should evolve between these regimes. For any finite $R$, $\mathcal{G}$ should ultimately crossover to the ballistic case, with the crossover region scaling closer and closer to the transition as $R$ increases.

In consequence, different finite size scaling predictions emerge in the two limits. Identifying $\eta^{-\gamma}=\xi_y^{\gamma/\nu_y}$ in the ballistic limit and $\eta^{-\gamma}$ as $\xi_x^{\gamma/\nu_x}$ in the incommensurate limit 
the scaling hypothesis can be re-written
\begin{eqnarray}
\chi_B&=& L^{\gamma/\nu_y}\mathcal{G}_B\left(\frac{\xi_y}{L}\right) \nonumber \\
\chi_I&=& L^{\gamma/\nu_x}\mathcal{G}_I\left(\frac{\xi_x}{L}\right)
\end{eqnarray}
where $\mathcal{G}_B(\xi_y/L)$ and $\mathcal{G}_I(\xi_x/L)$ are new scaling functions. 

\begin{figure}
\centering
\includegraphics[width=0.4\textwidth,trim=90 150 90 100]{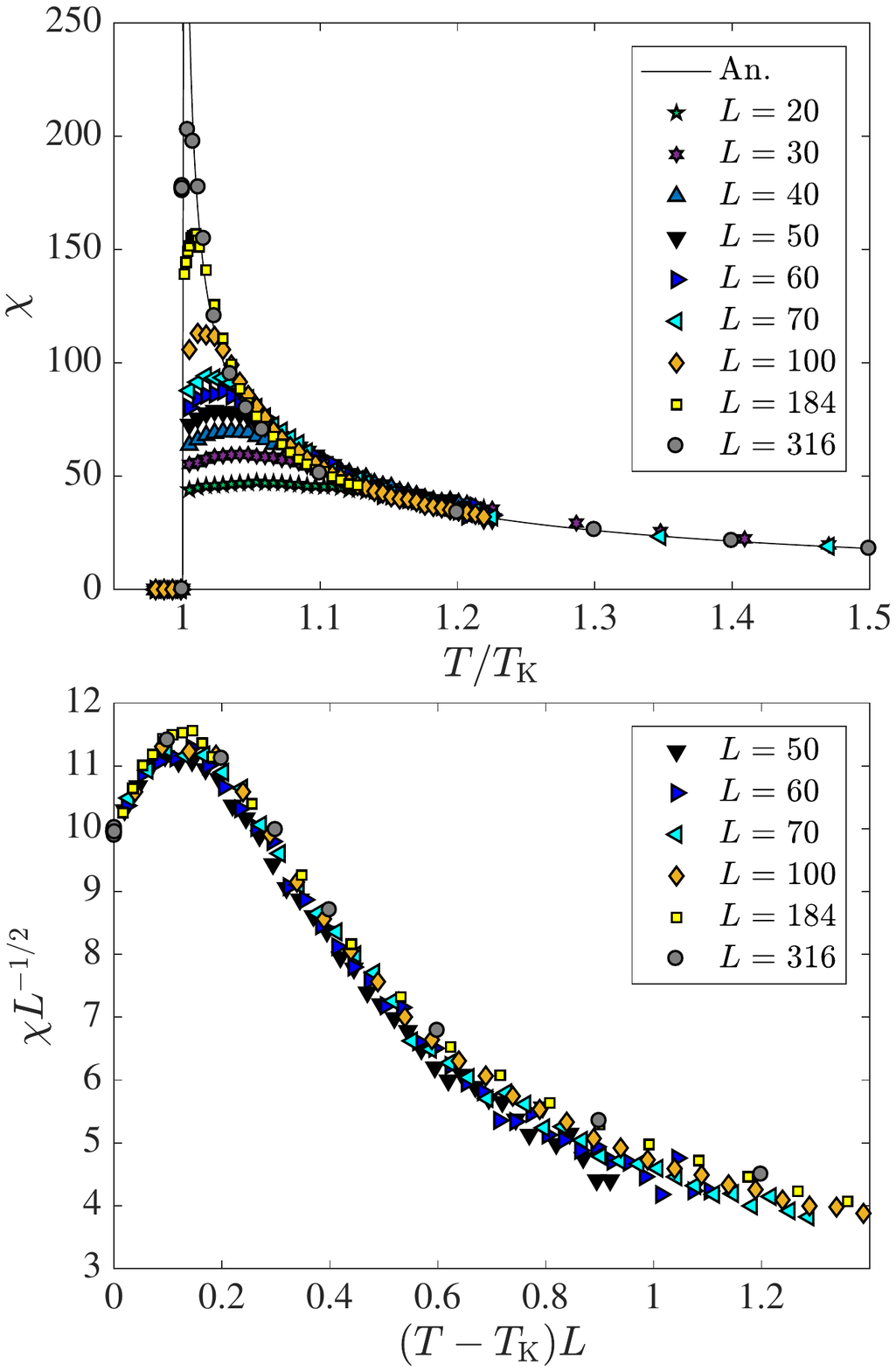}
    \caption{Top: simulated susceptibility of the kagome ice lattice with field at an angle $\phi = 0^{\circ}$ as a function of lattice size. Also shown is the analytical susceptibility in the thermodynamic limit (labeled `An.)~\cite{Moessner2003}. Bottom: The same data plotted as a function of scaling variables in the ballistic limit. The collapse of the data over a large range of lattice sizes validates the ballistic scaling controlled by the correlation length parallel to the direction of the ordered moment. }
    %{\color{cyan}{Peter: is this directly from M\&S?}} {\color{cyan}{Adam: is it possible to find the numerical data that appears in this figure?!}}
    \label{fig:adamfig56}
\end{figure}

\begin{figure}
\centering
\includegraphics[width=0.4\textwidth,trim=90 150 90 100]{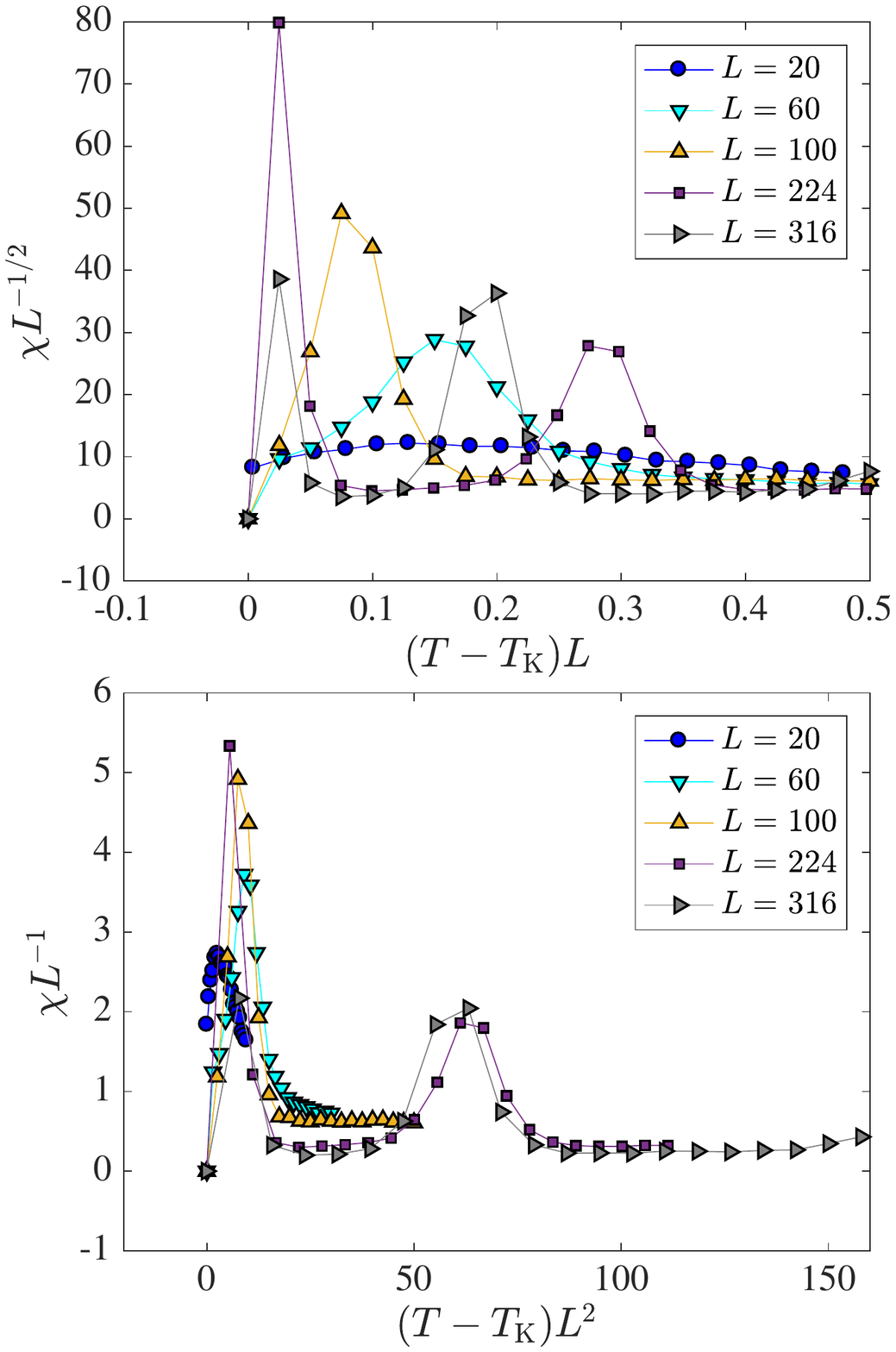}
    \caption{The simulated finite size scaling function with applied at an angle of $\phi=20^{\circ}$. In the top panel the data is plotted 
    as a function of scaling variables in the ballistic limit. In the bottom panel the scaling variables for the incommensurate limit are used. A better fit is clearly obtained in the lower panel illustrating the crossover with $\phi$ between the two scaling regimes governed by the correlation lengths parallel and perpendicular to the direction of the ordered moment. }
    \label{fig:adamfig57}
\end{figure}

This phenomenology is confirmed by simulation in Figs. \ref{fig:adamfig56} and \ref{fig:adamfig57}. In Fig. \ref{fig:adamfig56} we show the susceptibility near the transition for $\phi=0$ and for different system sizes. The divergence of the susceptibility at the transition is cut off by the finite size of the sample, as shown in the upper panel. The finite size scaling is tested in the lower panel where we plot  $\chi L^{-1/2}$ against $(T-T_{\rm K})L$ at fixed $H$, finding an excellent data collapse corresponding to the ballistic limit.

In Fig. \ref{fig:adamfig57} we show finite size scaling for data with $\phi=20^{\circ}$. In the upper panel we show that the ballistic scaling fails to give data collapse. However, in the lower panel we test the incommensurate scaling by plotting $\chi L^{-1}$ against $(T-T_{\rm K})L^2$ at fixed $H$, finding an encouraging  collapse of the numerical data. Although the simulation is quite challenging, and the data remains noisy, it seems that we have clear evidence of the crossover between the two scaling limits \cite{Bhattacharjee1983}. 

In the first instance the ballistic universal function shows a broad single peaked function, while in the incommensurate case two peaks are visible. This is because, in the ballistic case, the scaling is many body, with many simultaneous strings present in the large-$L$ limit. In the incommensurate case, despite reaching the scaling limit, one sees the individual effect of adding one string at a time. Our data shows the effect of adding a first string, and then a second, but more extensive simulations should reveal a comb of single string peaks stretching out from the transition \cite{Bhattacharjee1983}.

\begin{figure}
\centering
\includegraphics[width=0.4\textwidth,trim=1 1 1 1]{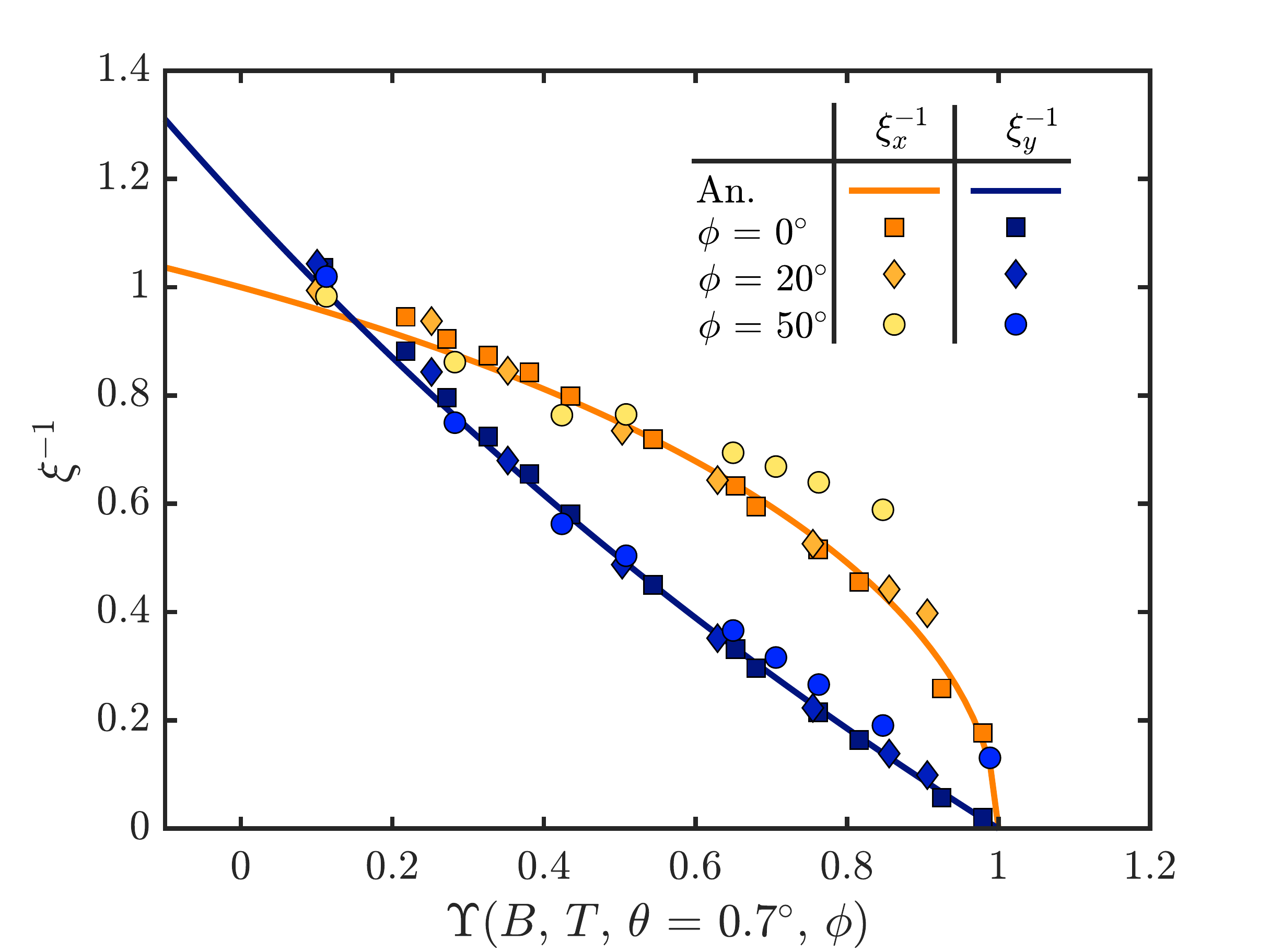}
    \caption{Simulated scaling of the correlation lengths in systems with $\phi=0$, $20$, and $50^{\circ}$. (compare with Fig. 6, where $\phi=0$), as obtained by tracking the position of the peak at $(\frac{\bar{4}}{3},\frac{4}{3},0)$ ($q_x=4/3$, $q_y=0$) (when $\Upsilon=0$) in the Monte Carlo simulation of the NSF channel, compared with the analytical predictions of Ref. [\onlinecite{Moessner2003}] (labelled `An.').}
    \label{fig:scaling_finite_phi}
\end{figure}

Finally, in Fig.~\ref{fig:scaling_finite_phi} we show the equivalent of Fig. 6 for $\phi= 20^{\circ}$ and $\phi= 50^{\circ}$, the evolution of the correlation lengths with $\Upsilon_{\rm K}$. We see that, despite the radical change in finite size scaling of the susceptibility as $\phi$ increases from zero, the behaviour of the correlation lengths as a function $\Upsilon_{\rm K}$ is independent of the finite size scaling regime reached.  

%{\color{green}{Peter: Effect of $\phi>0$, including finite size scaling of systems with general tilt}}

\section{\label{sec:Experimental_methods}Experiment}

\subsection{Method}

A large single crystal of \hto{} was grown by the floating zone method. The crystal approximated a long cylinder ($d \approx 7$ mm, $l \approx 60$ mm) with a visible $(1, 1, 1)$-type facet running the length of the boule.  It was previously used to measure diffuse scattering in zero field~\cite{Fennell2009}.  To ensure a precise alignment, a face perpendicular to the longest $\langle 1 1 1 \rangle$ direction was cut, so that when the crystal rests on the cut face, the $(\bar{h}, \bar{h}, 2h)-(\bar{h}, h, 0)$ plane lies in the horizontal scattering plane of a neutron spectrometer, with the $[111]$ direction vertical.  The long axis of the crystal boule makes an angle of $26^{\circ}$ with the vertical direction and is coplanar with the $[111]$ and $(\bar{h}, \bar{h}, 2h)$ directions (Fig. \ref{fig:Alignment}).  

The crystal was held in a copper clamp running the length of the boule, attached to a block with cut-outs that allow two adjustable orthogonal tilts and a continuous metallic path from the sample clamp to the mixing chamber of the dilution fridge.  Neutron Laue diffraction measurements were used to refine the alignment and showed that the crystal was mounted with $\theta\lessapprox 0.1^{\circ}$ and $\phi\lessapprox 0.2^{\circ}$ and the adjustable tilts were locked by opposing screws to prevent any movement of the sample by the applied field.  The sample was mounted in a dilution refrigerator insert, which itself was mounted in a 2.5 T vertical field cryomagnet.  The cryomagnet plus refrigerator insert and crystal were placed in the polarised neutron diffuse scattering spectrometer D7 at the Institut Laue Langevin (Grenoble, France)~\cite{Stewart:2008hw} with the $(\bar{h}, \bar{h}, 2h)$ defined by Laue diffraction approximately antialigned with the incident beam (i.e. with the tilted crystal boule approximately coplanar with the incident beam and vertical field and $[111]$ directions).

Diffuse scattering structure factors were measured in the  $(\bar{h}, \bar{h}, 2h)-(\bar{h}, h, 0)$ scattering plane using neutrons with wavelength $\lambda=4.8$ \AA.  The flipper currents were optimised using a 40 mm long `quartz' (amorphous silica) sample (matching the sample height) for selected fields from 0.1 to 2.5 T.  The cryomagnet provides the guide field at the sample position, and only $z$ (NSF) and $z'$ (SF) channels can be measured.  Empty sample holder measurements were used to subtract field independent background scattering.  Measurements of the quartz sample (for each field) and a vanadium cylinder (at 0.1 T) were used to calibrate polarisation efficiency and detector efficiency respectively.

Maps of the structure factor were made by measuring intensity in the $z$ and $z'$ channels while rotating the crystal about the vertical axis, also called an $\omega$-scan.  
%The angle $\omega$ has no special relation with the crystal axes and runs from 0 to 360 $\deg$.  
The data were transformed from the $\omega-2\theta$ frame to $q_x-q_y$, and an additional arbitrary offset angle, $\omega_0$, was used to rotate the scattering map to bring particular features to special values of $(q_x,q_y)$.  Usually this corresponded to placing two orthogonal Bragg peaks on $q_x$ and $q_y$ axes so that the scattering plane could be identified by its symmetry axes.  In this case, we also used $\omega_0$ to place a specific feature such as a pinch point parallel or perpendicular to $q_x$ or $q_y$ to facilitate cutting through the feature, and, by relating the initial crystal orientation with the $\omega$ angles at which Bragg peaks were observed, we could identify which of the $\langle\bar{h},\bar{h},2h\rangle$ axes were associated with the real shape of the crystal.

\subsection{\label{sec:Results}Results}

\begin{figure*}
\centering
\includegraphics[width=0.9\textwidth]{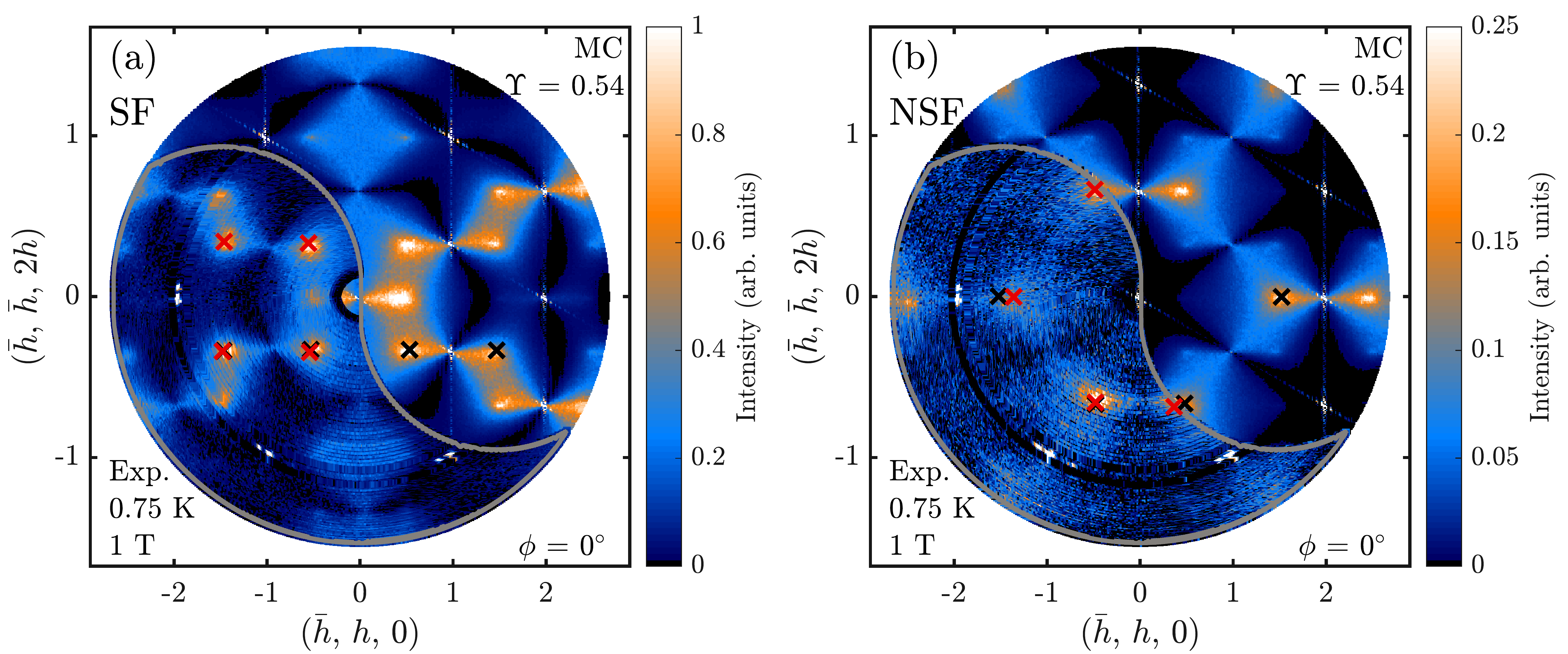}
\caption{\label{fig:KD2}Experimental data for the SF channel and NSF channel measured at 0.75 K and 1.0 T, compared with Monte Carlo simulations of kagome ice at $\Upsilon=0.54$ ($\phi=0$).  The experimental data is in the lower left quadrant of each panel, simulation (`MC') in the upper right. Crosses indicate the peak positions found by the peak tracking algorithm. }
 \label{fig:Diff_Data}
\end{figure*}

\subsubsection{Field alignment}

%\begin{figure}
%\includegraphics[scale = 0.2,trim=1 1 1 1 ]{Graphics/SF_NSF_4f_v2.png}
%\centering
%    \caption{Spin flip (a,b) and non-spin flip (c,d) diffuse scattering at 0.75 K {\color{cyan}{Just panel a and c, side by side, colour bar horizontal at the top}}}
 %  \label{fig:Diff_Data}
%\end{figure}

\begin{figure}
\centering
\includegraphics[scale=0.3]{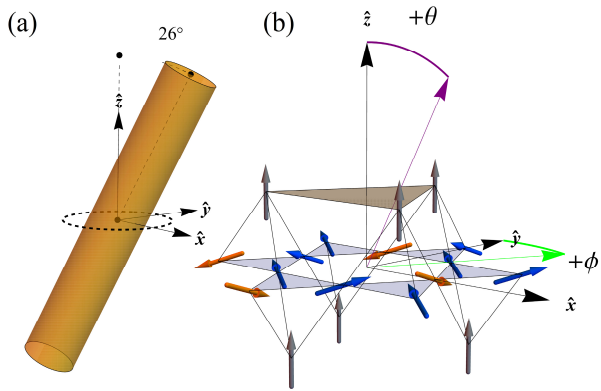}
\caption{\label{fig:Alignment}(a) Schematic of the crystal in direct space showing the axis vectors $\hat{x} = (\bar{1} 1 0)$, $\hat{y} = (\bar{1} \bar{1} 2)$ , $\hat{z} = (1 1 1)$. (b) The lattice in the same orientation, with the angles $\phi$ and $\theta$ defined in purple and green.}
\end{figure}

Although the crystal axes were initially aligned precisely with the applied field, comparison of the recorded data with simulation immediately suggests that the actual field within the sample was tilted. That is, the diffuse scattering data measured in the $(\bar{h}, \bar{h}, 2h)-(\bar{h}, h, 0)$ plane have distinctive features, characteristic of the kagome ice phase in tilted field.  As can be seen in Fig.~\ref{fig:Diff_Data}, the symmetry of the diffuse scattering in the SF and NSF channels is reduced from six-fold, to approximately two-fold.  This is particularly pronounced  in the SF channel, where the scattering is stronger, and the reduced symmetry can be clearly seen by comparing to a Monte Carlo simulation of kagome ice with a field tilt, $\theta>0,\phi=0$.  Quantitative comparisons with theory and simulation discussed below suggest that $\theta\approx0.7^{\circ}$ (and $\phi\approx0$). Note that while this crystal showed positive tilt, a previously studied crystal of a different shape~\cite{Fennell2007} showed negative tilt of similar magnitude.

\begin{figure*}
\includegraphics[width=0.9\textwidth]{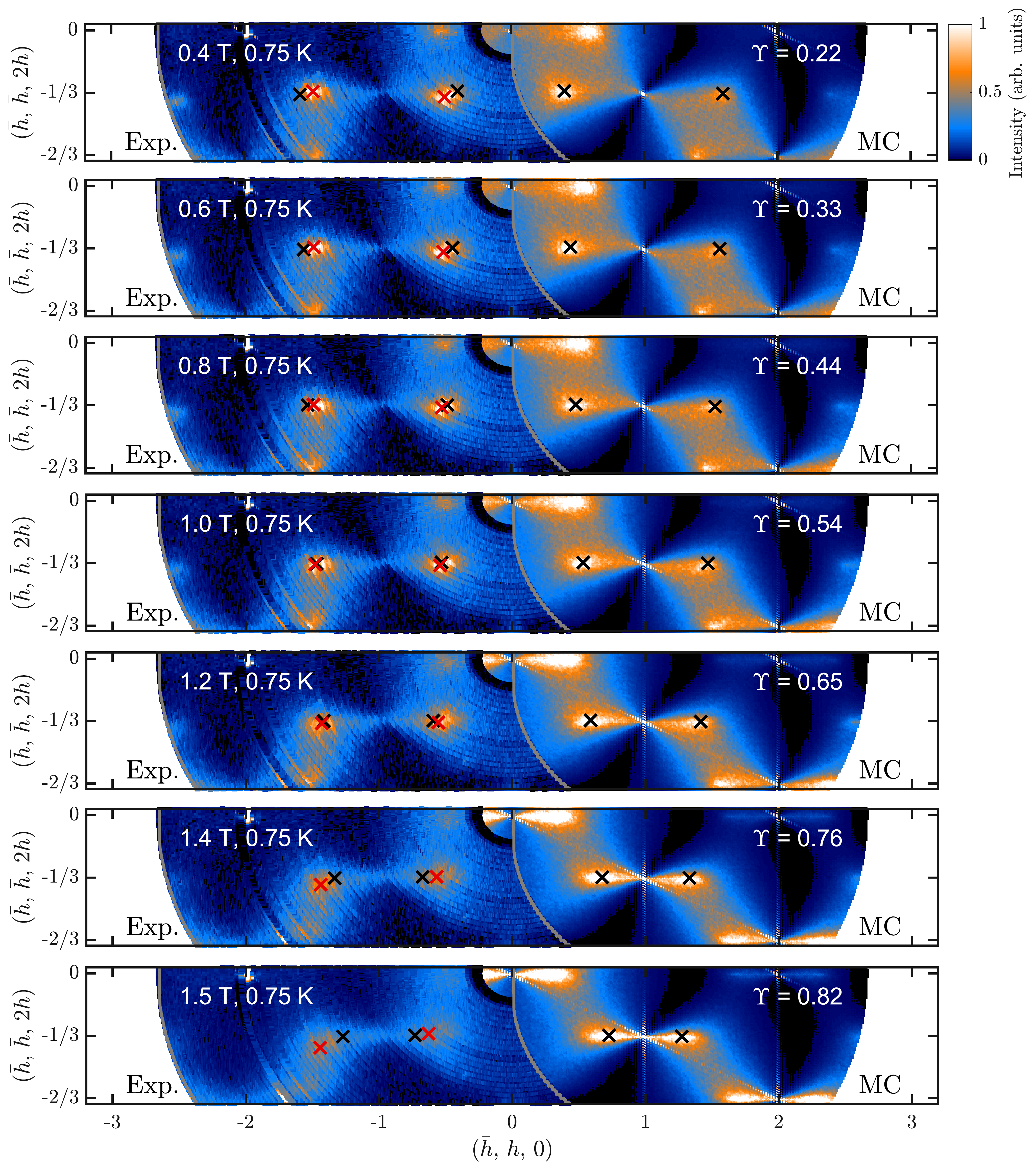}
\centering
    \caption{Diffuse scattering in the SF channel (i.e. due to kagome ice spin components) as a function of applied field at fixed temperature of 0.75 K.   Experimental data is shown in the left part of each panel, and Monte Carlo simulations in the right part of the panel.  The simulations are carried out at the value of $\Upsilon_{\rm K}$ determined by these field and temperature values that best match $\theta$.    Crosses show the experimental (red) and simulated (black) peak positions.}
        \label{fig:maps}
\end{figure*}

Given the precise alignment of the crystal, 
%it is at first sight surprising to see this signature of a tilted field in the structure factor. 
an obvious cause of the tilted field is the large demagnetizing effects in spin ice~\cite{Diepbook}.  If the sample is approximated as an ellipsoid whose unique axis is misaligned with the applied field, then the internal field will be uniform but not parallel to to the applied field. As noted by Morris {\it et al.}~\cite{Morris2009} the large, anisotropic shaped crystals of spin ice typically used in neutron scattering typically suffer from a significant misalignment of internal field and applied field.  The exceptionally large demagnetizing fields of  spin ice are well established, having been much discussed with respect to experimental corrections~\cite{Quilliam:2011df} and also as exemplifying departures from the usual textbook theory of demagnetizing factor~\cite{Bovo:2013fq,Twengstrom:2017df}.  The elongated shape of our sample and its tilted relation to the scattering plane therefore provides a convincing explanation of why the internal field is tilted and this may be safely assumed going forwards.    

The  $\omega$ angles at which the $(\bar{2},2,0)$ reflections were observed tell us that the uniquely selected $(h,h,\bar{2h})$ direction is not the one that is coplanar with the maximum tilt of the crystal boule as would be the case for a weakly magnetized sample, but is instead at $60^{\circ}$ to this direction.  We have chosen $\omega_0$ to place the uniquely selected direction parallel to $q_y$.
Fig.~\ref{fig:Alignment} shows its relation to the crystal in direct space.  

%\textcolor{red}{The following paragraph does not make sense with what has been said -- is this referring to two new crystals? If so, it should be stated. Or are these unreported experiments? If so, delete paragraph. }
%\textcolor{blue}{Our observations of the effect of a tilted field on kagome ice include several experiments on two samples with very different shapes.  The tilts of both samples, as measured by neutron diffractometers, were very small (less than $0.5 \deg$).  Signatures of negative tilt were obtained with one sample over the course of several experiments with intervening dismounting and remounting of the sample, such that the observed tilts of the scattering vectors were always similar and small, but not precisely the same~\cite{Fennell2007}.  This sample is shaped like a flattened lozenge, approximately 10 mm in length and 3 mm thick, giving a large flat face.  This face is a $\langle 111\rangle$-type direction, but in a vertical field experiment is not the direction of the applied field.  The face is inclined at an angle of $19 \deg$ with respect to the field, and the top and bottom of the sample are terminated by sharp $\langle110\rangle$-type edges.  With the second sample, which is described here, and which has a qualitatively different shape to the first, although the observed tilts were again very small, clear signatures of a positive tilt were obtained. } 

\subsubsection{Correlation functions}

As discussed above, the SF channel measures in plane spin components and hence in-plane kagome ice correlations, while the NSF channel detects out of plane components and hence  correlations of the pseudo-spins - a model kagome Ising antiferromagnet.  The latter cross section is much weaker, because the major part of the total $10$ $\mu_{\rm B}$ moment (8/9) contributes to the scattering process in the spin flip channel, while only a smaller projection (1/9) contributes to the NSF channel.  However, general comparison with simulations in Fig.~\ref{fig:Diff_Data} clearly shows the different forms of the scattering, and that descriptions in terms of either correlation function are warranted. 

Fig.~\ref{fig:maps} compares the SF scattering under different applied fields at a fixed temperature of 0.75 K. It shows that features in the diffuse scattering do clearly drift across the scattering plane as the field is changed~\cite{Moessner2003}, as was found with a more limited data set in Ref.~\onlinecite{Fennell2007}.  At fixed field, the features also move as a function of temperature.  However, we see that for $T\lessapprox 0.6$ K, the scattering pattern becomes independent of temperature and field within the kagome ice plateau.  

\begin{figure}
\includegraphics[width=0.5\textwidth]{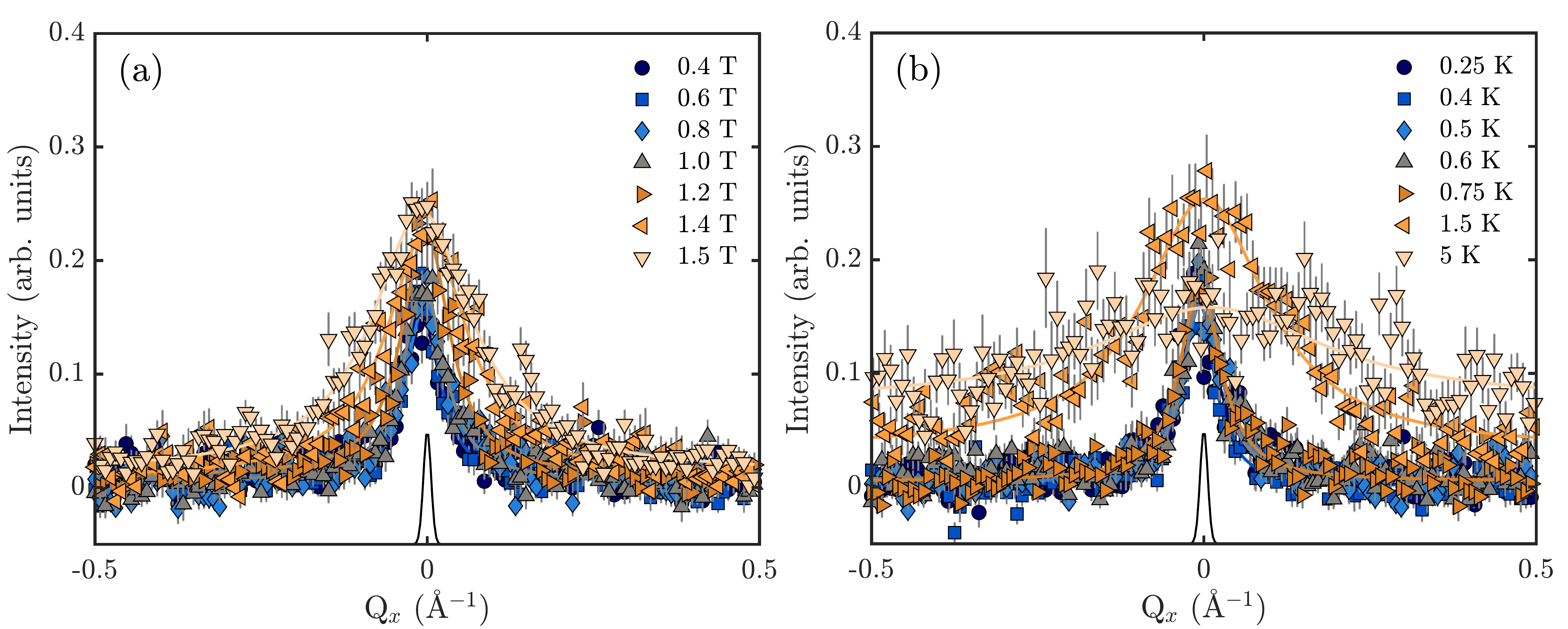}
\centering
    \caption{Cuts through the experimental pinch point at $(\frac{2}{3},\frac{2}{3},\frac{\bar{4}}{3})$ ($q_x=0$, $q_y=2/3$) as a function of field at constant temperature of 0.75 K (a) and as a function of temperature at constant field of 1 T (b).  At constant temperature ($\Upsilon_{\rm K}\propto B$) the pinch point broadening with field indicates a crossover out of the kagome ice phase. At constant field ($\Upsilon_{\rm K}^{-1}\propto T$), the increase in background at 1.5 K suggests an increase in monopole population and a more paramagnetic state. At low temperature the pinch point width does not seem to decrease below $T \approx 0.6$ K. }
        \label{fig:pp}
\end{figure}

\subsubsection{Aspects not addressed by theory}

The experimental data give access to two more aspects of kagome ice that are not addressed by the theory or the simulations described above: the behavior of the pinch points, and the approach to plateau termination.

Referring to Fig.~\ref{fig:pp}a, we see that, within the kagome ice plateau, at a constant applied field of 1 T, the pinch points sharpen as the temperature decreases to $T\lessapprox 0.6$ K, and below this temperature they remain of constant width, as with other parts of the diffuse scattering mentioned above.  At fixed temperature (0.75 K), the pinch points are similarly sharp within the kagome ice plateau ($0.4<H<1.4$ T), but as the plateau termination field approaches, they begin to broaden (Fig.~\ref{fig:pp}b).

\subsection{Scaling}

\begin{figure}
\includegraphics[width=0.4\textwidth]{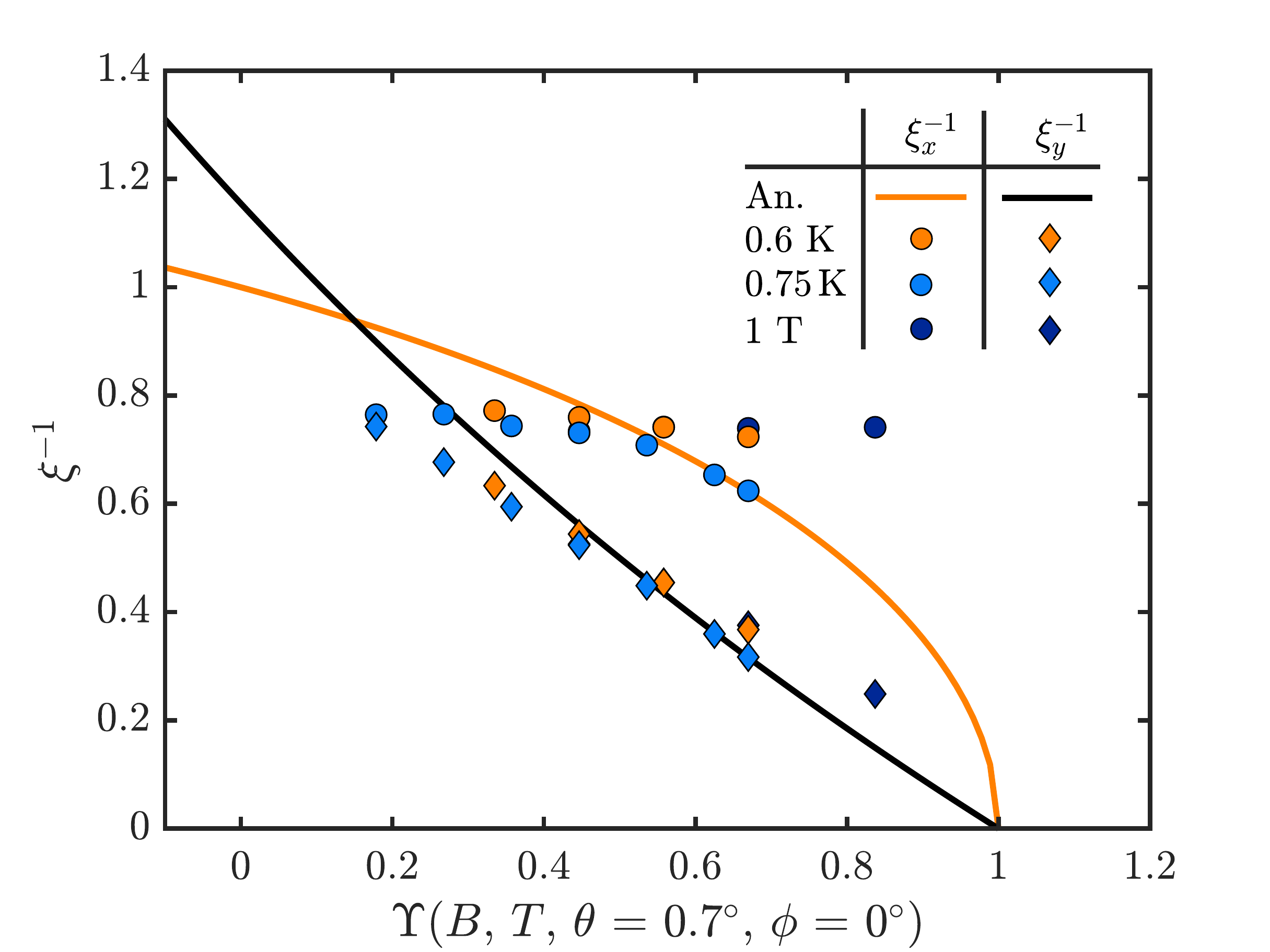}
\centering
    \caption{Comparison between the analytical Eq \ref{eq:InvCorrLength} (`An.') and experimental correlation lengths as derived from the peak positions in the neutron scattering data, either at constant field or at constant temperature.  The angle $\theta$ is chosen to be $0.7^{\circ}$ as this best scales the data, while $\phi$ is chosen to be $0^{\circ}$, as suggested by the symmetry of the diffuse scattering maps.  }
        \label{fig:InvCorr}
\end{figure}

From tracking peak locations in the diffuse neutron scattering patterns and fitting the wavevectors and intensities, we can reproduce the normalized inverse correlations $\xi^{-1}_x$ and $\xi^{-1}_y$  (see Section~\ref{sec:corr_func} and~\ref{sec:drift_sim}).  The location of a pair of diffuse scattering peaks on either side of the pinch point at $(\frac{\bar{2}}{3},\frac{4}{3},\frac{\bar{2}}{3})$ ($q_x=-1$, $q_y=-1/3$), as indicated on Fig.~\ref{fig:Diff_Data} and~\ref{fig:maps} by crosses (i.e. at $q_x\sim-2/3,-4/3$, $q_y\sim-1/3$ in Fig.~\ref{fig:Diff_Data} and~\ref{fig:maps}), were extracted with the peak tracking algorithm described in Appendix~\ref{peak_tracking_appendix}.  The reciprocal space distance from each diffuse peak to the kagome Brillouin zone center in $\hat{x}$ (as denoted by the white hexagons in Fig.~\ref{fig:Diff_Data} and Fig.~\ref{fig:maps}) was averaged to produce the inverse correlation length $\xi^{-1}_x$.  The result is shown in Fig.~\ref{fig:InvCorr}.

For $\xi^{-1}_x$, a square root critical exponent is consistent with the data measured at 0.75 K as the scaling parameter $\Upsilon(B$, $T$, $\theta)$ approaches the Kasteleyn transition.  It is apparent from the above discussion that without using an independent feature to extract it, $\xi^{-1}_y$ will also follow the predicted form.  The most effective scaling collapse (Fig.~\ref{fig:InvCorr}) incorporates data points in the window $0.4<B<1.4$ T and $0.6<T<1.0$ K, with values of $\theta\approx0.7^{\circ}$ and $\phi\approx0^{\circ}$.  We see that if the temperature is too low, the drift of the diffuse scattering features ceases: one plausible explanation being that the dynamics start to slow down, as commonly observed in spin ice at this temperature~\cite{Paulsen:2014cc}. Also, if the field is too large or too small, or the temperature too high, the data depart from the scaling form as the kagome ice phase is no longer well defined.  These two effects account for points at small or large $\Upsilon$ that depart from the scaling form.  

\section{Discussion}\label{sec:Discussion}

In the following, we discuss our analytical, numerical, and experimental results together.  We start in the kagome ice phase far from a Kasteleyn transition and follow the evolution of the system towards the transition.

\subsection{Critical correlations}

\subsubsection{Non-tilted case}

Although it proved difficult to approach the zero tilt condition exactly, both the in-kagome-plane spin correlations and the out of plane spin correlations that map to pseudo-spin degrees of freedom (see Fig. \ref{fig:KD2}) have been observed and characterised. Our experimental measurements broadly confirm that \hto{} gives an accurate realisation of the two dimensional near neighbour kagome ice Coulomb phase, in the region of the phase diagram that is far from the experimentally observed critical end point at high field field,~\cite{Castelnovo2008,Borzi:2016gj, Kadowaki2009}, or possible dipole driven ordering at low field \cite{Melko:2001el,Baez2017}.  Monopoles can be safely neglected in this regime, except insofar as they act as dynamical facilitators in the real system~\cite{Takatsu:2013fc,Otsuka:2014dg,Borzi:2013jo}.

\subsubsection{Tilted case} 

Our simulations and experiments clearly confirm that the applied tilted field can be used to quantitatively tune the Coulomb phase from isotropic to anisotropic.
We have determined correlation lengths parallel and perpendicular to the applied field that scale differently as one moves towards the Kasteleyn transition, with a resulting build up of anisotropic spin correlations. 

%As shown in our numerical simulations, the wavevectors at which these correlation lengths are extracted are not the same as those predicted in Ref. [\onlinecite{Moessner2003}], but nonetheless the structure factor of the near neighbour model contains features that yield anisotropically scaling correlation lengths that exactly follow those predictions.

%\textcolor{red}{A previous study of \hto{} by experiment and simulation~\cite{Fennell2007} already showed that features in the diffuse scattering do drift across the scattering plane as a function of field, with the Kasteleyn transition temperature falling to zero in the negative tilt case, highlighting difference between negative and positive tilt. Further critical scattering was observed in this case which was shown to be driven by a nearby dipolar ordering transition ~\cite{Kao2016}.}
%While the negative tilt of that study drives the system toward another critical point, with associated critical scattering, it coincides with the case described above where the Kasteleyn transition temperature is driven to zero and the ordering is driven by the strong dipolar interaction present in real spin ice materials~\cite{Kao2016}.  

When the methodology developed for the simulations is applied to the experimental data, we see that for moderate values of $\frac{\Upsilon}{\Upsilon_{\rm K}}$, the analytical prediction captures the scaling of the correlation lengths extracted from the experimental data.
This illustrates that these qualitative features survive the corrections to the simple model that are necessary for a quantitative description of real materials - dipolar interactions, demagnetizing effects, and ice rule violating monopole excitations.  We can firmly conclude that the framework of drifting peak positions encoding anisotropic scaling as the system is driven toward the Kasteleyn transition is relevant to a real material such as \hto{}.

\begin{figure} 
\includegraphics[width=0.52\textwidth]{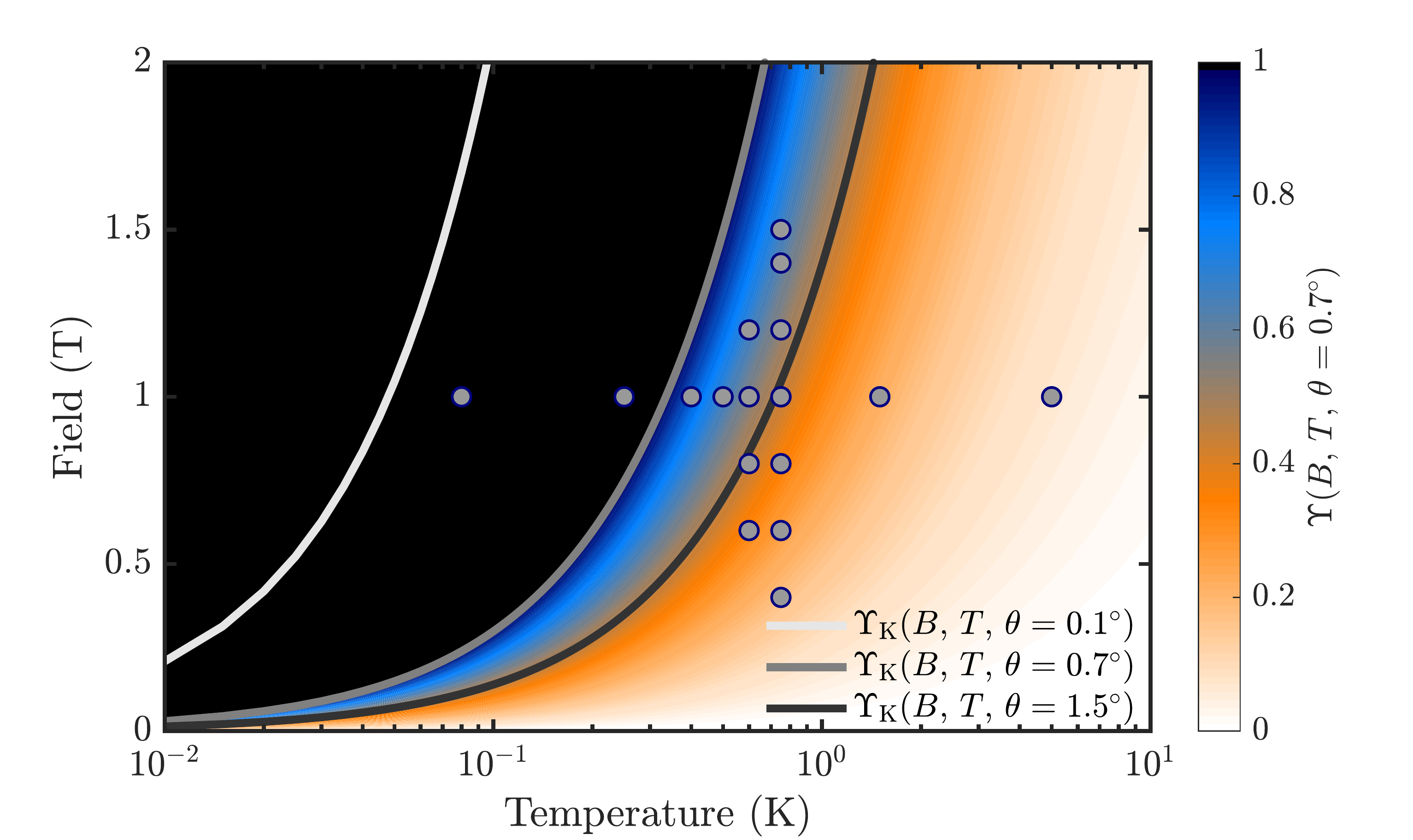}
\centering
    \caption{ Field and temperatures that satisfy the Kasteleyn transition criterion at $\theta = 0.1$, $0.7$ and $1.5^{\circ}$, with the experimentally measured data points shown by circles.  The solid lines show the loci of $\Upsilon_{\rm K}=1$ for three different field tilts; the color map shows $\Upsilon_{\rm K}$ for $\theta=0.7$. With the $\theta$ and $\phi$ derived from the crystal alignment by Laue diffraction, no point in the phase space is above the critical point for the Kasteleyn phase transition ($\Upsilon_{\rm K} = 1$).  For the values that best scale the data (i.e. $\theta=0.7^{\circ}$), only points with in the regime of slow dynamics fall below the transition.  To move the transition into a region where we could comfortably measure, the larger tilt of $\theta=1.5^{\circ}$ would be required. }
    \label{fig:PhaseDiagram}
\end{figure}

\subsubsection{A topological phase transition}

Our thermodynamic results have illustrated several unusual features of the Kasteleyn transition: how it has no  fluctuations below the transition, no symmetry breaking and yet satisfies all the thermodynamic and scaling criteria of a second order phase transition. It is associated with a single thermodynamic variable and a single independent critical exponent. A crossover exponent to the paramagnetic phase can also be introduced by allowing a finite concentration of magnetic monopoles and forcing the monopole concentration to zero at the critical field \cite{Powell:2013ct}.

Other examples of single exponent transitions are the athermal percolation transition and the Kosterlitz-Thouless transition \cite{Kosterlitz1973} which is also topological and which also fails to break a symmetry. Other hard particle transitions, such as the crystallisation of discs or spheres \cite{Alder60}, or certain liquid crystal transitions \cite{Frenkel94}, although also purely entropy driven, do break symmetries and so generate a second thermodynamic variable related to the symmetry breaking. 

The singular thermodynamics is driven by topological string excitations that are solely responsible for changing the magnetic moment of the sample. These strings distinguish the Coulomb phase from a paramagnet. At low field their magnetic responses differ only by a scale factor \cite{Isakov2004PRB,Jaubert:2013jz}, but while a paramagnet coasts towards a fully magnetized state in the limit $H/k_BT\rightarrow\infty$, the Coulomb phase crashes into it abruptly for a finite ratio.
The external field progressively favours system spanning string excitations, suppressing short loops and hence entropy, and lowering the symmetry of the observed scattering patterns.

In the critical region, the density of strings flipped against the ordered moment falls to zero, so that they can be considered as weakly interacting, mapping onto world lines for individual random walkers. In this limit, the in-plane field direction dictates the commensurability of the string looping on the torus. We have identified a crossover, with in-plane tilt $\phi$, to an incommensurate limit in which individual strings influence the total moment \cite{Bhattacharjee1983}. This remarkable topological property of the finite size scaling features strictly depends on the existence of a torus, so may be hard to realise in a real system, where extended strings must terminate in surface magnetic charge \cite{Jaubert2017} or defects \cite{Revell2013}.

\subsubsection{Kasteleyn Transition in Experiment}

While our numerical simulations can be driven right to a Kasteleyn transition, the experimental system could not be, being complicated by the slowing down of collective dynamics below $T\approx 0.6$ K and the onset (termination) of the kagome plateau at $H\approx0.4$ $(1.5)$ T. Within the kagome ice plateau, for fixed tilt, as in a typical experiment, to reach larger $\Upsilon$ requires larger fields or lower temperatures.  Increasing the field too much will proliferate monopoles and terminate the kagome ice plateau, but lowering the temperature will tend to cause  \hto{} tend to run out of dynamics as the monopoles disappear.  Real experimental examples of the transition are, in any case likely to be rounded by the eventual monopole contribution, as was shown for the three-dimensional example~\cite{Jaubert2008, Pili2021}, but the incorporation of monopoles can also lead to unconventional behaviour and scaling~\cite{Powell:2013ct,Hamp:2015tf}.  

However, rather than promoting the transition in classical spin ices such as \hto{} and \dto{}, into regions of phase space where other phenomena are crowded, perhaps it can be reached by studying a compound with faster monopole dynamics such as CdEr$_2$S$_4$~\cite{Gao:2018cu}.  In this case, the spin ice is classical and the thermal monopole population fairly similar to \dto{}, but hopping rates are much more rapid, suggesting that equilibrium can be maintained to lower temperature.  An alternative source of dynamics might be quantum fluctuations, and a quantum kagome ice phase could be a highly interesting system, as realized in a magnetization plateau of a quantum spin ice~\cite{Hermele:2004gg,Sibille:2016bd,Sibille:2018hca} or a two-dimensional kagome analog of spin ice~\cite{Paddison:2016bq}.  We are not aware of any theoretical study of the Kasteleyn transition in a quantum kagome ice~\cite{Carrasquilla:2015hu,Bojesen:2017gx}, which might also be a fascinating prospect, given the interest in quantum spin ice~\cite{Hermele:2004gg,Benton:2012ep,Gingras:2014ip}.

%\subsubsection{The effect of dipole interactions}
%Supposing that the transition temperature can be promoted with a suitably larger tilt as discussed above, there are further reasons that the transition may not occur.  The ordered spin structure that must develop is a so-called $\vec{Q}=0$ structure, in which every tetrahedron has the same `2-in--2-out' configuration.  Each tetrahedron will have a magnetic moment, which, in the current labelling, will be in the direction $[001]$, 54.5 $\deg$ from the direction of the applied magnetic field, and the macroscopic moment of the ordered structure will also be in this direction.  It seems that competition between the microscopic Zeeman energy and macroscopic demagnetization energy could frustrate the transition, or at least modify the approach to it.  Moreover, in real spin ice materials, the dipolar interaction is dominant, and the eventual ordered state is not the ground state of this interaction. Simulations of a model with dipolar interactions and positive tilt suggest that a Kasteleyn transition does not occur in this situation~\cite{Kao2016}, although the diffuse scattering is more accurately modelled and the correlations are still tuned as the field-induced ordered state is approached~\cite{turrinithesis}.

\section{Conclusions}\label{sec:Conclusion}

In conclusion, we have studied the Kasteleyn transition in the kagome ice phase of spin ice subjected to a field tilted away from the $[111]$ direction by a small angle, by analytical, numerical and experimental methods. We have
%.  We have sought to increase understanding of the Kasteleyn transition that occurs when this kagome ice phase is
 exposed   a
%Our analytical theory clarifies the nature of the transition, 
striking evolution of correlations and of topological properties of the Coulomb phase as the field is tuned towards the transition.  
%and underpins the construction of a loop algorithm for kagome ice.  Using this algorithim, our numerical simulations confirmed the form of the diffuse scattering of kagome ice, and show how this scattering evolves dramatically as the Coulomb phase correlations are tuned by field,  temperature and tilt.  

We find good qualitative agreement between experimental neutron scattering results from a single crystal of \hto~  and those provided by the nearest neighbour spin ice model. The agreement is perhaps better than 
%It is interesting to note that the diffuse scattering of kagome ice is much more closely approximated by a near neighbour model than is the case
in the three-dimensional, zero field regime of spin ice, where inclusion of the full dipolar interaction is essential to reproduce the broad features of the diffuse scattering~\cite{Bramwell2001PRB,Fennell:732176,Fennell2009,Henelius:2016ew,Giblin:2018cz,Twengstrom:2020jv} 
%(although see also Ref. ~\cite{Twengstrom:2020jv}). 
and it would be interesting to investigate this point further.
%to understand it quantitatively. 
Consequently, we may be very optimistic that yet more of the exotic behaviour of kagome ice will be experimentally  observable in spin ice materials. 
%Moreover, it gives hope that much of the exotic theoretical behavior of kagome ice can be searched for optimistically -- as we have shown, this is the case for scaling at small $\Upsilon$, and we have outlined some pathways for reaching the region where $\Upsilon\rightarrow\Upsilon_{\rm K}$.

\acknowledgements{We thank Ch. Ru\"egg for comments on this work and Ref. [\onlinecite{turrinithesis}] and L. Jaubert and T. Roscilde for useful discussions.  Neutron scattering experiments were carried out on D7 at the ILL, Grenoble, France.  Work at PSI was partly funded by the SNSF (Schweizerischer Nationalfonds zur F\"orderung der Wissenschaftlichen Forschung) (grant 200021\_140862 and 200020\_162626)}. P.C.W.H acknowledges financial support from ANR, France, Grant No. ANR-19-CE30-0040. I.G.W. acknowledges financial support from NERC Grant GR3/7497.

\appendix

\section{Relation with kagome spin ice}\label{Wills}

Prior to the discovery of kagome ice, Wills {\it et al.}~\cite{Wills2002} introduced a two-dimensional version of spin ice consisting of ferromagnetically coupled Ising-like spins on a kagome lattice, constrained to point `in' or `out' of the triangles. This state, which they called `kagome spin ice', is a very interesting system its own right and has a rich phase diagram when long range interactions are included~\cite{MoellerPRB2009,Chern2011,Zhang2013}. It is highly relevant to artificial spin ice arrays and has been much studied in this context~\cite{Mengotti:2010fo}. Here we reserve the term kagome ice for the state obtained when the magnetic field is applied along the $[111]$ direction of a spin ice and the term kagome spin ice for the model of Wills {\it et al.}   We identify similarities and differences between the two systems and to highlight the topological constraints of the kagome ice state \cite{Macdonald2011}.

In kagome spin ice the lowest energy configuration on each triangle satisfies an ice-like rule with either `2-in--1-out' or `1-in--2-out'. 
The odd number of in/out contributions leaves a net charge of $\pm Q/2$ associated with each triangle, where $Q$ is the monopole charge. The ground state of the nearest neighbour model is therefore a dense charge fluid with overall charge neutrality~\cite{MoellerPRL2006}, the so-called KI phase. Including long range dipolar interactions induces a
phase transition at finite temperature from the fully disordered KI phase to the partially ordered KII phase~\cite{MoellerPRB2009,Chern2011}, even in the absence of ice rule defects \cite{Sendetskyi2019}.  The transition is driven by a $Z_2$ symmetry breaking in which up and down oriented triangles select 
between  `2-in--1-out' or `1-in--2-out' configurations, lifting a topological degeneracy that allows for the formation of system spanning spin loops~\cite{Macdonald2011}. 
%The KII phase~\cite{MoellerPRB2009,Zhang2013} has reduced symmetry compared with the full phase space of kagome ice rule states because the $Z_2$ symmetry and topological degeneracy are lifted, with configurations on one class of triangle (up or down) restricted to `2-in--1-out' with the reverse sector restricted to `1-in--2-out'. 
The reduced symmetry corresponds to charge crystallisation~\cite{MoellerPRB2009,Chern2011,Zhang2013} but only partial magnetic ordering~\cite{BrooksBartlett:2014kf}. For perfect charge order, the spins effectively decouple, or fragment, into two independent parts, `longitudinal' and `transverse', with the transverse part forming a Coulomb phase with corresponding algebraic correlations.  
In dipolar kagome spin ice, corrections to the emergent Coulomb interaction between monopoles lead to a further transition at low temperature, to a fully ordered phase in analogy with dipolar spin ice~\cite{denHertog2000}. 

The application of the field along $[111]$ in a spin ice breaks this $Z_2$ symmetry for the spins on the kagome plane by pinning the apical spin of each tetrahedron (see below). Applying the ice rules with this constraint imposes the reduced choice of `2-in--1-out' or `1-in--2-out' for the three remaining spins. 

The kagome ice phase therefore corresponds to the KII phase of dipolar kagome spin ice.
%, because in kagome ice the magnetic field breaks the $Z_2$ symmetry through the interaction with the apical spins. The selection of a fixed orientation for this spin ensures that the three basal spins of the tetrahedra satisfy both the kagome ice rules and the topological constraints of the KII phase at low temperature, because the two possible kagome ice rules are no longer degenerate. 
In this case, the remnant charge of the three in-plane spins is neutralised by the charge on the apical spin giving charge neutrality. In this limit the spin ice Coulomb phase is split into decoupled planes, each of which has the configuration space of the KII phase.

\section{Kasteleyn transition temperature}\label{eqn3_deriv}

\begin{figure}
\includegraphics[width=0.5\textwidth]{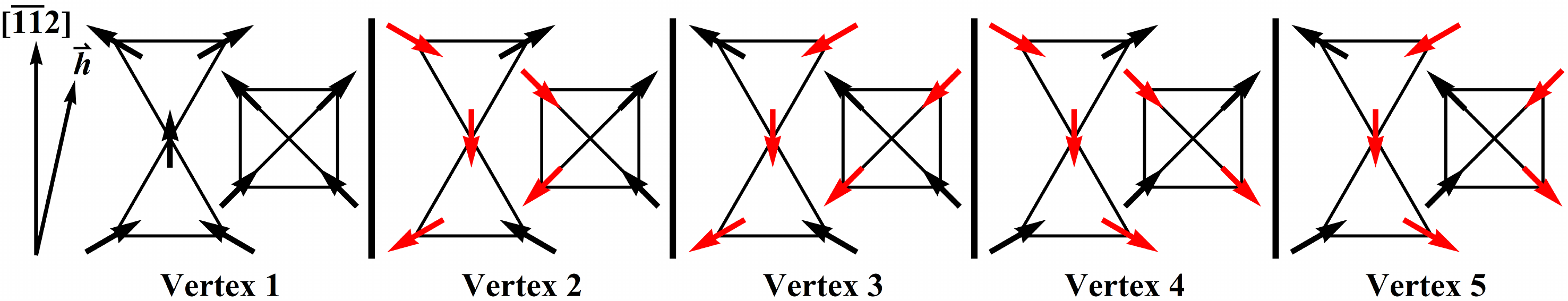}
\caption{\label{fig:KagomeVertices} {The five vertices relevant to the Kasteleyn transition on kagome lattice}}
\end{figure}

As described above, a kagome ice spin configuration with long range order is disordered at the Kasteleyn transition through the introduction of strings of reversed spins spanning the entire cell and passing through the periodic boundaries.  Here we present some details for this process for general  tilt $\theta$ and $\phi$.
%describe the construction of a ``worm'' algorithm, which stochastically introduces such a string in a system with applied field tilted in a general direction, as described above.

We start with the magnetically ordered state shown in Fig.~\ref{fig:KagomeVertices}, defined with all vertices in configuration 1. This ordering is along along the $[\bar{1}\bar{1}2]$, or $\hat{y}$ crystal axis, as defined for the underlying pyrochlore lattice (see also Fig.~\ref{fig:KI} and ~\ref{fig:DPD}). There is an in-plane field $\vec H$, placed at an angle $\phi$ with respect to $\hat{y}$.  The probability of introducing a string goes to zero at the Kasteleyn transition temperature $T_K$, which can be calculated by estimating the free energy change $\delta G= \delta \langle{\cal{H}}\rangle -T\delta S$ on introducing the string. $\delta \langle{\cal{H}}\rangle$ is the change in magnetic enthalpy, which in our constrained model is pure Zeeman energy, and where $\delta S$ is the change in entropy. The unit cell is taken to be  an `up' triangle of the kagome lattice. For a system spanning $L$ cells in the $\hat{y}$ direction we can define $\delta G=L\delta g$, $\delta \langle{\cal{H}}\rangle=L\delta \epsilon$ and $\delta S=L\delta s$. We are therefore looking for a change in sign in $\delta g= \delta \epsilon-T\delta s$: $\delta \epsilon=T_{\rm K}\delta s$.

At each step of the string construction, the virtual head of the string, or worm can be thought of as sitting in the centre of a down triangle. It advances by jumping from this site, through an up triangle to a neighbouring down triangle. The Zeeman energy to be considered is the change in energy coming from flipping the two spins on the up triangle through which the worm hops.  Considering the 5-spin, up-down triangle pair in Fig.~\ref{fig:KagomeVertices}, we calculate the probabilities that  vertex 1 changes to vertex 2-5. The worm head arrives in the down triangle at the end of the $(n-1)^{\rm th}$ step, and the probability of it arriving from left or right is taken care of during this step.  As this is a Markov process, the $n^{\rm th}$  step is independent of this, and in this step it jumps down and out of view to the left or right.  Hence there are only two probabilities, $P_L=P_{1\rightarrow 2}=P_{1\rightarrow 3}$ and $P_R=P_{1\rightarrow 4}=P_{1\rightarrow 5}$.

The changes in Zeeman energy associated with this move are:
\begin{equation}
\begin{aligned}
\epsilon_L&= 2HS_{\perp}\cos(\phi)+ 2HS_{\perp}\cos(\pi/3-\phi)\nonumber \\ 
              &= 3HS_{\perp}\cos(\phi)+\sqrt{3}HS_{\perp}\sin(\phi)\\
\epsilon_R&= 3HS_{\perp}\cos(\phi)-\sqrt{3}HS_{\perp}\sin(\phi).
\end{aligned}
\end{equation}

Given that $S_{\perp}=2\sqrt{2}/3$, we have
\begin{equation}
\epsilon_{L,R}=\epsilon_0\pm v, 
\end{equation}
where  $\epsilon_0=2\sqrt{2}H\cos(\phi)$ and $v=2\sqrt{\frac{2}{3}}H\sin(\phi)$.

The probabilities are then given by
\begin{equation}
P_{L,R}=\frac{{\exp{(\mp\beta v)}}}{\exp{(\beta v)} + \exp{(-\beta v)}},
%P_R&=&{\exp{(\beta x)}\over{\exp{(\beta x)} + \exp{(-\beta x)}}}
\end{equation}

$P_L+P_R=1$, and $P_L/P_R=\exp{(-\beta(\epsilon_L-\epsilon_R))}=\exp{(-2\beta v)}$.  Defining $P_{L,R}=(1/2)(1\mp q)$, we find $q=\tanh (\beta v)$. From this we can write $\delta \epsilon = P_L\epsilon_L + P_R\epsilon_R =\epsilon_0+v(P_L-P_R)=\epsilon_0 -qv $, and 

\begin{equation}
\begin{aligned}
\frac{1}{k_B}\delta s &= -P_L\log P_L -P_R\log P_R, \\
  &=-\left(\frac{1-q}{2}\right)\log {\left(\frac{1-q}{2}\right)}-\left(\frac{1+q}{2}\right)\log {\left(\frac{1+q}{2}\right)} \nonumber.
\end{aligned}
\end{equation}
We can now use the identities
\begin{equation}
\begin{aligned}
\frac{1}{4}(1+q)(1-q)&=\frac{1}{4\cosh^2(\beta v)},\\
 \frac{1+q}{1-q}&=\exp{(2\beta v)},\\
 \beta qv&=\frac{q}{2}\log{(\exp{2\beta v})},
\end{aligned}
\end{equation}

to show that, for $T=T_{\rm K}$
\begin{equation}
{\epsilon_0\over{k_BT}}=\log{(2\cosh{(\beta v)})}.
\end{equation}
Exponentiating both sides, changing units and using Eqn.~\ref{eq:HDef} we arrive at Eqn.~\ref{eq:KTCphi}.

%which is required result because it is the same expression as that found by defining fugacities $z_1,z_2,z_3$ for placing a dimer on one of the three bonds of the dual honeycomb lattice and setting $z_1=z_2+z_3$ (see below). 

\section{Loop algorithm for kagome ice} \label{loop_appendix}

%Previously, in the loop algorithm, once the probabilities $P_{1\rightarrow 2}$ and $P_{1\rightarrow 4}$ were defined, the reverse probabilities $P_{2\rightarrow 1}$ were defined such that detailed balance was satisified:
%%
%\begin{equation}
%{P_{2\rightarrow 1}\over{P_{1\rightarrow 2}}}={w_1\over{w_2}}=\exp{-\beta(\epsilon_1-\epsilon_2)},
%\end{equation}
%%
%where, for  $\phi=0$ and general $\phi$:
%\begin{IEEEeqnarray}{rCl}
%\epsilon_1&=&  -2hS_{\perp}\cos(\phi)\rightarrow\nonumber\\
% &&- hS_{\perp}\cos(\phi)- hS_{\perp}\cos(\pi/3-\phi)-hS_{\perp}\cos(\pi/3+\phi) \nonumber \\
%\epsilon_2&=&hS_{\perp}\cos(\phi)\rightarrow\nonumber \\
%&&+hS_{\perp}\cos(\phi)+ hS_{\perp}\cos(\pi/3-\phi)-hS_{\perp}\cos(\pi/3+\phi)) \nonumber \\
%\epsilon_4&=&hS_{\perp}\cos(\phi)\rightarrow\nonumber \\
%& &+hS_{\perp}\cos(\phi)- hS_{\perp}\cos(\pi/3-\phi)\nonumber \\
%&&+hS_{\perp}\cos(\pi/3+\phi)) .
%\end{IEEEeqnarray}
%%
%This gave $P_{2\rightarrow 1}={w_1\over{2w_2}}$. Now, f

%{\textcolor{red}{This needs to be checked as I don't quite understand it}}

The loop algorithm for numerical simulations is constructed in this spirit. Creating a worm requires the creation of a pair of oppositely charged topological defects which are considered as virtual, in that the Boltzmann weight for their creation is never taken into account. The worm makes a path of virtual hops which can be either forward, backwards or sideways until it returns to its starting position, destroying the defect pair. Once the probabilities $P_{1\rightarrow 2}$ and $P_{1\rightarrow 4}$ are defined, the reverse probabilities $P_{2\rightarrow 1}$ are defined such that detailed balance is satisified;
%.  Now, for general $\phi$, 
$P_{2\rightarrow 1} \rightarrow P_L {\exp(-\beta(\epsilon_1-\epsilon_2))}$, where $\epsilon_1$ and $\epsilon_2$ are the Boltzmann weights for vertices $1$ and $2$. As for general $\phi$,  $\epsilon_1-\epsilon_2=-(\epsilon_0 +v)$, it follows that 
\begin{equation}
P_{2\rightarrow 1}= \exp{\left(\beta \epsilon_0 - \log[2\cosh(\beta v)]\right)},
\end{equation}
so that $P_{2\rightarrow 1} \rightarrow 1$ at $T_{\rm K}$, which is the condition one needs for singular behaviour at the Kasteleyn transition. As, in addition one defines $P_{2\rightarrow 3}= 1 - P_{2\rightarrow 1}$, it follows that $P_{2\rightarrow 3}\rightarrow 0$ at $T_{\rm K}$. Similarly, one finds $P_{4\rightarrow 1}=P_R {\exp(-\beta(\epsilon_1-\epsilon_4))}$ and hence that $P_{4\rightarrow 1}= \exp{\left(\beta \epsilon_0 - \log[2\cosh(\beta v)]\right)}$, which is again the condition needed for the Kasteleyn transition.

\section{Dimer representation of the Kasteleyn transition for general $\phi$}\label{dimer_appendix}

%In the mapping to a dimer problem, a dimer is placed on the line joining the centres of the up and down triangles, for which the vertex spin points out of the up and into the down triangle (i.e. the $Z_2^+$ sector with `two-in--one-out' ice rule for each up triangle, an equivalent $Z_2^-$ sector with `one-in--two-out' exists in the parent pyrochlore if the main external field is reversed).  The dimer chemical potential is therefore the Zeeman energy required to take an outward pointing spin $1$, $2$ or $3$ and turn it inward, which corresponds to the energy required to take a $1$, $2$ or $3$ dimer and place it at an infinite distance from the on site potential. This definition gives an extra minus sign compared to the spin energy, but allows us to define the correct thermodynamic expressions for particle numbers and chemical potential. Hence, for the field defined as above we have the following chemical potentials for placing a dimer on sites 1,2 and 3:
%
%\begin{eqnarray}
%\tilde{\mu}_1&=&2hS_{\perp}\cos(\phi)={2\over{3}}\epsilon_0 \nonumber\\
%\tilde{\mu}_{2,3}&=&-2hS_{\perp}\cos(\pi/3\mp\phi)=-({1\over{3}}\epsilon_0\pm x),
%\end{eqnarray}
%
%and the fugacities, $z_i=\exp{\beta \tilde{\mu}_i}$ (it turns out that $w_1/w_2=z_1/z_2$). {\color{green}{Peter: What is $w_i$?}}

The dimer partition function can be written
\begin{equation}
Z=\rm{Tr}_{\{n_1,n_2,n_3\}} g(n_1,n_2,n_3)\exp \beta(n_1{\mu}_1+n_2{\mu}_2+n_3{\mu}_3),
\end{equation}
where $n_i$ is the number of dimers on sites $i$ and $g(n_1,n_2,n_3)$ is the number of configurations for fixed $n_i$. The total number of dimers, $n$, is fixed at one per triangle, so that we can define a semi-grand canonical free energy free energy \cite{Das06} (also the magnetic free energy) $G(T,n,{\mu}_1,{\mu}_2,{\mu}_3)$. 
%One of the chemical potentials fixes the energy scale, while the other two dictate the relative distribution of dimers on sites $1$, $2$, $3$. 
$Z$ can be calculated exactly by the Pfaffian method introduced by Kasteleyn~\cite{Kasteleyn1963}. Although Kasteleyn does not give an explicit expression for the honeycomb lattice, Wu~\cite{Wu} does, while at the same time showing how the dimers on honeycomb and therefore spins on the kagome lattice also map onto the five-vertex model on a square lattice.  The anisotropic six-vertex model was also treated exactly by Watson~\cite{Watson}. Wu's expression for $Z$ is:

\begin{eqnarray}
\log Z &=& {n\over{8\pi^2}}\int^{2\pi}_{0} d\tilde{\theta} \int^{2\pi}_{0} d\tilde{\phi} \log[z_1^2+z_2^2+z_3^2\nonumber \\
&&+2z_1z_2\cos{(\tilde{\theta})}+2z_1z_3\cos{(\tilde{\phi})}+2z_2z_3\cos{(\tilde{\theta} - \tilde{\phi})}],\nonumber \\
&&
\end{eqnarray}
where here $\tilde{\theta}$ and $\tilde{\phi}$ are dummy variables.

From the above one can calculate the mean number of dimers
\begin{equation} 
\langle n_{2,3}\rangle
= -{\partial G\over{\partial {\mu}_{2,3}}}
 = {1\over{\beta}}{\partial \log Z\over{\partial {\mu}_{2,3}}}.
\end{equation}
As the total number of dimers is fixed $\langle n_1\rangle$ is not independent: $\langle n_1\rangle=n-\langle n_2\rangle-\langle n_3\rangle$. Defining $\alpha_{i} = \langle n_i\rangle/n$, one finds from the exact partition function
\begin{eqnarray}
\alpha_2&=&{1\over{\pi}}\cos^{-1}\left({z_3^2-z_2^2+z_1^2\over{2z_1z_3}} \right)\nonumber\\
\alpha_3&=&{1\over{\pi}}\cos^{-1}\left({z_2^2-z_3^2+z_1^2\over{2z_1z_2}} \right),
\end{eqnarray}
with $\alpha_1=1-\alpha_2-\alpha_3$. To arrive at this expression one needs the identity
\begin{equation}
\int^{2\pi}_0 d\theta {1\over{A+B\cos(\theta)+C\sin(\theta)}} = {2\pi\over{(A^2-B^2-C^2)^{1/2}}},
\end{equation}
which is valid for $A^2> B^2+C^2$, corresponding to the disordered regime, $z_2+z_3> z_1$.
At high temperature $ \langle n_1\rangle = \langle n_2 \rangle = \langle n_3 \rangle =n/3$, while at the transition $ \langle n_1 \rangle =1$, $\langle n_2 \rangle =\langle n_3 \rangle=0$. As  the thermodynamic variable reaches the constraint, no further evolution can occur and the system is singular.

%A physical argument for the transition can be developed by considering the system in an ordered dimer configuration with $\langle n_1\rangle=n$. The lowest energy excitation is a string of dimers flipped from the $1$ position to the $2$ or $3$ positions randomly. Just as for the spins, once a single dimer has been moved it is necessary to complete a loop of dimer moves, hence the probability of flipping the whole string is $P^L$, where $P$ is the probability for a certain triangle. For a single triangle the probability of placing the dimer on site $i$ is 
%
%\begin{equation}
%p_i={z_i\over{z_1+z_2+z_3}}.
%\end{equation}
%So, if it is going to be fruitful to move the dimer we must have $p_2+p_3>p_1$ and we recover Kasteleyn's condition for the transition temperature, $z_1=z_2+z_3$.

%The standard energy expression, ${-\partial \log Z\over{\partial \beta}}$, gives a value for the ``particle enthalpy'', ${\cal{H}}=U - \sum_i \langle n_i\rangle  \tilde{\mu_i}$. Because the internal energy is zero in this problem, all we have is the contribution $-\mu N$, from which we recuperate the extra minus sign added above.
It follows straightforwardly that the mean energy per spin, which is $1/3$ of the mean energy per dimer is:
\begin{eqnarray}
\langle \epsilon\rangle  &=& - {1\over{3n}}({\mu}_1 n_1 +  {\mu}_2 n_2 + {\mu}_3 n_3 )     \nonumber \\
&=& - {1\over{3}}({\mu}_1 + \alpha_2({\mu}_{2}-{\mu}_1) + \alpha_3({\mu}_{3}-{\mu}_1)),
\end{eqnarray}

The in plane magnetization can also be calculated from the mean dimer numbers by considering each of the three spins separately: $\langle S_1^y\rangle = S_{\perp} $ when $\langle n_1\rangle =1$ and $\langle S_1^y\rangle = -S_{\perp}/3 $ when $\langle n_1\rangle =1/3$, from which it follows that $\langle S_1^y\rangle = S_{\perp}(2\alpha_1-1 )$. Similarly $\langle S_2^y\rangle =1/2 S_{\perp} $ when $\langle n_2\rangle =0$ and $\langle S_2^y\rangle = S_{\perp}/6 $ when $\langle n_2\rangle =1/3$, from which we find $\langle S_2^y\rangle = S_{\perp}(-\alpha_2+1/2 )$, with the equivalent expression for the third spin,  $\langle S_3^y\rangle = S_{\perp}(-\alpha_3+1/2 )$. The total $y$ component of the magnetization (per spin - hence the factor 1/3) is then $M_y=(S_{\perp}/3)(2\alpha_1-\alpha_2-\alpha_3)$. 

For the $x$ component, $\langle S_1^x\rangle =0$, $\langle S_2^x\rangle =(\sqrt{3}/2)S_{\perp}$ when $\langle n_2\rangle $=0 and $\langle S_2^x\rangle =(1/3)(\sqrt{3}/2)S_{\perp}$ when $\langle n_2\rangle =1/3$, leading to $\langle S_2^x\rangle =\sqrt{3}S_{\perp}(-\alpha_2+1/2)$ and to $\langle S_3^x\rangle =\sqrt{3}S_{\perp}(\alpha_3-1/2)$. This gives the average $x$ component per spin, $M_x= (1/3) \sqrt{3}S_{\perp}(\alpha_3-\alpha_2)$. Using the relation between the $\alpha_i$ and the value $S_{\perp}=2\sqrt{2}/3$ we finally find the correct expressions for $M_y$ and $M_x$ parallel and perpendicular to $[\bar{1}\bar{1}2]$:
\begin{eqnarray}
M_y&=& {4\sqrt{2}\over{9}}\left(1-{3\over{2}}(\alpha_2+\alpha_3)\right)\nonumber\\
M_x&=&{{2\over{3}}\sqrt{2\over{3}}}\left(\alpha_3-\alpha_2\right).
\end{eqnarray}
The dimensionless magnetic moment entering the thermodynamic discussion above would correspond to $M=NM_y$.

\section{Peak tracking algorithm \label{peak_tracking_appendix}}

\begin{figure}
\centering
\includegraphics[width=0.5\textwidth]{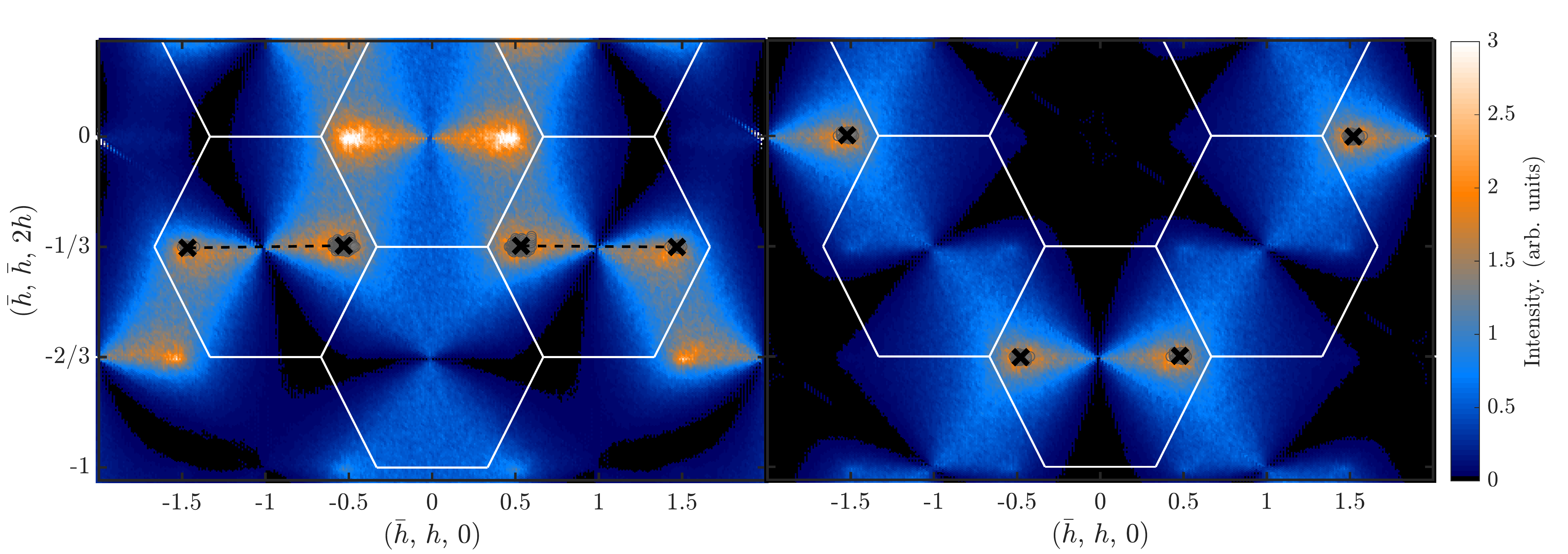}
\caption{\label{fig:PeakTracking} The results of the peak tracking algorithm in the spin flip (left) and non-spin flip (right) channel for $h_{\rm K}$(1 T, 0.75 K, $\theta$ $>$ 0, $\phi$ = 0), with included data points in grey dots. The centers of the asymmetrical ellipses as determined by the fuzzy cluster algorithm are marked with a black x and correspond to the actual location of the calculated local maxima, and the distance between the peaks used to compute $\xi_x$ is shown by a black dashed line in the SF channel. The mean membership grade for this peak tracking solution is $\mu^{m}$ = 0.98}
\end{figure}

From tracking the logarithmic peak locations in the diffuse neutron scattering patterns and determining their distance from the BZ center, we can reproduce the normalized inverse correlations $\xi^{-1}_x$ and $\xi^{-1}_y$ ~\cite{Moessner2003}.  The location of diffuse scattering peaks close to $h$ = $-$1/2 and $-$3/2, $k$ = $\pm$1/3 were extracted with a two dimensional peak tracking program based on the fuzzy cluster algorithm in MATLAB, in which an initial guess for the location of  four diffuse scattering peaks is refined by measuring the membership grades $\mu_{ij}$ of each point in reciprocal space with intensities higher than 5 times the background threshold\cite{Bezdek1981}. To get the cluster centers ($c$), the following objective function $J_m$ for $D$ data points and $N$ = 4 clusters is minimized: 

\begin{equation}
J_m = \sum_{i=1}^D \sum_{j=1}^N \mu^{m}_{ij}|x_i - c_j|^2.
\end{equation}

The membership grades of each point in the cluster demonstrates the relative uncertainty that the point is indeed in that cluster, with values approaching 1 for a collection of distinct spherical clusters. Although not as sharp as Bragg peaks, the logarithmic peaks indicative of short range ice-rule correlations are clearly distinguishable in both simulation and experimental data from the background with clear centers. Peaks from an experimental diffuse scattering patterns with an unknown misalignment can therefore be analyzed generically.

The distance from each diffuse peak to the kagome Brillouin zone center in $\hat{x}$ (as denoted by the white hexagons in Fig. \ref{fig:PeakTracking}) was averaged to produce the inverse correlation length $\xi^{-1}_x$, from which the analytical solution in Eq. \ref{eq:InvCorrLength} was used to derive $\xi^{-1}_y$. As the nearest neighbor Hamiltonian produces logarithmic peaks that have a small tail along the $(\bar{h}, h, 0)$, there is a slight deviation of the cluster center from the true maximum, which is less than 0.01 \AA$^{-1}$ for all simulations.

\bibliography{Spin_Correlations_v8_bib}

\begin{thebibliography}{91}%
\makeatletter
\providecommand \@ifxundefined [1]{%
 \@ifx{#1\undefined}
}%
\providecommand \@ifnum [1]{%
 \ifnum #1\expandafter \@firstoftwo
 \else \expandafter \@secondoftwo
 \fi
}%
\providecommand \@ifx [1]{%
 \ifx #1\expandafter \@firstoftwo
 \else \expandafter \@secondoftwo
 \fi
}%
\providecommand \natexlab [1]{#1}%
\providecommand \enquote  [1]{``#1''}%
\providecommand \bibnamefont  [1]{#1}%
\providecommand \bibfnamefont [1]{#1}%
\providecommand \citenamefont [1]{#1}%
\providecommand \href@noop [0]{\@secondoftwo}%
\providecommand \href [0]{\begingroup \@sanitize@url \@href}%
\providecommand \@href[1]{\@@startlink{#1}\@@href}%
\providecommand \@@href[1]{\endgroup#1\@@endlink}%
\providecommand \@sanitize@url [0]{\catcode `\\12\catcode `\$12\catcode
  `\&12\catcode `\#12\catcode `\^12\catcode `\_12\catcode `\%12\relax}%
\providecommand \@@startlink[1]{}%
\providecommand \@@endlink[0]{}%
\providecommand \url  [0]{\begingroup\@sanitize@url \@url }%
\providecommand \@url [1]{\endgroup\@href {#1}{\urlprefix }}%
\providecommand \urlprefix  [0]{URL }%
\providecommand \Eprint [0]{\href }%
\providecommand \doibase [0]{https://doi.org/}%
\providecommand \selectlanguage [0]{\@gobble}%
\providecommand \bibinfo  [0]{\@secondoftwo}%
\providecommand \bibfield  [0]{\@secondoftwo}%
\providecommand \translation [1]{[#1]}%
\providecommand \BibitemOpen [0]{}%
\providecommand \bibitemStop [0]{}%
\providecommand \bibitemNoStop [0]{.\EOS\space}%
\providecommand \EOS [0]{\spacefactor3000\relax}%
\providecommand \BibitemShut  [1]{\csname bibitem#1\endcsname}%
\let\auto@bib@innerbib\@empty
%</preamble>
\bibitem [{\citenamefont {Harris}\ \emph {et~al.}(1998)\citenamefont {Harris},
  \citenamefont {Bramwell}, \citenamefont {McMorrow}, \citenamefont {Zeiske},\
  and\ \citenamefont {King}}]{Harris1998}%
  \BibitemOpen
  \bibfield  {author} {\bibinfo {author} {\bibfnamefont {M.~J.}\ \bibnamefont
  {Harris}}, \bibinfo {author} {\bibfnamefont {S.~T.}\ \bibnamefont
  {Bramwell}}, \bibinfo {author} {\bibfnamefont {D.~F.}\ \bibnamefont
  {McMorrow}}, \bibinfo {author} {\bibfnamefont {T.}~\bibnamefont {Zeiske}},\
  and\ \bibinfo {author} {\bibfnamefont {P.~J.~C.}\ \bibnamefont {King}},\
  }\bibfield  {title} {\bibinfo {title} {Magnetic structures of highly
  frustrated pyrochlores},\ }\href@noop {} {\bibfield  {journal} {\bibinfo
  {journal} {International Conference on Magnetism (Part II)}\ }\textbf
  {\bibinfo {volume} {177-181}},\ \bibinfo {pages} {757} (\bibinfo {year}
  {1998})}\BibitemShut {NoStop}%
\bibitem [{\citenamefont {Matsuhira}\ \emph {et~al.}(2002)\citenamefont
  {Matsuhira}, \citenamefont {Hiroi}, \citenamefont {Tayama}, \citenamefont
  {Takagi},\ and\ \citenamefont {Sakakibara}}]{Matsuhira2002}%
  \BibitemOpen
  \bibfield  {author} {\bibinfo {author} {\bibfnamefont {K.}~\bibnamefont
  {Matsuhira}}, \bibinfo {author} {\bibfnamefont {Z.}~\bibnamefont {Hiroi}},
  \bibinfo {author} {\bibfnamefont {T.}~\bibnamefont {Tayama}}, \bibinfo
  {author} {\bibfnamefont {S.}~\bibnamefont {Takagi}},\ and\ \bibinfo {author}
  {\bibfnamefont {T.}~\bibnamefont {Sakakibara}},\ }\bibfield  {title}
  {\bibinfo {title} {A new macroscopically degenerate ground state in the spin
  ice compound dy$_2$ti$_2$o$_7$ under a magnetic field},\ }\href@noop {}
  {\bibfield  {journal} {\bibinfo  {journal} {J. Phys.: Condens. Matter}\
  }\textbf {\bibinfo {volume} {14}},\ \bibinfo {pages} {L559} (\bibinfo {year}
  {2002})}\BibitemShut {NoStop}%
\bibitem [{\citenamefont {Moessner}\ and\ \citenamefont
  {Sondhi}(2003)}]{Moessner2003}%
  \BibitemOpen
  \bibfield  {author} {\bibinfo {author} {\bibfnamefont {R.}~\bibnamefont
  {Moessner}}\ and\ \bibinfo {author} {\bibfnamefont {S.~L.}\ \bibnamefont
  {Sondhi}},\ }\bibfield  {title} {\bibinfo {title} {Theory of the [111]
  magnetization plateau in spin ice},\ }\href@noop {} {\bibfield  {journal}
  {\bibinfo  {journal} {Phys. Rev. B}\ }\textbf {\bibinfo {volume} {68}},\
  \bibinfo {pages} {064411} (\bibinfo {year} {2003})}\BibitemShut {NoStop}%
\bibitem [{\citenamefont {Macdonald}\ \emph {et~al.}(2011)\citenamefont
  {Macdonald}, \citenamefont {Holdsworth},\ and\ \citenamefont
  {Melko}}]{Macdonald2011}%
  \BibitemOpen
  \bibfield  {author} {\bibinfo {author} {\bibfnamefont {A.~J.}\ \bibnamefont
  {Macdonald}}, \bibinfo {author} {\bibfnamefont {P.~C.~W.}\ \bibnamefont
  {Holdsworth}},\ and\ \bibinfo {author} {\bibfnamefont {R.~G.}\ \bibnamefont
  {Melko}},\ }\bibfield  {title} {\bibinfo {title} {Classical topological order
  in kagome ice},\ }\href {https://doi.org/10.1088/0953-8984/23/16/164208}
  {\bibfield  {journal} {\bibinfo  {journal} {Journal of Physics: Condensed
  Matter}\ }\textbf {\bibinfo {volume} {23}},\ \bibinfo {pages} {164208}
  (\bibinfo {year} {2011})}\BibitemShut {NoStop}%
\bibitem [{\citenamefont {Nagle}(1966)}]{Nagle1973}%
  \BibitemOpen
  \bibfield  {author} {\bibinfo {author} {\bibfnamefont {J.~F.}\ \bibnamefont
  {Nagle}},\ }\bibfield  {title} {\bibinfo {title} {Lipid bilayer phase
  transition: density measurements and theory},\ }\href@noop {} {\bibfield
  {journal} {\bibinfo  {journal} {Proceedings of the National Academy of
  Sciences}\ }\textbf {\bibinfo {volume} {70}},\ \bibinfo {pages} {3443}
  (\bibinfo {year} {1966})}\BibitemShut {NoStop}%
\bibitem [{\citenamefont {Alet}\ \emph {et~al.}(2006)\citenamefont {Alet},
  \citenamefont {Misguich}, \citenamefont {Pasquier}, \citenamefont
  {Moessner},\ and\ \citenamefont {Jacobsen}}]{Alet2006}%
  \BibitemOpen
  \bibfield  {author} {\bibinfo {author} {\bibfnamefont {F.}~\bibnamefont
  {Alet}}, \bibinfo {author} {\bibfnamefont {G.}~\bibnamefont {Misguich}},
  \bibinfo {author} {\bibfnamefont {V.}~\bibnamefont {Pasquier}}, \bibinfo
  {author} {\bibfnamefont {R.}~\bibnamefont {Moessner}},\ and\ \bibinfo
  {author} {\bibfnamefont {J.~L.}\ \bibnamefont {Jacobsen}},\ }\bibfield
  {title} {\bibinfo {title} {Unconventional {Continuous} {Phase} {Transition}
  in a {Three}-{Dimensional} {Dimer} {Model}},\ }\href@noop {} {\bibfield
  {journal} {\bibinfo  {journal} {Physical Review Letters}\ }\textbf {\bibinfo
  {volume} {97}},\ \bibinfo {pages} {030403} (\bibinfo {year}
  {2006})}\BibitemShut {NoStop}%
\bibitem [{\citenamefont {Powell}(2011)}]{Powell:2011hda}%
  \BibitemOpen
  \bibfield  {author} {\bibinfo {author} {\bibfnamefont {S.}~\bibnamefont
  {Powell}},\ }\bibfield  {title} {\bibinfo {title} {{Higgs transitions of spin
  ice}},\ }\href@noop {} {\bibfield  {journal} {\bibinfo  {journal} {Physical
  Review B}\ }\textbf {\bibinfo {volume} {84}},\ \bibinfo {pages} {094437}
  (\bibinfo {year} {2011})}\BibitemShut {NoStop}%
\bibitem [{\citenamefont {Baxter}(1982)}]{baxter}%
  \BibitemOpen
  \bibfield  {author} {\bibinfo {author} {\bibfnamefont {R.~J.}\ \bibnamefont
  {Baxter}},\ }\href@noop {} {\emph {\bibinfo {title} {Exactly Solved Models in
  Statistical Mechanics}}}\ (\bibinfo  {publisher} {Academic Press, London},\
  \bibinfo {year} {1982})\BibitemShut {NoStop}%
\bibitem [{\citenamefont {Sakakibara}\ \emph {et~al.}(2003)\citenamefont
  {Sakakibara}, \citenamefont {Tayama}, \citenamefont {Hiroi}, \citenamefont
  {Matsuhira},\ and\ \citenamefont {Takagi}}]{Sakakibara2003}%
  \BibitemOpen
  \bibfield  {author} {\bibinfo {author} {\bibfnamefont {T.}~\bibnamefont
  {Sakakibara}}, \bibinfo {author} {\bibfnamefont {T.}~\bibnamefont {Tayama}},
  \bibinfo {author} {\bibfnamefont {Z.}~\bibnamefont {Hiroi}}, \bibinfo
  {author} {\bibfnamefont {K.}~\bibnamefont {Matsuhira}},\ and\ \bibinfo
  {author} {\bibfnamefont {S.}~\bibnamefont {Takagi}},\ }\bibfield  {title}
  {\bibinfo {title} {Observation of a {Liquid}-{Gas}-{Type} {Transition} in the
  {Pyrochlore} {Spin} {Ice} {Compound} {D} y 2 {T} i 2 {O} 7 in a {Magnetic}
  {Field}},\ }\href@noop {} {\bibfield  {journal} {\bibinfo  {journal}
  {Physical Review Letters}\ }\textbf {\bibinfo {volume} {90}},\ \bibinfo
  {pages} {207205} (\bibinfo {year} {2003})}\BibitemShut {NoStop}%
\bibitem [{\citenamefont {Fukazawa}\ \emph {et~al.}(2002)\citenamefont
  {Fukazawa}, \citenamefont {Melko}, \citenamefont {Higashinaka}, \citenamefont
  {Maeno},\ and\ \citenamefont {Gingras}}]{Fukazawa2002}%
  \BibitemOpen
  \bibfield  {author} {\bibinfo {author} {\bibfnamefont {H.}~\bibnamefont
  {Fukazawa}}, \bibinfo {author} {\bibfnamefont {R.~G.}\ \bibnamefont {Melko}},
  \bibinfo {author} {\bibfnamefont {R.}~\bibnamefont {Higashinaka}}, \bibinfo
  {author} {\bibfnamefont {Y.}~\bibnamefont {Maeno}},\ and\ \bibinfo {author}
  {\bibfnamefont {M.~J.~P.}\ \bibnamefont {Gingras}},\ }\bibfield  {title}
  {\bibinfo {title} {Magnetic anisotropy of the spin-ice compound
  ${\mathrm{dy}}_{2}{\mathrm{ti}}_{2}{\mathrm{o}}_{7}$},\ }\href@noop {}
  {\bibfield  {journal} {\bibinfo  {journal} {Phys. Rev. B}\ }\textbf {\bibinfo
  {volume} {65}},\ \bibinfo {pages} {054410} (\bibinfo {year}
  {2002})}\BibitemShut {NoStop}%
\bibitem [{\citenamefont {Hiroi}\ \emph
  {et~al.}(2003{\natexlab{a}})\citenamefont {Hiroi}, \citenamefont {Matsuhira},
  \citenamefont {Tayama}, \citenamefont {Takagi},\ and\ \citenamefont
  {Sakakibara}}]{Hiroi2003111}%
  \BibitemOpen
  \bibfield  {author} {\bibinfo {author} {\bibfnamefont {Z.}~\bibnamefont
  {Hiroi}}, \bibinfo {author} {\bibfnamefont {K.}~\bibnamefont {Matsuhira}},
  \bibinfo {author} {\bibfnamefont {T.}~\bibnamefont {Tayama}}, \bibinfo
  {author} {\bibfnamefont {S.}~\bibnamefont {Takagi}},\ and\ \bibinfo {author}
  {\bibfnamefont {T.}~\bibnamefont {Sakakibara}},\ }\bibfield  {title}
  {\bibinfo {title} {Specific heat of kagome ice in the pyrochlore oxide
  dy$_2$ti$_2$o$_7$},\ }\href@noop {} {\bibfield  {journal} {\bibinfo
  {journal} {Journal of the Physical Society of Japan}\ }\textbf {\bibinfo
  {volume} {72}},\ \bibinfo {pages} {411} (\bibinfo {year}
  {2003}{\natexlab{a}})}\BibitemShut {NoStop}%
\bibitem [{\citenamefont {Aoki}\ \emph {et~al.}(2004)\citenamefont {Aoki},
  \citenamefont {Sakakibara}, \citenamefont {Matsuhira},\ and\ \citenamefont
  {Hiroi}}]{Aoki2004}%
  \BibitemOpen
  \bibfield  {author} {\bibinfo {author} {\bibfnamefont {H.}~\bibnamefont
  {Aoki}}, \bibinfo {author} {\bibfnamefont {T.}~\bibnamefont {Sakakibara}},
  \bibinfo {author} {\bibfnamefont {K.}~\bibnamefont {Matsuhira}},\ and\
  \bibinfo {author} {\bibfnamefont {Z.}~\bibnamefont {Hiroi}},\ }\bibfield
  {title} {\bibinfo {title} {Magnetocaloric effect study on the pyrochlore spin
  ice compound dy$_2$ti$_2$o$_7$ in a [111] magnetic field},\ }\href@noop {}
  {\bibfield  {journal} {\bibinfo  {journal} {Journal of the Physical Society
  of Japan}\ }\textbf {\bibinfo {volume} {73}},\ \bibinfo {pages} {2851}
  (\bibinfo {year} {2004})}\BibitemShut {NoStop}%
\bibitem [{\citenamefont {Tabata}\ \emph {et~al.}(2006)\citenamefont {Tabata},
  \citenamefont {Kadowaki}, \citenamefont {Matsuhira}, \citenamefont {Hiroi},
  \citenamefont {Aso}, \citenamefont {Ressouche},\ and\ \citenamefont
  {F\aa{}k}}]{Tabata2006}%
  \BibitemOpen
  \bibfield  {author} {\bibinfo {author} {\bibfnamefont {Y.}~\bibnamefont
  {Tabata}}, \bibinfo {author} {\bibfnamefont {H.}~\bibnamefont {Kadowaki}},
  \bibinfo {author} {\bibfnamefont {K.}~\bibnamefont {Matsuhira}}, \bibinfo
  {author} {\bibfnamefont {Z.}~\bibnamefont {Hiroi}}, \bibinfo {author}
  {\bibfnamefont {N.}~\bibnamefont {Aso}}, \bibinfo {author} {\bibfnamefont
  {E.}~\bibnamefont {Ressouche}},\ and\ \bibinfo {author} {\bibfnamefont
  {B.}~\bibnamefont {F\aa{}k}},\ }\bibfield  {title} {\bibinfo {title} {Kagomé
  ice state in the dipolar spin ice
  ${\mathrm{d}\mathrm{y}}_{2}{\mathrm{t}\mathrm{i}}_{2}{\mathrm{o}}_{7}$},\
  }\href@noop {} {\bibfield  {journal} {\bibinfo  {journal} {Phys. Rev. Lett.}\
  }\textbf {\bibinfo {volume} {97}},\ \bibinfo {pages} {257205} (\bibinfo
  {year} {2006})}\BibitemShut {NoStop}%
\bibitem [{\citenamefont {Fennell}\ \emph {et~al.}(2007)\citenamefont
  {Fennell}, \citenamefont {Bramwell}, \citenamefont {McMorrow}, \citenamefont
  {Manuel},\ and\ \citenamefont {Wildes}}]{Fennell2007}%
  \BibitemOpen
  \bibfield  {author} {\bibinfo {author} {\bibfnamefont {T.}~\bibnamefont
  {Fennell}}, \bibinfo {author} {\bibfnamefont {S.~T.}\ \bibnamefont
  {Bramwell}}, \bibinfo {author} {\bibfnamefont {D.~F.}\ \bibnamefont
  {McMorrow}}, \bibinfo {author} {\bibfnamefont {P.}~\bibnamefont {Manuel}},\
  and\ \bibinfo {author} {\bibfnamefont {A.~R.}\ \bibnamefont {Wildes}},\
  }\bibfield  {title} {\bibinfo {title} {Pinch points and kasteleyn transitions
  in kagome ice},\ }\href@noop {} {\bibfield  {journal} {\bibinfo  {journal}
  {Nature Phyiscs}\ }\textbf {\bibinfo {volume} {3}},\ \bibinfo {pages} {566}
  (\bibinfo {year} {2007})}\BibitemShut {NoStop}%
\bibitem [{\citenamefont {Kadowaki}\ \emph {et~al.}(2009)\citenamefont
  {Kadowaki}, \citenamefont {Doi}, \citenamefont {Aoki}, \citenamefont
  {Tabata}, \citenamefont {J.~Sato}, \citenamefont {W.~Lynn}, \citenamefont
  {Matsuhira},\ and\ \citenamefont {Hiroi}}]{Kadowaki2009}%
  \BibitemOpen
  \bibfield  {author} {\bibinfo {author} {\bibfnamefont {H.}~\bibnamefont
  {Kadowaki}}, \bibinfo {author} {\bibfnamefont {N.}~\bibnamefont {Doi}},
  \bibinfo {author} {\bibfnamefont {Y.}~\bibnamefont {Aoki}}, \bibinfo {author}
  {\bibfnamefont {Y.}~\bibnamefont {Tabata}}, \bibinfo {author} {\bibfnamefont
  {T.}~\bibnamefont {J.~Sato}}, \bibinfo {author} {\bibfnamefont
  {J.}~\bibnamefont {W.~Lynn}}, \bibinfo {author} {\bibfnamefont
  {K.}~\bibnamefont {Matsuhira}},\ and\ \bibinfo {author} {\bibfnamefont
  {Z.}~\bibnamefont {Hiroi}},\ }\bibfield  {title} {\bibinfo {title}
  {Observation of {Magnetic} {Monopoles} in {Spin} {Ice}},\ }\href
  {https://doi.org/10.1143/JPSJ.78.103706} {\bibfield  {journal} {\bibinfo
  {journal} {Journal of the Physical Society of Japan}\ }\textbf {\bibinfo
  {volume} {78}},\ \bibinfo {pages} {103706} (\bibinfo {year}
  {2009})}\BibitemShut {NoStop}%
\bibitem [{\citenamefont {Bernal}\ and\ \citenamefont
  {Fowler}(1933)}]{Bernal1933}%
  \BibitemOpen
  \bibfield  {author} {\bibinfo {author} {\bibfnamefont {J.~D.}\ \bibnamefont
  {Bernal}}\ and\ \bibinfo {author} {\bibfnamefont {R.~H.}\ \bibnamefont
  {Fowler}},\ }\bibfield  {title} {\bibinfo {title} {A theory of water and
  ionic solution, with particular reference to hydrogen and hydroxyl ions},\
  }\href@noop {} {\bibfield  {journal} {\bibinfo  {journal} {The Journal of
  Chemical Physics}\ }\textbf {\bibinfo {volume} {1}},\ \bibinfo {pages} {515}
  (\bibinfo {year} {1933})}\BibitemShut {NoStop}%
\bibitem [{\citenamefont {Pauling}(1935)}]{Pauling1935}%
  \BibitemOpen
  \bibfield  {author} {\bibinfo {author} {\bibfnamefont {L.}~\bibnamefont
  {Pauling}},\ }\bibfield  {title} {\bibinfo {title} {The structure and entropy
  of ice and of other crystals with some randomness of atomic arrangement},\
  }\href@noop {} {\bibfield  {journal} {\bibinfo  {journal} {J. Am. Chem.
  Soc.}\ }\textbf {\bibinfo {volume} {57}},\ \bibinfo {pages} {2680} (\bibinfo
  {year} {1935})}\BibitemShut {NoStop}%
\bibitem [{\citenamefont {Bramwell}\ and\ \citenamefont
  {Harris}(1998)}]{Bramwell1998}%
  \BibitemOpen
  \bibfield  {author} {\bibinfo {author} {\bibfnamefont {S.~T.}\ \bibnamefont
  {Bramwell}}\ and\ \bibinfo {author} {\bibfnamefont {M.~J.}\ \bibnamefont
  {Harris}},\ }\bibfield  {title} {\bibinfo {title} {Frustration in ising-type
  spin models on the pyrochlore lattice},\ }\href@noop {} {\bibfield  {journal}
  {\bibinfo  {journal} {J. Phys.: Condens. Matter}\ }\textbf {\bibinfo {volume}
  {10}},\ \bibinfo {pages} {L215} (\bibinfo {year} {1998})}\BibitemShut
  {NoStop}%
\bibitem [{\citenamefont {Henley}(2010)}]{Henley2010}%
  \BibitemOpen
  \bibfield  {author} {\bibinfo {author} {\bibfnamefont {C.~L.}\ \bibnamefont
  {Henley}},\ }\bibfield  {title} {\bibinfo {title} {The 'coulomb phase' in
  frustrated systems},\ }\href@noop {} {\bibfield  {journal} {\bibinfo
  {journal} {Annual Review of Condensed Matter Physics}\ }\textbf {\bibinfo
  {volume} {1}},\ \bibinfo {pages} {179} (\bibinfo {year} {2010})}\BibitemShut
  {NoStop}%
\bibitem [{\citenamefont {Fennell}\ \emph {et~al.}(2009)\citenamefont
  {Fennell}, \citenamefont {Deen}, \citenamefont {Wildes}, \citenamefont
  {Schmalzl}, \citenamefont {Prabhakaran}, \citenamefont {Boothroyd},
  \citenamefont {Aldus}, \citenamefont {McMorrow},\ and\ \citenamefont
  {Bramwell}}]{Fennell2009}%
  \BibitemOpen
  \bibfield  {author} {\bibinfo {author} {\bibfnamefont {T.}~\bibnamefont
  {Fennell}}, \bibinfo {author} {\bibfnamefont {P.~P.}\ \bibnamefont {Deen}},
  \bibinfo {author} {\bibfnamefont {A.~R.}\ \bibnamefont {Wildes}}, \bibinfo
  {author} {\bibfnamefont {K.}~\bibnamefont {Schmalzl}}, \bibinfo {author}
  {\bibfnamefont {D.}~\bibnamefont {Prabhakaran}}, \bibinfo {author}
  {\bibfnamefont {A.~T.}\ \bibnamefont {Boothroyd}}, \bibinfo {author}
  {\bibfnamefont {R.~J.}\ \bibnamefont {Aldus}}, \bibinfo {author}
  {\bibfnamefont {D.~F.}\ \bibnamefont {McMorrow}},\ and\ \bibinfo {author}
  {\bibfnamefont {S.~T.}\ \bibnamefont {Bramwell}},\ }\bibfield  {title}
  {\bibinfo {title} {Magnetic coulomb phase in the spin ice
  ho$_2$ti$_2$o$_7$},\ }\href@noop {} {\bibfield  {journal} {\bibinfo
  {journal} {Science}\ }\textbf {\bibinfo {volume} {326}},\ \bibinfo {pages}
  {415} (\bibinfo {year} {2009})}\BibitemShut {NoStop}%
\bibitem [{\citenamefont {Krey}\ \emph {et~al.}(2012)\citenamefont {Krey},
  \citenamefont {Legl}, \citenamefont {Dunsiger}, \citenamefont {Meven},
  \citenamefont {Gardner}, \citenamefont {Roper},\ and\ \citenamefont
  {Pfleiderer}}]{Krey2012}%
  \BibitemOpen
  \bibfield  {author} {\bibinfo {author} {\bibfnamefont {C.}~\bibnamefont
  {Krey}}, \bibinfo {author} {\bibfnamefont {S.}~\bibnamefont {Legl}}, \bibinfo
  {author} {\bibfnamefont {S.~R.}\ \bibnamefont {Dunsiger}}, \bibinfo {author}
  {\bibfnamefont {M.}~\bibnamefont {Meven}}, \bibinfo {author} {\bibfnamefont
  {J.~S.}\ \bibnamefont {Gardner}}, \bibinfo {author} {\bibfnamefont {J.~M.}\
  \bibnamefont {Roper}},\ and\ \bibinfo {author} {\bibfnamefont
  {C.}~\bibnamefont {Pfleiderer}},\ }\bibfield  {title} {\bibinfo {title}
  {First {Order} {Metamagnetic} {Transition} in {Ho} 2 {Ti} 2 {O} 7 {Observed}
  by {Vibrating} {Coil} {Magnetometry} at {Milli}-{Kelvin} {Temperatures}},\
  }\href {https://doi.org/10.1103/PhysRevLett.108.257204} {\bibfield  {journal}
  {\bibinfo  {journal} {Physical Review Letters}\ }\textbf {\bibinfo {volume}
  {108}},\ \bibinfo {pages} {257204} (\bibinfo {year} {2012})}\BibitemShut
  {NoStop}%
\bibitem [{\citenamefont {Castelnovo}\ \emph {et~al.}(2008)\citenamefont
  {Castelnovo}, \citenamefont {Moessner},\ and\ \citenamefont
  {Sondhi}}]{Castelnovo2008}%
  \BibitemOpen
  \bibfield  {author} {\bibinfo {author} {\bibfnamefont {C.}~\bibnamefont
  {Castelnovo}}, \bibinfo {author} {\bibfnamefont {R.}~\bibnamefont
  {Moessner}},\ and\ \bibinfo {author} {\bibfnamefont {S.~L.}\ \bibnamefont
  {Sondhi}},\ }\bibfield  {title} {\bibinfo {title} {Magnetic monopoles in spin
  ice},\ }\href@noop {} {\bibfield  {journal} {\bibinfo  {journal} {Nature}\
  }\textbf {\bibinfo {volume} {451}},\ \bibinfo {pages} {42} (\bibinfo {year}
  {2008})}\BibitemShut {NoStop}%
\bibitem [{\citenamefont {Raban}\ \emph {et~al.}(2019)\citenamefont {Raban},
  \citenamefont {Suen}, \citenamefont {Berthier},\ and\ \citenamefont
  {Holdsworth}}]{Raban:2019bk}%
  \BibitemOpen
  \bibfield  {author} {\bibinfo {author} {\bibfnamefont {V.}~\bibnamefont
  {Raban}}, \bibinfo {author} {\bibfnamefont {C.~T.}\ \bibnamefont {Suen}},
  \bibinfo {author} {\bibfnamefont {L.}~\bibnamefont {Berthier}},\ and\
  \bibinfo {author} {\bibfnamefont {P.~C.~W.}\ \bibnamefont {Holdsworth}},\
  }\bibfield  {title} {\bibinfo {title} {{Multiple symmetry sustaining phase
  transitions in spin ice}},\ }\href@noop {} {\bibfield  {journal} {\bibinfo
  {journal} {Physical Review B}\ }\textbf {\bibinfo {volume} {99}},\ \bibinfo
  {pages} {224425} (\bibinfo {year} {2019})}\BibitemShut {NoStop}%
\bibitem [{\citenamefont {Huse}\ \emph {et~al.}(2003)\citenamefont {Huse},
  \citenamefont {Krauth}, \citenamefont {Moessner},\ and\ \citenamefont
  {Sondhi}}]{Huse2003}%
  \BibitemOpen
  \bibfield  {author} {\bibinfo {author} {\bibfnamefont {D.~A.}\ \bibnamefont
  {Huse}}, \bibinfo {author} {\bibfnamefont {W.}~\bibnamefont {Krauth}},
  \bibinfo {author} {\bibfnamefont {R.}~\bibnamefont {Moessner}},\ and\
  \bibinfo {author} {\bibfnamefont {S.~L.}\ \bibnamefont {Sondhi}},\ }\bibfield
   {title} {\bibinfo {title} {Coulomb and {Liquid} {Dimer} {Models} in {Three}
  {Dimensions}},\ }\href@noop {} {\bibfield  {journal} {\bibinfo  {journal}
  {Physical Review Letters}\ }\textbf {\bibinfo {volume} {91}},\ \bibinfo
  {pages} {167004} (\bibinfo {year} {2003})}\BibitemShut {NoStop}%
\bibitem [{\citenamefont {Powell}\ and\ \citenamefont
  {Chalker}(2008{\natexlab{a}})}]{Powell2008PRB}%
  \BibitemOpen
  \bibfield  {author} {\bibinfo {author} {\bibfnamefont {S.}~\bibnamefont
  {Powell}}\ and\ \bibinfo {author} {\bibfnamefont {J.~T.}\ \bibnamefont
  {Chalker}},\ }\bibfield  {title} {\bibinfo {title} {Classical to quantum
  mappings for geometrically frustrated systems: Spin-ice in a [100] field},\
  }\href@noop {} {\bibfield  {journal} {\bibinfo  {journal} {Phys. Rev. B}\
  }\textbf {\bibinfo {volume} {78}},\ \bibinfo {pages} {024422} (\bibinfo
  {year} {2008}{\natexlab{a}})}\BibitemShut {NoStop}%
\bibitem [{\citenamefont {Powell}\ and\ \citenamefont
  {Chalker}(2008{\natexlab{b}})}]{Powell2008PRL}%
  \BibitemOpen
  \bibfield  {author} {\bibinfo {author} {\bibfnamefont {S.}~\bibnamefont
  {Powell}}\ and\ \bibinfo {author} {\bibfnamefont {J.~T.}\ \bibnamefont
  {Chalker}},\ }\bibfield  {title} {\bibinfo {title} {Su(2)-invariant continuum
  theory for an unconventional phase transition in a three-dimensional
  classical dimer model},\ }\href@noop {} {\bibfield  {journal} {\bibinfo
  {journal} {Phys. Rev. Lett.}\ }\textbf {\bibinfo {volume} {101}},\ \bibinfo
  {pages} {155702} (\bibinfo {year} {2008}{\natexlab{b}})}\BibitemShut
  {NoStop}%
\bibitem [{\citenamefont {Kasteleyn}(1963)}]{Kasteleyn1963}%
  \BibitemOpen
  \bibfield  {author} {\bibinfo {author} {\bibfnamefont {P.~W.}\ \bibnamefont
  {Kasteleyn}},\ }\bibfield  {title} {\bibinfo {title} {Dimer statistics and
  phase transitions},\ }\href@noop {} {\bibfield  {journal} {\bibinfo
  {journal} {Journal of Mathematical Physics}\ }\textbf {\bibinfo {volume}
  {4}},\ \bibinfo {pages} {287} (\bibinfo {year} {1963})}\BibitemShut {NoStop}%
\bibitem [{\citenamefont {Laeuchli}\ \emph {et~al.}(2008)\citenamefont
  {Laeuchli}, \citenamefont {Capponi},\ and\ \citenamefont
  {Assaad}}]{Laeuchli:2008cd}%
  \BibitemOpen
  \bibfield  {author} {\bibinfo {author} {\bibfnamefont {A.~M.}\ \bibnamefont
  {Laeuchli}}, \bibinfo {author} {\bibfnamefont {S.}~\bibnamefont {Capponi}},\
  and\ \bibinfo {author} {\bibfnamefont {F.~F.}\ \bibnamefont {Assaad}},\
  }\bibfield  {title} {\bibinfo {title} {{Dynamical dimer correlations at
  bipartite and non-bipartite Rokhsar-Kivelson points}},\ }\href@noop {}
  {\bibfield  {journal} {\bibinfo  {journal} {Journal Of Statistical Mechanics:
  Theory And Experiment}\ }\textbf {\bibinfo {volume} {2008}},\ \bibinfo
  {pages} {P01010} (\bibinfo {year} {2008})}\BibitemShut {NoStop}%
\bibitem [{\citenamefont {Kao}\ \emph {et~al.}(2016)\citenamefont {Kao},
  \citenamefont {Holdsworth},\ and\ \citenamefont {Kao}}]{Kao2016}%
  \BibitemOpen
  \bibfield  {author} {\bibinfo {author} {\bibfnamefont {W.~H.}\ \bibnamefont
  {Kao}}, \bibinfo {author} {\bibfnamefont {P.~C.~W.}\ \bibnamefont
  {Holdsworth}},\ and\ \bibinfo {author} {\bibfnamefont {Y.~J.}\ \bibnamefont
  {Kao}},\ }\bibfield  {title} {\bibinfo {title} {Field-induced ordering in
  dipolar spin ice'},\ }\href@noop {} {\bibfield  {journal} {\bibinfo
  {journal} {Phys. Rev. B}\ }\textbf {\bibinfo {volume} {93}},\ \bibinfo
  {pages} {180410(R)} (\bibinfo {year} {2016})}\BibitemShut {NoStop}%
\bibitem [{\citenamefont {Jaubert}\ \emph {et~al.}(2008)\citenamefont
  {Jaubert}, \citenamefont {Chalker}, \citenamefont {Holdsworth},\ and\
  \citenamefont {Moessner}}]{Jaubert2008}%
  \BibitemOpen
  \bibfield  {author} {\bibinfo {author} {\bibfnamefont {L.~D.~C.}\
  \bibnamefont {Jaubert}}, \bibinfo {author} {\bibfnamefont {J.~T.}\
  \bibnamefont {Chalker}}, \bibinfo {author} {\bibfnamefont {P.~C.~W.}\
  \bibnamefont {Holdsworth}},\ and\ \bibinfo {author} {\bibfnamefont
  {R.}~\bibnamefont {Moessner}},\ }\bibfield  {title} {\bibinfo {title}
  {Three-dimensional kasteleyn transition: Spin ice in a [100] field},\
  }\href@noop {} {\bibfield  {journal} {\bibinfo  {journal} {Phys. Rev. Lett.}\
  }\textbf {\bibinfo {volume} {100}} (\bibinfo {year} {2008})}\BibitemShut
  {NoStop}%
\bibitem [{\citenamefont {Jaubert}\ and\ \citenamefont
  {Holdsworth}(2009)}]{Jaubert2009}%
  \BibitemOpen
  \bibfield  {author} {\bibinfo {author} {\bibfnamefont {L.~D.~C.}\
  \bibnamefont {Jaubert}}\ and\ \bibinfo {author} {\bibfnamefont {P.~C.~W.}\
  \bibnamefont {Holdsworth}},\ }\bibfield  {title} {\bibinfo {title} {Signature
  of magnetic monopole and dirac string dynamics in spin ice},\ }\href@noop {}
  {\bibfield  {journal} {\bibinfo  {journal} {Nat. Phys.}\ }\textbf {\bibinfo
  {volume} {5}},\ \bibinfo {pages} {258} (\bibinfo {year} {2009})}\BibitemShut
  {NoStop}%
\bibitem [{\citenamefont {Brooks-Bartlett}\ \emph {et~al.}(2014)\citenamefont
  {Brooks-Bartlett}, \citenamefont {Banks}, \citenamefont {Jaubert},
  \citenamefont {Harman-Clarke},\ and\ \citenamefont
  {Holdsworth}}]{BrooksBartlett:2014kf}%
  \BibitemOpen
  \bibfield  {author} {\bibinfo {author} {\bibfnamefont {M.~E.}\ \bibnamefont
  {Brooks-Bartlett}}, \bibinfo {author} {\bibfnamefont {S.~T.}\ \bibnamefont
  {Banks}}, \bibinfo {author} {\bibfnamefont {L.~D.~C.}\ \bibnamefont
  {Jaubert}}, \bibinfo {author} {\bibfnamefont {A.}~\bibnamefont
  {Harman-Clarke}},\ and\ \bibinfo {author} {\bibfnamefont {P.~C.~W.}\
  \bibnamefont {Holdsworth}},\ }\bibfield  {title} {\bibinfo {title}
  {{Magnetic-Moment Fragmentation and Monopole Crystallization}},\ }\href@noop
  {} {\bibfield  {journal} {\bibinfo  {journal} {Physical Review X}\ }\textbf
  {\bibinfo {volume} {4}},\ \bibinfo {pages} {011007} (\bibinfo {year}
  {2014})}\BibitemShut {NoStop}%
\bibitem [{\citenamefont {Baez}\ and\ \citenamefont {Borzi}(2017)}]{Baez2017}%
  \BibitemOpen
  \bibfield  {author} {\bibinfo {author} {\bibfnamefont {M.~L.}\ \bibnamefont
  {Baez}}\ and\ \bibinfo {author} {\bibfnamefont {R.~A.}\ \bibnamefont
  {Borzi}},\ }\bibfield  {title} {\bibinfo {title} {The 3d kasteleyn transition
  in dipolar spin ice: a numerical study with the conserved monopoles
  algorithm},\ }\href@noop {} {\bibfield  {journal} {\bibinfo  {journal} {J.
  Phys.: Condens. Matter}\ }\textbf {\bibinfo {volume} {29}},\ \bibinfo {pages}
  {055806} (\bibinfo {year} {2017})}\BibitemShut {NoStop}%
\bibitem [{\citenamefont {Powell}(2013)}]{Powell:2013ct}%
  \BibitemOpen
  \bibfield  {author} {\bibinfo {author} {\bibfnamefont {S.}~\bibnamefont
  {Powell}},\ }\bibfield  {title} {\bibinfo {title} {{Confinement of monopoles
  and scaling theory near unconventional critical points}},\ }\href@noop {}
  {\bibfield  {journal} {\bibinfo  {journal} {Physical Review B}\ }\textbf
  {\bibinfo {volume} {87}},\ \bibinfo {pages} {064414} (\bibinfo {year}
  {2013})}\BibitemShut {NoStop}%
\bibitem [{\citenamefont {M\"oller}\ and\ \citenamefont
  {Moessner}(2009)}]{MoellerPRB2009}%
  \BibitemOpen
  \bibfield  {author} {\bibinfo {author} {\bibfnamefont {G.}~\bibnamefont
  {M\"oller}}\ and\ \bibinfo {author} {\bibfnamefont {R.}~\bibnamefont
  {Moessner}},\ }\bibfield  {title} {\bibinfo {title} {Magnetic multipole
  analysis of kagome and artificial spin-ice dipolar arrays},\ }\href
  {https://doi.org/10.1103/PhysRevB.80.140409} {\bibfield  {journal} {\bibinfo
  {journal} {Phys. Rev. B}\ }\textbf {\bibinfo {volume} {80}},\ \bibinfo
  {pages} {140409(R)} (\bibinfo {year} {2009})}\BibitemShut {NoStop}%
\bibitem [{\citenamefont {Chern}\ \emph {et~al.}(2011)\citenamefont {Chern},
  \citenamefont {Mellado},\ and\ \citenamefont {Tchernyshyov}}]{Chern2011}%
  \BibitemOpen
  \bibfield  {author} {\bibinfo {author} {\bibfnamefont {G.-W.}\ \bibnamefont
  {Chern}}, \bibinfo {author} {\bibfnamefont {P.}~\bibnamefont {Mellado}},\
  and\ \bibinfo {author} {\bibfnamefont {O.}~\bibnamefont {Tchernyshyov}},\
  }\bibfield  {title} {\bibinfo {title} {{Two-Stage Ordering of Spins in
  Dipolar Spin Ice on the Kagome Lattice}},\ }\href@noop {} {\bibfield
  {journal} {\bibinfo  {journal} {Physical Review Letters}\ }\textbf {\bibinfo
  {volume} {106}},\ \bibinfo {pages} {207202} (\bibinfo {year}
  {2011})}\BibitemShut {NoStop}%
\bibitem [{\citenamefont {Anderson}(1956)}]{Anderson56}%
  \BibitemOpen
  \bibfield  {author} {\bibinfo {author} {\bibfnamefont {P.~W.}\ \bibnamefont
  {Anderson}},\ }\bibfield  {title} {\bibinfo {title} {{Ordering and
  antiferromagnetism in ferrites}},\ }\href@noop {} {\bibfield  {journal}
  {\bibinfo  {journal} {Physical Review}\ }\textbf {\bibinfo {volume} {102}},\
  \bibinfo {pages} {1008} (\bibinfo {year} {1956})}\BibitemShut {NoStop}%
\bibitem [{\citenamefont {Harman-Clarke}(2010)}]{HarmanClarke2012}%
  \BibitemOpen
  \bibfield  {author} {\bibinfo {author} {\bibfnamefont {A.}~\bibnamefont
  {Harman-Clarke}},\ }\emph {\bibinfo {title} {Topological Constraints and
  Ordering in Model Frustrated Magnets}},\ \href@noop {} {Ph.D. thesis},\
  \bibinfo  {school} {University College London} (\bibinfo {year}
  {2010})\BibitemShut {NoStop}%
\bibitem [{\citenamefont {Hiroi}\ \emph
  {et~al.}(2003{\natexlab{b}})\citenamefont {Hiroi}, \citenamefont
  {Matsuhira},\ and\ \citenamefont {Ogata}}]{Hiroi2003110}%
  \BibitemOpen
  \bibfield  {author} {\bibinfo {author} {\bibfnamefont {Z.}~\bibnamefont
  {Hiroi}}, \bibinfo {author} {\bibfnamefont {K.}~\bibnamefont {Matsuhira}},\
  and\ \bibinfo {author} {\bibfnamefont {M.}~\bibnamefont {Ogata}},\ }\bibfield
   {title} {\bibinfo {title} {Ferromagnetic {Ising} {Spin} {Chains} {Emerging}
  from the {Spin} {Ice} under {Magnetic} {Field}},\ }\href
  {https://doi.org/10.1143/JPSJ.72.3045} {\bibfield  {journal} {\bibinfo
  {journal} {J. Phys. Soc. Jpn.}\ }\textbf {\bibinfo {volume} {72}},\ \bibinfo
  {pages} {3045} (\bibinfo {year} {2003}{\natexlab{b}})}\BibitemShut {NoStop}%
\bibitem [{\citenamefont {Fennell}\ \emph {et~al.}(2005)\citenamefont
  {Fennell}, \citenamefont {Petrenko}, \citenamefont {F\aa{}k}, \citenamefont
  {Gardner}, \citenamefont {Bramwell},\ and\ \citenamefont
  {Ouladdiaf}}]{Fennell2005}%
  \BibitemOpen
  \bibfield  {author} {\bibinfo {author} {\bibfnamefont {T.}~\bibnamefont
  {Fennell}}, \bibinfo {author} {\bibfnamefont {O.~A.}\ \bibnamefont
  {Petrenko}}, \bibinfo {author} {\bibfnamefont {B.}~\bibnamefont {F\aa{}k}},
  \bibinfo {author} {\bibfnamefont {J.~S.}\ \bibnamefont {Gardner}}, \bibinfo
  {author} {\bibfnamefont {S.~T.}\ \bibnamefont {Bramwell}},\ and\ \bibinfo
  {author} {\bibfnamefont {B.}~\bibnamefont {Ouladdiaf}},\ }\bibfield  {title}
  {\bibinfo {title} {Neutron scattering studies of the spin ices
  ho$_2$ti$_2$o$_7$ and dy$_2$ti$_2$o$_7$ in applied magnetic field},\
  }\href@noop {} {\bibfield  {journal} {\bibinfo  {journal} {Phys. Rev. B}\
  }\textbf {\bibinfo {volume} {72}},\ \bibinfo {pages} {224411} (\bibinfo
  {year} {2005})}\BibitemShut {NoStop}%
\bibitem [{\citenamefont {Ruff}\ \emph {et~al.}(2005)\citenamefont {Ruff},
  \citenamefont {Melko},\ and\ \citenamefont {Gingras}}]{Ruff2005}%
  \BibitemOpen
  \bibfield  {author} {\bibinfo {author} {\bibfnamefont {J.~P.~C.}\
  \bibnamefont {Ruff}}, \bibinfo {author} {\bibfnamefont {R.~G.}\ \bibnamefont
  {Melko}},\ and\ \bibinfo {author} {\bibfnamefont {M.~J.~P.}\ \bibnamefont
  {Gingras}},\ }\bibfield  {title} {\bibinfo {title} {Finite-{Temperature}
  {Transitions} in {Dipolar} {Spin} {Ice} in a {Large} {Magnetic} {Field}},\
  }\href@noop {} {\bibfield  {journal} {\bibinfo  {journal} {Physical Review
  Letters}\ }\textbf {\bibinfo {volume} {95}},\ \bibinfo {pages} {097202}
  (\bibinfo {year} {2005})}\BibitemShut {NoStop}%
\bibitem [{\citenamefont {Jaubert}\ \emph {et~al.}(2013)\citenamefont
  {Jaubert}, \citenamefont {Harris}, \citenamefont {Fennell}, \citenamefont
  {Melko}, \citenamefont {Bramwell},\ and\ \citenamefont
  {Holdsworth}}]{Jaubert:2013jz}%
  \BibitemOpen
  \bibfield  {author} {\bibinfo {author} {\bibfnamefont {L.~D.~C.}\
  \bibnamefont {Jaubert}}, \bibinfo {author} {\bibfnamefont {M.~J.}\
  \bibnamefont {Harris}}, \bibinfo {author} {\bibfnamefont {T.}~\bibnamefont
  {Fennell}}, \bibinfo {author} {\bibfnamefont {R.~G.}\ \bibnamefont {Melko}},
  \bibinfo {author} {\bibfnamefont {S.~T.}\ \bibnamefont {Bramwell}},\ and\
  \bibinfo {author} {\bibfnamefont {P.~C.~W.}\ \bibnamefont {Holdsworth}},\
  }\bibfield  {title} {\bibinfo {title} {{Topological-Sector Fluctuations and
  Curie-Law Crossover in Spin Ice}},\ }\href@noop {} {\bibfield  {journal}
  {\bibinfo  {journal} {Physical Review X}\ }\textbf {\bibinfo {volume} {3}},\
  \bibinfo {pages} {011014} (\bibinfo {year} {2013})}\BibitemShut {NoStop}%
\bibitem [{\citenamefont {Bhattacharjee}\ \emph {et~al.}(1983)\citenamefont
  {Bhattacharjee}, \citenamefont {Nagle}, \citenamefont {Huse},\ and\
  \citenamefont {Fisher}}]{Bhattacharjee1983}%
  \BibitemOpen
  \bibfield  {author} {\bibinfo {author} {\bibfnamefont {S.~M.}\ \bibnamefont
  {Bhattacharjee}}, \bibinfo {author} {\bibfnamefont {J.~F.}\ \bibnamefont
  {Nagle}}, \bibinfo {author} {\bibfnamefont {D.~A.}\ \bibnamefont {Huse}},\
  and\ \bibinfo {author} {\bibfnamefont {M.~E.}\ \bibnamefont {Fisher}},\
  }\bibfield  {title} {\bibinfo {title} {Critical behavior of a
  three-dimensional dimer model},\ }\href@noop {} {\bibfield  {journal}
  {\bibinfo  {journal} {Journal of Statistical Physics}\ }\textbf {\bibinfo
  {volume} {32}},\ \bibinfo {pages} {361} (\bibinfo {year} {1983})}\BibitemShut
  {NoStop}%
\bibitem [{\citenamefont {Alder}\ and\ \citenamefont {E}(1960)}]{Alder60}%
  \BibitemOpen
  \bibfield  {author} {\bibinfo {author} {\bibfnamefont {B.~J.}\ \bibnamefont
  {Alder}}\ and\ \bibinfo {author} {\bibfnamefont {W.~T.}\ \bibnamefont {E}},\
  }\href@noop {} {\bibfield  {journal} {\bibinfo  {journal} {Journal of
  Chemical Physics}\ }\textbf {\bibinfo {volume} {33}},\ \bibinfo {pages}
  {1439} (\bibinfo {year} {1960})}\BibitemShut {NoStop}%
\bibitem [{\citenamefont {Dijkstra}\ and\ \citenamefont
  {Frenkel}(1994)}]{Frenkel94}%
  \BibitemOpen
  \bibfield  {author} {\bibinfo {author} {\bibfnamefont {M.}~\bibnamefont
  {Dijkstra}}\ and\ \bibinfo {author} {\bibfnamefont {D.}~\bibnamefont
  {Frenkel}},\ }\href@noop {} {\bibfield  {journal} {\bibinfo  {journal}
  {Physical Review Letters}\ }\textbf {\bibinfo {volume} {72}},\ \bibinfo
  {pages} {298} (\bibinfo {year} {1994})}\BibitemShut {NoStop}%
\bibitem [{\citenamefont {Jaubert}(2009)}]{Jaubert:2009to}%
  \BibitemOpen
  \bibfield  {author} {\bibinfo {author} {\bibfnamefont {L.~D.~C.}\
  \bibnamefont {Jaubert}},\ }\href@noop {} {\emph {\bibinfo {title}
  {{Topological Constraints and Defects in Spin Ice}}}}\ (\bibinfo  {publisher}
  {ENS Lyon},\ \bibinfo {year} {2009})\BibitemShut {NoStop}%
\bibitem [{\citenamefont {Goldenfeld}(1994)}]{Goldenfeld94}%
  \BibitemOpen
  \bibfield  {author} {\bibinfo {author} {\bibfnamefont {N.}~\bibnamefont
  {Goldenfeld}},\ }\href@noop {} {\emph {\bibinfo {title} {Renormalization
  group in critical phenomena}}}\ (\bibinfo  {publisher} {Addison-Wesley},\
  \bibinfo {address} {Reading},\ \bibinfo {year} {1994})\BibitemShut {NoStop}%
\bibitem [{\citenamefont {Powell}(2015)}]{Powell:2015ut}%
  \BibitemOpen
  \bibfield  {author} {\bibinfo {author} {\bibfnamefont {S.}~\bibnamefont
  {Powell}},\ }\bibfield  {title} {\bibinfo {title} {{Ferromagnetic Coulomb
  phase in classical spin ice}},\ }\href@noop {} {\bibfield  {journal}
  {\bibinfo  {journal} {Physical Review B}\ }\textbf {\bibinfo {volume} {91}},\
  \bibinfo {pages} {094431} (\bibinfo {year} {2015})}\BibitemShut {NoStop}%
\bibitem [{\citenamefont {Stewart}\ \emph {et~al.}(2008)\citenamefont
  {Stewart}, \citenamefont {Deen}, \citenamefont {Andersen}, \citenamefont
  {Schober}, \citenamefont {Barth{\'e}l{\'e}my}, \citenamefont {Hillier},
  \citenamefont {Murani}, \citenamefont {Hayes},\ and\ \citenamefont
  {Lindenau}}]{Stewart:2008hw}%
  \BibitemOpen
  \bibfield  {author} {\bibinfo {author} {\bibfnamefont {J.~R.}\ \bibnamefont
  {Stewart}}, \bibinfo {author} {\bibfnamefont {P.~P.}\ \bibnamefont {Deen}},
  \bibinfo {author} {\bibfnamefont {K.~H.}\ \bibnamefont {Andersen}}, \bibinfo
  {author} {\bibfnamefont {H.}~\bibnamefont {Schober}}, \bibinfo {author}
  {\bibfnamefont {J.-F.}\ \bibnamefont {Barth{\'e}l{\'e}my}}, \bibinfo {author}
  {\bibfnamefont {J.~M.}\ \bibnamefont {Hillier}}, \bibinfo {author}
  {\bibfnamefont {A.~P.}\ \bibnamefont {Murani}}, \bibinfo {author}
  {\bibfnamefont {T.}~\bibnamefont {Hayes}},\ and\ \bibinfo {author}
  {\bibfnamefont {B.}~\bibnamefont {Lindenau}},\ }\bibfield  {title} {\bibinfo
  {title} {{Disordered materials studied using neutron polarization analysis on
  the multi-detector spectrometer, D7}},\ }\href@noop {} {\bibfield  {journal}
  {\bibinfo  {journal} {Journal of Applied Crystallography}\ }\textbf {\bibinfo
  {volume} {42}},\ \bibinfo {pages} {69} (\bibinfo {year} {2008})}\BibitemShut
  {NoStop}%
\bibitem [{\citenamefont {Diep}(2013)}]{Diepbook}%
  \BibitemOpen
  \bibfield  {author} {\bibinfo {author} {\bibfnamefont {H.~T.}\ \bibnamefont
  {Diep}},\ }\href@noop {} {\emph {\bibinfo {title} {Frustrated Spin
  Systems}}},\ \bibinfo {edition} {2nd}\ ed.\ (\bibinfo  {publisher} {World
  Scientific Publishing Co.},\ \bibinfo {address} {Singapore},\ \bibinfo {year}
  {2013})\BibitemShut {NoStop}%
\bibitem [{\citenamefont {Morris}\ \emph {et~al.}(2009)\citenamefont {Morris},
  \citenamefont {Tennant}, \citenamefont {Grigera}, \citenamefont {Klemke},
  \citenamefont {Castelnovo}, \citenamefont {Moessner}, \citenamefont
  {Czternasty}, \citenamefont {Meissner}, \citenamefont {Rule}, \citenamefont
  {Hoffman}, \citenamefont {Kiever}, \citenamefont {Gerischer}, \citenamefont
  {Slobinsky},\ and\ \citenamefont {Perry}}]{Morris2009}%
  \BibitemOpen
  \bibfield  {author} {\bibinfo {author} {\bibfnamefont {D.~J.~P.}\
  \bibnamefont {Morris}}, \bibinfo {author} {\bibfnamefont {D.~A.}\
  \bibnamefont {Tennant}}, \bibinfo {author} {\bibfnamefont {S.~A.}\
  \bibnamefont {Grigera}}, \bibinfo {author} {\bibfnamefont {B.}~\bibnamefont
  {Klemke}}, \bibinfo {author} {\bibfnamefont {C.}~\bibnamefont {Castelnovo}},
  \bibinfo {author} {\bibfnamefont {R.}~\bibnamefont {Moessner}}, \bibinfo
  {author} {\bibfnamefont {C.}~\bibnamefont {Czternasty}}, \bibinfo {author}
  {\bibfnamefont {M.}~\bibnamefont {Meissner}}, \bibinfo {author}
  {\bibfnamefont {K.~C.}\ \bibnamefont {Rule}}, \bibinfo {author}
  {\bibfnamefont {J.~U.}\ \bibnamefont {Hoffman}}, \bibinfo {author}
  {\bibfnamefont {K.}~\bibnamefont {Kiever}}, \bibinfo {author} {\bibfnamefont
  {S.}~\bibnamefont {Gerischer}}, \bibinfo {author} {\bibfnamefont
  {D.}~\bibnamefont {Slobinsky}},\ and\ \bibinfo {author} {\bibfnamefont
  {R.~S.}\ \bibnamefont {Perry}},\ }\bibfield  {title} {\bibinfo {title} {Dirac
  strings and magnetic monopoles in the spin ice dy$_2$ti$_2$o$_7$},\
  }\href@noop {} {\bibfield  {journal} {\bibinfo  {journal} {Science}\ }\textbf
  {\bibinfo {volume} {326}},\ \bibinfo {pages} {411} (\bibinfo {year}
  {2009})}\BibitemShut {NoStop}%
\bibitem [{\citenamefont {Quilliam}\ \emph {et~al.}(2011)\citenamefont
  {Quilliam}, \citenamefont {Yaraskavitch}, \citenamefont {Dabkowska},
  \citenamefont {Gaulin},\ and\ \citenamefont {Kycia}}]{Quilliam:2011df}%
  \BibitemOpen
  \bibfield  {author} {\bibinfo {author} {\bibfnamefont {J.~A.}\ \bibnamefont
  {Quilliam}}, \bibinfo {author} {\bibfnamefont {L.~R.}\ \bibnamefont
  {Yaraskavitch}}, \bibinfo {author} {\bibfnamefont {H.~A.}\ \bibnamefont
  {Dabkowska}}, \bibinfo {author} {\bibfnamefont {B.~D.}\ \bibnamefont
  {Gaulin}},\ and\ \bibinfo {author} {\bibfnamefont {J.~B.}\ \bibnamefont
  {Kycia}},\ }\bibfield  {title} {\bibinfo {title} {{Dynamics of the magnetic
  susceptibility deep in the Coulomb phase of the dipolar spin ice material
  Ho$_{2}$Ti$_{2}$O$_{7}$}},\ }\href@noop {} {\bibfield  {journal} {\bibinfo
  {journal} {Physical Review B}\ }\textbf {\bibinfo {volume} {83}},\ \bibinfo
  {pages} {094424} (\bibinfo {year} {2011})}\BibitemShut {NoStop}%
\bibitem [{\citenamefont {Bovo}\ \emph {et~al.}(2013)\citenamefont {Bovo},
  \citenamefont {Jaubert}, \citenamefont {Holdsworth},\ and\ \citenamefont
  {Bramwell}}]{Bovo:2013fq}%
  \BibitemOpen
  \bibfield  {author} {\bibinfo {author} {\bibfnamefont {L.}~\bibnamefont
  {Bovo}}, \bibinfo {author} {\bibfnamefont {L.~D.~C.}\ \bibnamefont
  {Jaubert}}, \bibinfo {author} {\bibfnamefont {P.~C.~W.}\ \bibnamefont
  {Holdsworth}},\ and\ \bibinfo {author} {\bibfnamefont {S.~T.}\ \bibnamefont
  {Bramwell}},\ }\bibfield  {title} {\bibinfo {title} {{Crystal shape-dependent
  magnetic susceptibility and Curie law crossover in the spin ices Dy2Ti2O7 and
  Ho2Ti2O7}},\ }\href@noop {} {\bibfield  {journal} {\bibinfo  {journal}
  {Journal Of Physics-Condensed Matter}\ }\textbf {\bibinfo {volume} {25}},\
  \bibinfo {pages} {386002} (\bibinfo {year} {2013})}\BibitemShut {NoStop}%
\bibitem [{\citenamefont {Twengstr{\"o}m}\ \emph {et~al.}(2017)\citenamefont
  {Twengstr{\"o}m}, \citenamefont {Bovo}, \citenamefont {Gingras},
  \citenamefont {Bramwell},\ and\ \citenamefont
  {Henelius}}]{Twengstrom:2017df}%
  \BibitemOpen
  \bibfield  {author} {\bibinfo {author} {\bibfnamefont {M.}~\bibnamefont
  {Twengstr{\"o}m}}, \bibinfo {author} {\bibfnamefont {L.}~\bibnamefont
  {Bovo}}, \bibinfo {author} {\bibfnamefont {M.~J.~P.}\ \bibnamefont
  {Gingras}}, \bibinfo {author} {\bibfnamefont {S.~T.}\ \bibnamefont
  {Bramwell}},\ and\ \bibinfo {author} {\bibfnamefont {P.}~\bibnamefont
  {Henelius}},\ }\bibfield  {title} {\bibinfo {title} {{Microscopic aspects of
  magnetic lattice demagnetizing factors}},\ }\href@noop {} {\bibfield
  {journal} {\bibinfo  {journal} {Physical Review Materials}\ }\textbf
  {\bibinfo {volume} {1}},\ \bibinfo {pages} {044406} (\bibinfo {year}
  {2017})}\BibitemShut {NoStop}%
\bibitem [{\citenamefont {Paulsen}\ \emph {et~al.}(2014)\citenamefont
  {Paulsen}, \citenamefont {Jackson}, \citenamefont {Lhotel}, \citenamefont
  {Canals}, \citenamefont {Prabhakaran}, \citenamefont {Matsuhira},
  \citenamefont {Giblin},\ and\ \citenamefont {Bramwell}}]{Paulsen:2014cc}%
  \BibitemOpen
  \bibfield  {author} {\bibinfo {author} {\bibfnamefont {C.}~\bibnamefont
  {Paulsen}}, \bibinfo {author} {\bibfnamefont {M.~J.}\ \bibnamefont
  {Jackson}}, \bibinfo {author} {\bibfnamefont {E.}~\bibnamefont {Lhotel}},
  \bibinfo {author} {\bibfnamefont {B.}~\bibnamefont {Canals}}, \bibinfo
  {author} {\bibfnamefont {D.}~\bibnamefont {Prabhakaran}}, \bibinfo {author}
  {\bibfnamefont {K.}~\bibnamefont {Matsuhira}}, \bibinfo {author}
  {\bibfnamefont {S.~R.}\ \bibnamefont {Giblin}},\ and\ \bibinfo {author}
  {\bibfnamefont {S.~T.}\ \bibnamefont {Bramwell}},\ }\bibfield  {title}
  {\bibinfo {title} {{Far-from-equilibrium monopole dynamics in spin ice}},\
  }\href@noop {} {\bibfield  {journal} {\bibinfo  {journal} {Nature Physics}\
  }\textbf {\bibinfo {volume} {10}},\ \bibinfo {pages} {135} (\bibinfo {year}
  {2014})}\BibitemShut {NoStop}%
\bibitem [{\citenamefont {Borzi}\ \emph {et~al.}(2016)\citenamefont {Borzi},
  \citenamefont {G{\'o}mez~Albarrac{\'\i}n}, \citenamefont {Rosales},
  \citenamefont {Rossini}, \citenamefont {Steppke}, \citenamefont
  {Prabhakaran}, \citenamefont {Mackenzie}, \citenamefont {Cabra},\ and\
  \citenamefont {Grigera}}]{Borzi:2016gj}%
  \BibitemOpen
  \bibfield  {author} {\bibinfo {author} {\bibfnamefont {R.~A.}\ \bibnamefont
  {Borzi}}, \bibinfo {author} {\bibfnamefont {F.~A.}\ \bibnamefont
  {G{\'o}mez~Albarrac{\'\i}n}}, \bibinfo {author} {\bibfnamefont {H.~D.}\
  \bibnamefont {Rosales}}, \bibinfo {author} {\bibfnamefont {G.~L.}\
  \bibnamefont {Rossini}}, \bibinfo {author} {\bibfnamefont {A.}~\bibnamefont
  {Steppke}}, \bibinfo {author} {\bibfnamefont {D.}~\bibnamefont
  {Prabhakaran}}, \bibinfo {author} {\bibfnamefont {A.~P.}\ \bibnamefont
  {Mackenzie}}, \bibinfo {author} {\bibfnamefont {D.~C.}\ \bibnamefont
  {Cabra}},\ and\ \bibinfo {author} {\bibfnamefont {S.~A.}\ \bibnamefont
  {Grigera}},\ }\bibfield  {title} {\bibinfo {title} {{Intermediate
  magnetization state and competing orders in Dy2Ti2O7 and Ho2Ti2O7}},\
  }\href@noop {} {\bibfield  {journal} {\bibinfo  {journal} {Nature
  Communications}\ }\textbf {\bibinfo {volume} {7}},\ \bibinfo {pages} {2554}
  (\bibinfo {year} {2016})}\BibitemShut {NoStop}%
\bibitem [{\citenamefont {Melko}\ \emph {et~al.}(2001)\citenamefont {Melko},
  \citenamefont {denHertog},\ and\ \citenamefont {Gingras}}]{Melko:2001el}%
  \BibitemOpen
  \bibfield  {author} {\bibinfo {author} {\bibfnamefont {R.~G.}\ \bibnamefont
  {Melko}}, \bibinfo {author} {\bibfnamefont {B.~C.}\ \bibnamefont
  {denHertog}},\ and\ \bibinfo {author} {\bibfnamefont {M.~J.~P.}\ \bibnamefont
  {Gingras}},\ }\bibfield  {title} {\bibinfo {title} {{Long-range order at low
  temperatures in dipolar spin ice}},\ }\href@noop {} {\bibfield  {journal}
  {\bibinfo  {journal} {Physical Review Letters}\ }\textbf {\bibinfo {volume}
  {87}},\ \bibinfo {pages} {067203} (\bibinfo {year} {2001})}\BibitemShut
  {NoStop}%
\bibitem [{\citenamefont {Takatsu}\ \emph {et~al.}(2013)\citenamefont
  {Takatsu}, \citenamefont {Goto}, \citenamefont {Otsuka}, \citenamefont
  {Higashinaka}, \citenamefont {Matsubayashi}, \citenamefont {Uwatoko},\ and\
  \citenamefont {Kadowaki}}]{Takatsu:2013fc}%
  \BibitemOpen
  \bibfield  {author} {\bibinfo {author} {\bibfnamefont {H.}~\bibnamefont
  {Takatsu}}, \bibinfo {author} {\bibfnamefont {K.}~\bibnamefont {Goto}},
  \bibinfo {author} {\bibfnamefont {H.}~\bibnamefont {Otsuka}}, \bibinfo
  {author} {\bibfnamefont {R.}~\bibnamefont {Higashinaka}}, \bibinfo {author}
  {\bibfnamefont {K.}~\bibnamefont {Matsubayashi}}, \bibinfo {author}
  {\bibfnamefont {Y.}~\bibnamefont {Uwatoko}},\ and\ \bibinfo {author}
  {\bibfnamefont {H.}~\bibnamefont {Kadowaki}},\ }\bibfield  {title} {\bibinfo
  {title} {{Two-Dimensional Monopole Dynamics in the Dipolar Spin Ice Dy 2Ti 2O
  7}},\ }\href@noop {} {\bibfield  {journal} {\bibinfo  {journal} {Journal Of
  The Physical Society Of Japan}\ }\textbf {\bibinfo {volume} {82}},\ \bibinfo
  {pages} {073707} (\bibinfo {year} {2013})}\BibitemShut {NoStop}%
\bibitem [{\citenamefont {Otsuka}\ \emph {et~al.}(2014)\citenamefont {Otsuka},
  \citenamefont {Takatsu}, \citenamefont {Goto},\ and\ \citenamefont
  {Kadowaki}}]{Otsuka:2014dg}%
  \BibitemOpen
  \bibfield  {author} {\bibinfo {author} {\bibfnamefont {H.}~\bibnamefont
  {Otsuka}}, \bibinfo {author} {\bibfnamefont {H.}~\bibnamefont {Takatsu}},
  \bibinfo {author} {\bibfnamefont {K.}~\bibnamefont {Goto}},\ and\ \bibinfo
  {author} {\bibfnamefont {H.}~\bibnamefont {Kadowaki}},\ }\bibfield  {title}
  {\bibinfo {title} {{Scaling ansatz for the ac magnetic response in
  two-dimensional spin ice}},\ }\href@noop {} {\bibfield  {journal} {\bibinfo
  {journal} {Physical Review B}\ }\textbf {\bibinfo {volume} {90}},\ \bibinfo
  {pages} {144428} (\bibinfo {year} {2014})}\BibitemShut {NoStop}%
\bibitem [{\citenamefont {Borzi}\ \emph {et~al.}(2013)\citenamefont {Borzi},
  \citenamefont {Slobinsky},\ and\ \citenamefont {Grigera}}]{Borzi:2013jo}%
  \BibitemOpen
  \bibfield  {author} {\bibinfo {author} {\bibfnamefont {R.~A.}\ \bibnamefont
  {Borzi}}, \bibinfo {author} {\bibfnamefont {D.}~\bibnamefont {Slobinsky}},\
  and\ \bibinfo {author} {\bibfnamefont {S.~A.}\ \bibnamefont {Grigera}},\
  }\bibfield  {title} {\bibinfo {title} {{Charge Ordering in a Pure Spin Model:
  Dipolar Spin Ice}},\ }\href@noop {} {\bibfield  {journal} {\bibinfo
  {journal} {Physical Review Letters}\ }\textbf {\bibinfo {volume} {111}},\
  \bibinfo {pages} {147204} (\bibinfo {year} {2013})}\BibitemShut {NoStop}%
\bibitem [{\citenamefont {Kosterlitz}\ and\ \citenamefont
  {Thouless}(1973)}]{Kosterlitz1973}%
  \BibitemOpen
  \bibfield  {author} {\bibinfo {author} {\bibfnamefont {J.~M.}\ \bibnamefont
  {Kosterlitz}}\ and\ \bibinfo {author} {\bibfnamefont {D.~J.}\ \bibnamefont
  {Thouless}},\ }\bibfield  {title} {\bibinfo {title} {Ordering, metastability
  and phase transitions in two-dimensional systems},\ }\href
  {https://doi.org/10.1088/0022-3719/6/7/010} {\bibfield  {journal} {\bibinfo
  {journal} {Journal of Physics C: Solid State Physics}\ }\textbf {\bibinfo
  {volume} {6}},\ \bibinfo {pages} {1181} (\bibinfo {year} {1973})}\BibitemShut
  {NoStop}%
\bibitem [{\citenamefont {Isakov}\ \emph {et~al.}(2004)\citenamefont {Isakov},
  \citenamefont {Raman}, \citenamefont {Moessner},\ and\ \citenamefont
  {Sondhi}}]{Isakov2004PRB}%
  \BibitemOpen
  \bibfield  {author} {\bibinfo {author} {\bibfnamefont {S.~V.}\ \bibnamefont
  {Isakov}}, \bibinfo {author} {\bibfnamefont {K.~S.}\ \bibnamefont {Raman}},
  \bibinfo {author} {\bibfnamefont {R.}~\bibnamefont {Moessner}},\ and\
  \bibinfo {author} {\bibfnamefont {S.~L.}\ \bibnamefont {Sondhi}},\ }\bibfield
   {title} {\bibinfo {title} {Magnetization curve of spin ice in a [111]
  magnetic field},\ }\href {https://doi.org/10.1103/PhysRevB.70.104418}
  {\bibfield  {journal} {\bibinfo  {journal} {Phys. Rev. B}\ }\textbf {\bibinfo
  {volume} {70}},\ \bibinfo {pages} {104418} (\bibinfo {year}
  {2004})}\BibitemShut {NoStop}%
\bibitem [{\citenamefont {Jaubert}\ \emph {et~al.}(2017)\citenamefont
  {Jaubert}, \citenamefont {Lin}, \citenamefont {Opel}, \citenamefont
  {Holdsworth},\ and\ \citenamefont {Gingras}}]{Jaubert2017}%
  \BibitemOpen
  \bibfield  {author} {\bibinfo {author} {\bibfnamefont {L.~D.~C.}\
  \bibnamefont {Jaubert}}, \bibinfo {author} {\bibfnamefont {T.}~\bibnamefont
  {Lin}}, \bibinfo {author} {\bibfnamefont {T.~S.}\ \bibnamefont {Opel}},
  \bibinfo {author} {\bibfnamefont {P.~C.~W.}\ \bibnamefont {Holdsworth}},\
  and\ \bibinfo {author} {\bibfnamefont {M.~J.~P.}\ \bibnamefont {Gingras}},\
  }\bibfield  {title} {\bibinfo {title} {Spin ice thin film: Surface ordering,
  emergent square ice, and strain effects},\ }\href
  {https://doi.org/10.1103/PhysRevLett.118.207206} {\bibfield  {journal}
  {\bibinfo  {journal} {Phys. Rev. Lett.}\ }\textbf {\bibinfo {volume} {118}},\
  \bibinfo {pages} {207206} (\bibinfo {year} {2017})}\BibitemShut {NoStop}%
\bibitem [{\citenamefont {Revell}\ \emph {et~al.}(2013)\citenamefont {Revell},
  \citenamefont {Yaraskavitch}, \citenamefont {Mason}, \citenamefont {Ross},
  \citenamefont {Noad}, \citenamefont {Dabkowska}, \citenamefont {Gaulin},
  \citenamefont {Henelius},\ and\ \citenamefont {Kycia}}]{Revell2013}%
  \BibitemOpen
  \bibfield  {author} {\bibinfo {author} {\bibfnamefont {H.~M.}\ \bibnamefont
  {Revell}}, \bibinfo {author} {\bibfnamefont {L.~R.}\ \bibnamefont
  {Yaraskavitch}}, \bibinfo {author} {\bibfnamefont {J.~D.}\ \bibnamefont
  {Mason}}, \bibinfo {author} {\bibfnamefont {K.~A.}\ \bibnamefont {Ross}},
  \bibinfo {author} {\bibfnamefont {H.~M.~L.}\ \bibnamefont {Noad}}, \bibinfo
  {author} {\bibfnamefont {H.~A.}\ \bibnamefont {Dabkowska}}, \bibinfo {author}
  {\bibfnamefont {B.~D.}\ \bibnamefont {Gaulin}}, \bibinfo {author}
  {\bibfnamefont {P.}~\bibnamefont {Henelius}},\ and\ \bibinfo {author}
  {\bibfnamefont {J.~B.}\ \bibnamefont {Kycia}},\ }\bibfield  {title} {\bibinfo
  {title} {Evidence of impurity and boundary effects on magnetic monopole
  dynamics in spin ice},\ }\href {https://doi.org/10.1038/nphys2466} {\bibfield
   {journal} {\bibinfo  {journal} {Nature Physics}\ }\textbf {\bibinfo {volume}
  {9}},\ \bibinfo {pages} {34} (\bibinfo {year} {2013})}\BibitemShut {NoStop}%
\bibitem [{\citenamefont {Pili}\ \emph {et~al.}(2021)\citenamefont {Pili},
  \citenamefont {Grigera}, \citenamefont {Borzi}, \citenamefont {Steppke},
  \citenamefont {Hicks},\ and\ \citenamefont {Mackenzie}}]{Pili2021}%
  \BibitemOpen
  \bibfield  {author} {\bibinfo {author} {\bibfnamefont {L.}~\bibnamefont
  {Pili}}, \bibinfo {author} {\bibfnamefont {S.~A.}\ \bibnamefont {Grigera}},
  \bibinfo {author} {\bibfnamefont {R.~A.}\ \bibnamefont {Borzi}}, \bibinfo
  {author} {\bibfnamefont {A.}~\bibnamefont {Steppke}}, \bibinfo {author}
  {\bibfnamefont {C.}~\bibnamefont {Hicks}},\ and\ \bibinfo {author}
  {\bibfnamefont {A.~P.}\ \bibnamefont {Mackenzie}},\ }\href@noop {} {\bibfield
   {journal} {\bibinfo  {journal} {in preparation}\ } (\bibinfo {year}
  {2021})}\BibitemShut {NoStop}%
\bibitem [{\citenamefont {Hamp}\ \emph {et~al.}(2015)\citenamefont {Hamp},
  \citenamefont {Chandran}, \citenamefont {Moessner},\ and\ \citenamefont
  {Castelnovo}}]{Hamp:2015tf}%
  \BibitemOpen
  \bibfield  {author} {\bibinfo {author} {\bibfnamefont {J.}~\bibnamefont
  {Hamp}}, \bibinfo {author} {\bibfnamefont {A.}~\bibnamefont {Chandran}},
  \bibinfo {author} {\bibfnamefont {R.}~\bibnamefont {Moessner}},\ and\
  \bibinfo {author} {\bibfnamefont {C.}~\bibnamefont {Castelnovo}},\ }\bibfield
   {title} {\bibinfo {title} {{Emergent Coulombic criticality and Kibble-Zurek
  scaling in a topological magnet}},\ }\href@noop {} {\bibfield  {journal}
  {\bibinfo  {journal} {Physical Review B}\ }\textbf {\bibinfo {volume} {92}},\
  \bibinfo {pages} {075142} (\bibinfo {year} {2015})}\BibitemShut {NoStop}%
\bibitem [{\citenamefont {Gao}\ \emph {et~al.}(2018)\citenamefont {Gao},
  \citenamefont {Zaharko}, \citenamefont {Tsurkan}, \citenamefont {Prodan},
  \citenamefont {Riordan}, \citenamefont {Lago}, \citenamefont {Fak},
  \citenamefont {Wildes}, \citenamefont {Koza}, \citenamefont {Ritter},
  \citenamefont {Fouquet}, \citenamefont {Keller}, \citenamefont {Can{\'e}vet},
  \citenamefont {Medarde}, \citenamefont {Blomgren}, \citenamefont {Johansson},
  \citenamefont {Giblin}, \citenamefont {Vrtnik}, \citenamefont {Luzar},
  \citenamefont {Loidl}, \citenamefont {R{\"u}egg},\ and\ \citenamefont
  {Fennell}}]{Gao:2018cu}%
  \BibitemOpen
  \bibfield  {author} {\bibinfo {author} {\bibfnamefont {S.}~\bibnamefont
  {Gao}}, \bibinfo {author} {\bibfnamefont {O.}~\bibnamefont {Zaharko}},
  \bibinfo {author} {\bibfnamefont {V.}~\bibnamefont {Tsurkan}}, \bibinfo
  {author} {\bibfnamefont {L.}~\bibnamefont {Prodan}}, \bibinfo {author}
  {\bibfnamefont {E.}~\bibnamefont {Riordan}}, \bibinfo {author} {\bibfnamefont
  {J.}~\bibnamefont {Lago}}, \bibinfo {author} {\bibfnamefont {B.}~\bibnamefont
  {Fak}}, \bibinfo {author} {\bibfnamefont {A.~R.}\ \bibnamefont {Wildes}},
  \bibinfo {author} {\bibfnamefont {M.~M.}\ \bibnamefont {Koza}}, \bibinfo
  {author} {\bibfnamefont {C.}~\bibnamefont {Ritter}}, \bibinfo {author}
  {\bibfnamefont {P.}~\bibnamefont {Fouquet}}, \bibinfo {author} {\bibfnamefont
  {L.}~\bibnamefont {Keller}}, \bibinfo {author} {\bibfnamefont
  {E.}~\bibnamefont {Can{\'e}vet}}, \bibinfo {author} {\bibfnamefont
  {M.}~\bibnamefont {Medarde}}, \bibinfo {author} {\bibfnamefont
  {J.}~\bibnamefont {Blomgren}}, \bibinfo {author} {\bibfnamefont
  {C.}~\bibnamefont {Johansson}}, \bibinfo {author} {\bibfnamefont {S.~R.}\
  \bibnamefont {Giblin}}, \bibinfo {author} {\bibfnamefont {S.}~\bibnamefont
  {Vrtnik}}, \bibinfo {author} {\bibfnamefont {J.}~\bibnamefont {Luzar}},
  \bibinfo {author} {\bibfnamefont {A.}~\bibnamefont {Loidl}}, \bibinfo
  {author} {\bibfnamefont {C.}~\bibnamefont {R{\"u}egg}},\ and\ \bibinfo
  {author} {\bibfnamefont {T.}~\bibnamefont {Fennell}},\ }\href@noop {}
  {\bibfield  {journal} {\bibinfo  {journal} {Physical Review Letters}\
  }\textbf {\bibinfo {volume} {120}},\ \bibinfo {pages} {137201} (\bibinfo
  {year} {2018})}\BibitemShut {NoStop}%
\bibitem [{\citenamefont {Hermele}\ \emph {et~al.}(2004)\citenamefont
  {Hermele}, \citenamefont {Fisher},\ and\ \citenamefont
  {Balents}}]{Hermele:2004gg}%
  \BibitemOpen
  \bibfield  {author} {\bibinfo {author} {\bibfnamefont {M.}~\bibnamefont
  {Hermele}}, \bibinfo {author} {\bibfnamefont {M.~P.~A.}\ \bibnamefont
  {Fisher}},\ and\ \bibinfo {author} {\bibfnamefont {L.}~\bibnamefont
  {Balents}},\ }\bibfield  {title} {\bibinfo {title} {{Pyrochlore photons: The
  U(1) spin liquid in a S=12 three-dimensional frustrated magnet}},\
  }\href@noop {} {\bibfield  {journal} {\bibinfo  {journal} {Physical Review
  B}\ }\textbf {\bibinfo {volume} {69}},\ \bibinfo {pages} {064404} (\bibinfo
  {year} {2004})}\BibitemShut {NoStop}%
\bibitem [{\citenamefont {Sibille}\ \emph {et~al.}(2016)\citenamefont
  {Sibille}, \citenamefont {Lhotel}, \citenamefont {Hatnean}, \citenamefont
  {Balakrishnan}, \citenamefont {F{\aa}k}, \citenamefont {Gauthier},
  \citenamefont {Fennell},\ and\ \citenamefont {Kenzelmann}}]{Sibille:2016bd}%
  \BibitemOpen
  \bibfield  {author} {\bibinfo {author} {\bibfnamefont {R.}~\bibnamefont
  {Sibille}}, \bibinfo {author} {\bibfnamefont {E.}~\bibnamefont {Lhotel}},
  \bibinfo {author} {\bibfnamefont {M.~C.}\ \bibnamefont {Hatnean}}, \bibinfo
  {author} {\bibfnamefont {G.}~\bibnamefont {Balakrishnan}}, \bibinfo {author}
  {\bibfnamefont {B.}~\bibnamefont {F{\aa}k}}, \bibinfo {author} {\bibfnamefont
  {N.}~\bibnamefont {Gauthier}}, \bibinfo {author} {\bibfnamefont
  {T.}~\bibnamefont {Fennell}},\ and\ \bibinfo {author} {\bibfnamefont
  {M.}~\bibnamefont {Kenzelmann}},\ }\bibfield  {title} {\bibinfo {title}
  {{Candidate quantum spin ice in the pyrochlore Pr2Hf2O7}},\ }\href@noop {}
  {\bibfield  {journal} {\bibinfo  {journal} {Physical Review B}\ }\textbf
  {\bibinfo {volume} {94}},\ \bibinfo {pages} {024436} (\bibinfo {year}
  {2016})}\BibitemShut {NoStop}%
\bibitem [{\citenamefont {Sibille}\ \emph {et~al.}(2018)\citenamefont
  {Sibille}, \citenamefont {Gauthier}, \citenamefont {Yan}, \citenamefont
  {Hatnean}, \citenamefont {Ollivier}, \citenamefont {Winn}, \citenamefont
  {Filges}, \citenamefont {Balakrishnan}, \citenamefont {Kenzelmann},
  \citenamefont {Shannon},\ and\ \citenamefont {Fennell}}]{Sibille:2018hca}%
  \BibitemOpen
  \bibfield  {author} {\bibinfo {author} {\bibfnamefont {R.}~\bibnamefont
  {Sibille}}, \bibinfo {author} {\bibfnamefont {N.}~\bibnamefont {Gauthier}},
  \bibinfo {author} {\bibfnamefont {H.}~\bibnamefont {Yan}}, \bibinfo {author}
  {\bibfnamefont {M.~C.}\ \bibnamefont {Hatnean}}, \bibinfo {author}
  {\bibfnamefont {J.}~\bibnamefont {Ollivier}}, \bibinfo {author}
  {\bibfnamefont {B.}~\bibnamefont {Winn}}, \bibinfo {author} {\bibfnamefont
  {U.}~\bibnamefont {Filges}}, \bibinfo {author} {\bibfnamefont
  {G.}~\bibnamefont {Balakrishnan}}, \bibinfo {author} {\bibfnamefont
  {M.}~\bibnamefont {Kenzelmann}}, \bibinfo {author} {\bibfnamefont
  {N.}~\bibnamefont {Shannon}},\ and\ \bibinfo {author} {\bibfnamefont
  {T.}~\bibnamefont {Fennell}},\ }\bibfield  {title} {\bibinfo {title}
  {{Experimental signatures of emergent quantum electrodynamics in Pr2Hf2O7}},\
  }\href@noop {} {\bibfield  {journal} {\bibinfo  {journal} {Nature Physics}\
  }\textbf {\bibinfo {volume} {14}},\ \bibinfo {pages} {711} (\bibinfo {year}
  {2018})}\BibitemShut {NoStop}%
\bibitem [{\citenamefont {Paddison}\ \emph {et~al.}(2016)\citenamefont
  {Paddison}, \citenamefont {Ong}, \citenamefont {Hamp}, \citenamefont
  {Mukherjee}, \citenamefont {Bai}, \citenamefont {Tucker}, \citenamefont
  {Butch}, \citenamefont {Castelnovo}, \citenamefont {Mourigal},\ and\
  \citenamefont {Dutton}}]{Paddison:2016bq}%
  \BibitemOpen
  \bibfield  {author} {\bibinfo {author} {\bibfnamefont {J.~A.~M.}\
  \bibnamefont {Paddison}}, \bibinfo {author} {\bibfnamefont {H.~S.}\
  \bibnamefont {Ong}}, \bibinfo {author} {\bibfnamefont {J.~O.}\ \bibnamefont
  {Hamp}}, \bibinfo {author} {\bibfnamefont {P.}~\bibnamefont {Mukherjee}},
  \bibinfo {author} {\bibfnamefont {X.}~\bibnamefont {Bai}}, \bibinfo {author}
  {\bibfnamefont {M.~G.}\ \bibnamefont {Tucker}}, \bibinfo {author}
  {\bibfnamefont {N.~P.}\ \bibnamefont {Butch}}, \bibinfo {author}
  {\bibfnamefont {C.}~\bibnamefont {Castelnovo}}, \bibinfo {author}
  {\bibfnamefont {M.}~\bibnamefont {Mourigal}},\ and\ \bibinfo {author}
  {\bibfnamefont {S.~E.}\ \bibnamefont {Dutton}},\ }\bibfield  {title}
  {\bibinfo {title} {{Emergent order in the kagome Ising magnet
  Dy3Mg2Sb3O14}},\ }\href@noop {} {\bibfield  {journal} {\bibinfo  {journal}
  {Nature Communications}\ }\textbf {\bibinfo {volume} {7}},\ \bibinfo {pages}
  {13842} (\bibinfo {year} {2016})}\BibitemShut {NoStop}%
\bibitem [{\citenamefont {Carrasquilla}\ \emph {et~al.}(2015)\citenamefont
  {Carrasquilla}, \citenamefont {Hao},\ and\ \citenamefont
  {Melko}}]{Carrasquilla:2015hu}%
  \BibitemOpen
  \bibfield  {author} {\bibinfo {author} {\bibfnamefont {J.}~\bibnamefont
  {Carrasquilla}}, \bibinfo {author} {\bibfnamefont {Z.}~\bibnamefont {Hao}},\
  and\ \bibinfo {author} {\bibfnamefont {R.~G.}\ \bibnamefont {Melko}},\
  }\bibfield  {title} {\bibinfo {title} {{A two-dimensional spin liquid in
  quantum kagome ice}},\ }\href@noop {} {\bibfield  {journal} {\bibinfo
  {journal} {Nature Communications}\ }\textbf {\bibinfo {volume} {6}},\
  \bibinfo {pages} {7421} (\bibinfo {year} {2015})}\BibitemShut {NoStop}%
\bibitem [{\citenamefont {Bojesen}\ and\ \citenamefont
  {Onoda}(2017)}]{Bojesen:2017gx}%
  \BibitemOpen
  \bibfield  {author} {\bibinfo {author} {\bibfnamefont {T.~A.}\ \bibnamefont
  {Bojesen}}\ and\ \bibinfo {author} {\bibfnamefont {S.}~\bibnamefont
  {Onoda}},\ }\bibfield  {title} {\bibinfo {title} {{Quantum Spin Ice under a
  [111] Magnetic Field: From Pyrochlore to Kagome}},\ }\href@noop {} {\bibfield
   {journal} {\bibinfo  {journal} {Physical Review Letters}\ }\textbf {\bibinfo
  {volume} {119}},\ \bibinfo {pages} {227204} (\bibinfo {year}
  {2017})}\BibitemShut {NoStop}%
\bibitem [{\citenamefont {Benton}\ \emph {et~al.}(2012)\citenamefont {Benton},
  \citenamefont {Sikora},\ and\ \citenamefont {Shannon}}]{Benton:2012ep}%
  \BibitemOpen
  \bibfield  {author} {\bibinfo {author} {\bibfnamefont {O.}~\bibnamefont
  {Benton}}, \bibinfo {author} {\bibfnamefont {O.}~\bibnamefont {Sikora}},\
  and\ \bibinfo {author} {\bibfnamefont {N.}~\bibnamefont {Shannon}},\
  }\bibfield  {title} {\bibinfo {title} {{Seeing the light: Experimental
  signatures of emergent electromagnetism in a quantum spin ice}},\ }\href@noop
  {} {\bibfield  {journal} {\bibinfo  {journal} {Physical Review B}\ }\textbf
  {\bibinfo {volume} {86}},\ \bibinfo {pages} {075154} (\bibinfo {year}
  {2012})}\BibitemShut {NoStop}%
\bibitem [{\citenamefont {Gingras}\ and\ \citenamefont
  {McClarty}(2014)}]{Gingras:2014ip}%
  \BibitemOpen
  \bibfield  {author} {\bibinfo {author} {\bibfnamefont {M.~J.~P.}\
  \bibnamefont {Gingras}}\ and\ \bibinfo {author} {\bibfnamefont {P.~A.}\
  \bibnamefont {McClarty}},\ }\bibfield  {title} {\bibinfo {title} {{Quantum
  spin ice: a search for gapless quantum spin liquids in pyrochlore magnets}},\
  }\href@noop {} {\bibfield  {journal} {\bibinfo  {journal} {Reports on
  progress in physics}\ }\textbf {\bibinfo {volume} {77}},\ \bibinfo {pages}
  {056501} (\bibinfo {year} {2014})}\BibitemShut {NoStop}%
\bibitem [{\citenamefont {Bramwell}\ \emph {et~al.}(2001)\citenamefont
  {Bramwell}, \citenamefont {Harris}, \citenamefont {denHertog}, \citenamefont
  {Gingras}, \citenamefont {Gardner}, \citenamefont {McMorrow}, \citenamefont
  {Wildes}, \citenamefont {Cornelius}, \citenamefont {Champion}, \citenamefont
  {Melko.},\ and\ \citenamefont {Fennell.}}]{Bramwell2001PRB}%
  \BibitemOpen
  \bibfield  {author} {\bibinfo {author} {\bibfnamefont {S.~T.}\ \bibnamefont
  {Bramwell}}, \bibinfo {author} {\bibfnamefont {M.~J.}\ \bibnamefont
  {Harris}}, \bibinfo {author} {\bibfnamefont {B.~C.}\ \bibnamefont
  {denHertog}}, \bibinfo {author} {\bibfnamefont {M.~J.~P.}\ \bibnamefont
  {Gingras}}, \bibinfo {author} {\bibfnamefont {J.~S.}\ \bibnamefont
  {Gardner}}, \bibinfo {author} {\bibfnamefont {D.~F.}\ \bibnamefont
  {McMorrow}}, \bibinfo {author} {\bibfnamefont {A.~R.}\ \bibnamefont
  {Wildes}}, \bibinfo {author} {\bibfnamefont {A.}~\bibnamefont {Cornelius}},
  \bibinfo {author} {\bibfnamefont {J.~D.~M.}\ \bibnamefont {Champion}},
  \bibinfo {author} {\bibfnamefont {R.~G.}\ \bibnamefont {Melko.}},\ and\
  \bibinfo {author} {\bibfnamefont {T.}~\bibnamefont {Fennell.}},\ }\bibfield
  {title} {\bibinfo {title} {Spin correlations in ho$_2$ti$_2$o$_7$: A dipolar
  spin ice system},\ }\href@noop {} {\bibfield  {journal} {\bibinfo  {journal}
  {Phys. Rev. Lett.}\ }\textbf {\bibinfo {volume} {87}},\ \bibinfo {pages}
  {047205} (\bibinfo {year} {2001})}\BibitemShut {NoStop}%
\bibitem [{\citenamefont {Fennell}\ \emph {et~al.}(2004)\citenamefont
  {Fennell}, \citenamefont {Petrenko}, \citenamefont {Fak}, \citenamefont
  {Bramwell}, \citenamefont {Enjalran}, \citenamefont
  {Yavors{\textquoteright}kii}, \citenamefont {Gingras}, \citenamefont
  {Melko},\ and\ \citenamefont {Balakrishnan}}]{Fennell:732176}%
  \BibitemOpen
  \bibfield  {author} {\bibinfo {author} {\bibfnamefont {T.}~\bibnamefont
  {Fennell}}, \bibinfo {author} {\bibfnamefont {O.~A.}\ \bibnamefont
  {Petrenko}}, \bibinfo {author} {\bibfnamefont {B.}~\bibnamefont {Fak}},
  \bibinfo {author} {\bibfnamefont {S.~T.}\ \bibnamefont {Bramwell}}, \bibinfo
  {author} {\bibfnamefont {M.}~\bibnamefont {Enjalran}}, \bibinfo {author}
  {\bibfnamefont {T.}~\bibnamefont {Yavors{\textquoteright}kii}}, \bibinfo
  {author} {\bibfnamefont {M.~J.~P.}\ \bibnamefont {Gingras}}, \bibinfo
  {author} {\bibfnamefont {R.~G.}\ \bibnamefont {Melko}},\ and\ \bibinfo
  {author} {\bibfnamefont {G.}~\bibnamefont {Balakrishnan}},\ }\bibfield
  {title} {\bibinfo {title} {{Neutron Scattering Investigation of the Spin Ice
  State in Dy2Ti2O7}},\ }\href@noop {} {\bibfield  {journal} {\bibinfo
  {journal} {Physical Review B}\ }\textbf {\bibinfo {volume} {70}},\ \bibinfo
  {pages} {134408} (\bibinfo {year} {2004})}\BibitemShut {NoStop}%
\bibitem [{\citenamefont {Henelius}\ \emph {et~al.}(2016)\citenamefont
  {Henelius}, \citenamefont {Lin}, \citenamefont {Enjalran}, \citenamefont
  {Hao}, \citenamefont {Rau}, \citenamefont {Altosaar}, \citenamefont
  {Flicker}, \citenamefont {Yavors{\textquoteright}kii},\ and\ \citenamefont
  {Gingras}}]{Henelius:2016ew}%
  \BibitemOpen
  \bibfield  {author} {\bibinfo {author} {\bibfnamefont {P.}~\bibnamefont
  {Henelius}}, \bibinfo {author} {\bibfnamefont {T.}~\bibnamefont {Lin}},
  \bibinfo {author} {\bibfnamefont {M.}~\bibnamefont {Enjalran}}, \bibinfo
  {author} {\bibfnamefont {Z.}~\bibnamefont {Hao}}, \bibinfo {author}
  {\bibfnamefont {J.~G.}\ \bibnamefont {Rau}}, \bibinfo {author} {\bibfnamefont
  {J.}~\bibnamefont {Altosaar}}, \bibinfo {author} {\bibfnamefont
  {F.}~\bibnamefont {Flicker}}, \bibinfo {author} {\bibfnamefont
  {T.}~\bibnamefont {Yavors{\textquoteright}kii}},\ and\ \bibinfo {author}
  {\bibfnamefont {M.~J.~P.}\ \bibnamefont {Gingras}},\ }\bibfield  {title}
  {\bibinfo {title} {{Refrustration and competing orders in the prototypical
  Dy2Ti2O7spin ice material}},\ }\href@noop {} {\bibfield  {journal} {\bibinfo
  {journal} {Physical Review B}\ }\textbf {\bibinfo {volume} {93}},\ \bibinfo
  {pages} {024402} (\bibinfo {year} {2016})}\BibitemShut {NoStop}%
\bibitem [{\citenamefont {Giblin}\ \emph {et~al.}(2018)\citenamefont {Giblin},
  \citenamefont {Twengstr{\"o}m}, \citenamefont {Bovo}, \citenamefont {Ruminy},
  \citenamefont {Bartkowiak}, \citenamefont {Manuel}, \citenamefont {Andresen},
  \citenamefont {Prabhakaran}, \citenamefont {Balakrishnan}, \citenamefont
  {Pomjakushina}, \citenamefont {Paulsen}, \citenamefont {Lhotel},
  \citenamefont {Keller}, \citenamefont {Frontzek}, \citenamefont {Capelli},
  \citenamefont {Zaharko}, \citenamefont {McClarty}, \citenamefont {Bramwell},
  \citenamefont {Henelius},\ and\ \citenamefont {Fennell}}]{Giblin:2018cz}%
  \BibitemOpen
  \bibfield  {author} {\bibinfo {author} {\bibfnamefont {S.~R.}\ \bibnamefont
  {Giblin}}, \bibinfo {author} {\bibfnamefont {M.}~\bibnamefont
  {Twengstr{\"o}m}}, \bibinfo {author} {\bibfnamefont {L.}~\bibnamefont
  {Bovo}}, \bibinfo {author} {\bibfnamefont {M.}~\bibnamefont {Ruminy}},
  \bibinfo {author} {\bibfnamefont {M.}~\bibnamefont {Bartkowiak}}, \bibinfo
  {author} {\bibfnamefont {P.}~\bibnamefont {Manuel}}, \bibinfo {author}
  {\bibfnamefont {J.~C.}\ \bibnamefont {Andresen}}, \bibinfo {author}
  {\bibfnamefont {D.}~\bibnamefont {Prabhakaran}}, \bibinfo {author}
  {\bibfnamefont {G.}~\bibnamefont {Balakrishnan}}, \bibinfo {author}
  {\bibfnamefont {E.}~\bibnamefont {Pomjakushina}}, \bibinfo {author}
  {\bibfnamefont {C.}~\bibnamefont {Paulsen}}, \bibinfo {author} {\bibfnamefont
  {E.}~\bibnamefont {Lhotel}}, \bibinfo {author} {\bibfnamefont
  {L.}~\bibnamefont {Keller}}, \bibinfo {author} {\bibfnamefont
  {M.}~\bibnamefont {Frontzek}}, \bibinfo {author} {\bibfnamefont {S.~C.}\
  \bibnamefont {Capelli}}, \bibinfo {author} {\bibfnamefont {O.}~\bibnamefont
  {Zaharko}}, \bibinfo {author} {\bibfnamefont {P.~A.}\ \bibnamefont
  {McClarty}}, \bibinfo {author} {\bibfnamefont {S.~T.}\ \bibnamefont
  {Bramwell}}, \bibinfo {author} {\bibfnamefont {P.}~\bibnamefont {Henelius}},\
  and\ \bibinfo {author} {\bibfnamefont {T.}~\bibnamefont {Fennell}},\
  }\bibfield  {title} {\bibinfo {title} {{Pauling Entropy, Metastability, and
  Equilibrium in Dy$_2$Ti$_2$O$_7$ Spin Ice}},\ }\href@noop {} {\bibfield
  {journal} {\bibinfo  {journal} {Physical Review Letters}\ }\textbf {\bibinfo
  {volume} {121}},\ \bibinfo {pages} {067202} (\bibinfo {year}
  {2018})}\BibitemShut {NoStop}%
\bibitem [{\citenamefont {Twengstr{\"o}m}\ \emph {et~al.}(2020)\citenamefont
  {Twengstr{\"o}m}, \citenamefont {Henelius},\ and\ \citenamefont
  {Bramwell}}]{Twengstrom:2020jv}%
  \BibitemOpen
  \bibfield  {author} {\bibinfo {author} {\bibfnamefont {M.}~\bibnamefont
  {Twengstr{\"o}m}}, \bibinfo {author} {\bibfnamefont {P.}~\bibnamefont
  {Henelius}},\ and\ \bibinfo {author} {\bibfnamefont {S.~T.}\ \bibnamefont
  {Bramwell}},\ }\bibfield  {title} {\bibinfo {title} {{Screening and the pinch
  point paradox in spin ice}},\ }\href@noop {} {\bibfield  {journal} {\bibinfo
  {journal} {Physical Review Research}\ }\textbf {\bibinfo {volume} {2}},\
  \bibinfo {pages} {013305} (\bibinfo {year} {2020})}\BibitemShut {NoStop}%
\bibitem [{\citenamefont {Turrini}(2020)}]{turrinithesis}%
  \BibitemOpen
  \bibfield  {author} {\bibinfo {author} {\bibfnamefont {A.~A.}\ \bibnamefont
  {Turrini}},\ }\href@noop {} {\emph {\bibinfo {title} {{Thermodynamic Behavior
  of Rare Earth Pyrochlores}}}}\ (\bibinfo  {publisher} {Universit{\'e} de
  Gen{\`e}ve},\ \bibinfo {year} {2020})\BibitemShut {NoStop}%
\bibitem [{\citenamefont {Wills}\ \emph {et~al.}(2002)\citenamefont {Wills},
  \citenamefont {Ballou},\ and\ \citenamefont {Lacroix}}]{Wills2002}%
  \BibitemOpen
  \bibfield  {author} {\bibinfo {author} {\bibfnamefont {A.~S.}\ \bibnamefont
  {Wills}}, \bibinfo {author} {\bibfnamefont {R.}~\bibnamefont {Ballou}},\ and\
  \bibinfo {author} {\bibfnamefont {C.}~\bibnamefont {Lacroix}},\ }\bibfield
  {title} {\bibinfo {title} {Model of localized highly frustrated
  ferromagnetism: The kagom\'e spin ice'},\ }\href@noop {} {\bibfield
  {journal} {\bibinfo  {journal} {Phys. Rev. B}\ }\textbf {\bibinfo {volume}
  {66}},\ \bibinfo {pages} {144407} (\bibinfo {year} {2002})}\BibitemShut
  {NoStop}%
\bibitem [{\citenamefont {Zhang}\ \emph {et~al.}(2013)\citenamefont {Zhang},
  \citenamefont {Gilbert}, \citenamefont {Nisoli}, \citenamefont {Chern},
  \citenamefont {Erickson}, \citenamefont {O’Brien}, \citenamefont
  {Leighton}, \citenamefont {Lammert}, \citenamefont {Crespi},\ and\
  \citenamefont {Schiffer}}]{Zhang2013}%
  \BibitemOpen
  \bibfield  {author} {\bibinfo {author} {\bibfnamefont {S.}~\bibnamefont
  {Zhang}}, \bibinfo {author} {\bibfnamefont {I.}~\bibnamefont {Gilbert}},
  \bibinfo {author} {\bibfnamefont {C.}~\bibnamefont {Nisoli}}, \bibinfo
  {author} {\bibfnamefont {G.-W.}\ \bibnamefont {Chern}}, \bibinfo {author}
  {\bibfnamefont {M.~J.}\ \bibnamefont {Erickson}}, \bibinfo {author}
  {\bibfnamefont {L.}~\bibnamefont {O’Brien}}, \bibinfo {author}
  {\bibfnamefont {C.}~\bibnamefont {Leighton}}, \bibinfo {author}
  {\bibfnamefont {P.~E.}\ \bibnamefont {Lammert}}, \bibinfo {author}
  {\bibfnamefont {V.~H.}\ \bibnamefont {Crespi}},\ and\ \bibinfo {author}
  {\bibfnamefont {P.}~\bibnamefont {Schiffer}},\ }\bibfield  {title} {\bibinfo
  {title} {Crystallites of magnetic charges in artificial spin ice},\ }\href
  {https://doi.org/10.1038/nature12399} {\bibfield  {journal} {\bibinfo
  {journal} {Nature}\ }\textbf {\bibinfo {volume} {500}},\ \bibinfo {pages}
  {553} (\bibinfo {year} {2013})}\BibitemShut {NoStop}%
\bibitem [{\citenamefont {Mengotti}\ \emph {et~al.}(2010)\citenamefont
  {Mengotti}, \citenamefont {Heyderman}, \citenamefont {Rodr{\'\i}guez},
  \citenamefont {Nolting}, \citenamefont {H{\"u}gli},\ and\ \citenamefont
  {Braun}}]{Mengotti:2010fo}%
  \BibitemOpen
  \bibfield  {author} {\bibinfo {author} {\bibfnamefont {E.}~\bibnamefont
  {Mengotti}}, \bibinfo {author} {\bibfnamefont {L.~J.}\ \bibnamefont
  {Heyderman}}, \bibinfo {author} {\bibfnamefont {A.~F.}\ \bibnamefont
  {Rodr{\'\i}guez}}, \bibinfo {author} {\bibfnamefont {F.}~\bibnamefont
  {Nolting}}, \bibinfo {author} {\bibfnamefont {R.~V.}\ \bibnamefont
  {H{\"u}gli}},\ and\ \bibinfo {author} {\bibfnamefont {H.-B.}\ \bibnamefont
  {Braun}},\ }\bibfield  {title} {\bibinfo {title} {{Real-space observation of
  emergent magnetic monopoles and associated Dirac strings in artificial kagome
  spin ice}},\ }\href@noop {} {\bibfield  {journal} {\bibinfo  {journal}
  {Nature Physics}\ }\textbf {\bibinfo {volume} {7}},\ \bibinfo {pages} {68}
  (\bibinfo {year} {2010})}\BibitemShut {NoStop}%
\bibitem [{\citenamefont {M\"oller}\ and\ \citenamefont
  {Moessner}(2006)}]{MoellerPRL2006}%
  \BibitemOpen
  \bibfield  {author} {\bibinfo {author} {\bibfnamefont {G.}~\bibnamefont
  {M\"oller}}\ and\ \bibinfo {author} {\bibfnamefont {R.}~\bibnamefont
  {Moessner}},\ }\bibfield  {title} {\bibinfo {title} {Artificial {Square}
  {Ice} and {Related} {Dipolar} {Nanoarrays}},\ }\href
  {https://doi.org/10.1103/PhysRevLett.96.237202} {\bibfield  {journal}
  {\bibinfo  {journal} {Phys. Rev. Lett.}\ }\textbf {\bibinfo {volume} {96}},\
  \bibinfo {pages} {237202} (\bibinfo {year} {2006})}\BibitemShut {NoStop}%
\bibitem [{\citenamefont {Sendetskyi}\ \emph {et~al.}(2019)\citenamefont
  {Sendetskyi}, \citenamefont {Scagnoli}, \citenamefont {Leo}, \citenamefont
  {Anghinolfi}, \citenamefont {Alberca}, \citenamefont {Luning}, \citenamefont
  {Staub}, \citenamefont {Derlet},\ and\ \citenamefont
  {Heyderman}}]{Sendetskyi2019}%
  \BibitemOpen
  \bibfield  {author} {\bibinfo {author} {\bibfnamefont {O.}~\bibnamefont
  {Sendetskyi}}, \bibinfo {author} {\bibfnamefont {V.}~\bibnamefont
  {Scagnoli}}, \bibinfo {author} {\bibfnamefont {N.}~\bibnamefont {Leo}},
  \bibinfo {author} {\bibfnamefont {L.}~\bibnamefont {Anghinolfi}}, \bibinfo
  {author} {\bibfnamefont {A.}~\bibnamefont {Alberca}}, \bibinfo {author}
  {\bibfnamefont {J.}~\bibnamefont {Luning}}, \bibinfo {author} {\bibfnamefont
  {U.}~\bibnamefont {Staub}}, \bibinfo {author} {\bibfnamefont {P.~M.}\
  \bibnamefont {Derlet}},\ and\ \bibinfo {author} {\bibfnamefont {L.~J.}\
  \bibnamefont {Heyderman}},\ }\bibfield  {title} {\bibinfo {title} {Continuous
  magnetic phase transition in artificial square ice},\ }\href@noop {}
  {\bibfield  {journal} {\bibinfo  {journal} {Physical Review B}\ }\textbf
  {\bibinfo {volume} {99}},\ \bibinfo {pages} {214430} (\bibinfo {year}
  {2019})}\BibitemShut {NoStop}%
\bibitem [{\citenamefont {denHertog}\ and\ \citenamefont
  {Gingras}(2000)}]{denHertog2000}%
  \BibitemOpen
  \bibfield  {author} {\bibinfo {author} {\bibfnamefont {B.~C.}\ \bibnamefont
  {denHertog}}\ and\ \bibinfo {author} {\bibfnamefont {M.~J.~P.}\ \bibnamefont
  {Gingras}},\ }\bibfield  {title} {\bibinfo {title} {Dipolar interactions and
  origin of spin ice in ising pyrochlore magnets},\ }\href@noop {} {\bibfield
  {journal} {\bibinfo  {journal} {Phys. Rev. Lett.}\ }\textbf {\bibinfo
  {volume} {84}},\ \bibinfo {pages} {3430} (\bibinfo {year}
  {2000})}\BibitemShut {NoStop}%
\bibitem [{\citenamefont {Das}\ \emph {et~al.}(2006)\citenamefont {Das},
  \citenamefont {Horbach}, \citenamefont {Binder}, \citenamefont {Fisher},\
  and\ \citenamefont {Sengers}}]{Das06}%
  \BibitemOpen
  \bibfield  {author} {\bibinfo {author} {\bibfnamefont {S.~K.}\ \bibnamefont
  {Das}}, \bibinfo {author} {\bibfnamefont {J.}~\bibnamefont {Horbach}},
  \bibinfo {author} {\bibfnamefont {K.}~\bibnamefont {Binder}}, \bibinfo
  {author} {\bibfnamefont {M.~E.}\ \bibnamefont {Fisher}},\ and\ \bibinfo
  {author} {\bibfnamefont {J.~V.}\ \bibnamefont {Sengers}},\ }\href@noop {}
  {\bibfield  {journal} {\bibinfo  {journal} {The Journal of Chemical Physics}\
  }\textbf {\bibinfo {volume} {125}},\ \bibinfo {pages} {024506} (\bibinfo
  {year} {2006})}\BibitemShut {NoStop}%
\bibitem [{\citenamefont {Wu}(1968)}]{Wu}%
  \BibitemOpen
  \bibfield  {author} {\bibinfo {author} {\bibfnamefont {F.~Y.}\ \bibnamefont
  {Wu}},\ }\href@noop {} {\bibfield  {journal} {\bibinfo  {journal} {Phys.
  Rev.}\ }\textbf {\bibinfo {volume} {168}},\ \bibinfo {pages} {539} (\bibinfo
  {year} {1968})}\BibitemShut {NoStop}%
\bibitem [{\citenamefont {Watson}(1999)}]{Watson}%
  \BibitemOpen
  \bibfield  {author} {\bibinfo {author} {\bibfnamefont {G.~I.}\ \bibnamefont
  {Watson}},\ }\href@noop {} {\bibfield  {journal} {\bibinfo  {journal} {J.
  Stat. Phys.}\ }\textbf {\bibinfo {volume} {94}},\ \bibinfo {pages} {1045}
  (\bibinfo {year} {1999})}\BibitemShut {NoStop}%
\bibitem [{\citenamefont {Bezdek}(1981)}]{Bezdek1981}%
  \BibitemOpen
  \bibfield  {author} {\bibinfo {author} {\bibfnamefont {J.~C.}\ \bibnamefont
  {Bezdek}},\ }\href@noop {} {\emph {\bibinfo {title} {Pattern Recognition with
  Fuzzy Objective Function Algorithms}}}\ (\bibinfo  {publisher} {Springer
  US},\ \bibinfo {year} {1981})\BibitemShut {NoStop}%
\end{thebibliography}%

\end{document}